\documentclass[twocolumn,twocolappendix,usenatbib,useAMS,fleqn,usenames,dvipsnames, tighten]{aastex63}
\usepackage{amsmath}
\hypersetup{linkcolor=Aquamarine,citecolor=MidnightBlue,filecolor=cyan,urlcolor=MidnightBlue}
\graphicspath{{./}{plots/}}
\accepted{September 21, 2020}
\submitjournal{ApJ}

\usepackage{enumitem, amsmath,amstext, mathtools}

\newcommand{\alfven}{Alfv{\'e}n }

\newcommand{\K}{{\rm K}}
\newcommand{\tcool}{t_{\rm cool}}
\newcommand{\tff}{t_{\rm ff}}
\newcommand{\tcff}{t_{\rm cool} / t_{\rm ff}}
\newcommand{\flux}{$M_{\odot}\mathrm{kpc}^{-2}\mathrm{yr}^{-1}$}

\shorttitle{Thermal Instability with Cosmic Rays}
\shortauthors{Butsky et al.}

\begin{document}

\title{The Impact of Cosmic Rays on Thermal Instability in the Circumgalactic Medium}

\correspondingauthor{Iryna S. Butsky}
\email{ibutsky@uw.edu}

\author[0000-0003-1257-5007]{Iryna S. Butsky}
\affiliation{Astronomy Department, University of Washington, Seattle, WA 98195, USA}
\affiliation{Center for Computational Astrophysics, Flatiron Institute, 162 Fifth Avenue, New York, NY 10010, USA}
\author[0000-0003-3806-8548]{Drummond B. Fielding}
\affiliation{Center for Computational Astrophysics, Flatiron Institute, 162 Fifth Avenue, New York, NY 10010, USA}
\author[0000-0003-4073-3236]{Christopher C. Hayward}
\affiliation{Center for Computational Astrophysics, Flatiron Institute, 162 Fifth Avenue, New York, NY 10010, USA}
\author[0000-0002-3817-8133]{Cameron B. Hummels}
\affiliation{TAPIR, California Institute of Technology, Pasadena, CA 91125, USA}
\author{Thomas R. Quinn}
\author[0000-0002-0355-0134]{Jessica K. Werk}
\affiliation{Astronomy Department, University of Washington, Seattle, WA 98195, USA}


\begin{abstract}
Large reservoirs of cold ($\sim 10^4$ K) gas exist out to and beyond the virial radius in the circumgalactic medium (CGM) of all types of galaxies. Photoionization modeling suggests that cold CGM gas has significantly lower densities than expected by theoretical predictions based on thermal pressure equilibrium with hot CGM gas. In this work, we investigate the impact of cosmic ray physics on the formation of cold gas via thermal instability. We use idealized three-dimensional magnetohydrodynamic simulations to follow the evolution of thermally unstable gas in a gravitationally stratified medium. We find that cosmic ray pressure lowers the density and increases the size of cold gas clouds formed through thermal instability. We develop a simple model for how the cold cloud sizes and the relative densities of cold and hot gas depend on cosmic ray pressure. Cosmic ray pressure can help counteract gravity to keep cold gas in the CGM for longer, thereby increasing the predicted cold mass fraction and decreasing the predicted cold gas inflow rates. Efficient cosmic ray transport, by streaming or diffusion, redistributes cosmic ray pressure from the cold gas to the background medium, resulting in cold gas properties that are in-between those predicted by simulations with inefficient transport and simulations without cosmic rays. We show that cosmic rays can significantly reduce galactic accretion rates and resolve the tension between theoretical models and observational constraints on the properties of cold CGM gas.
\end{abstract}

\section{Introduction}
Observations of H I Lyman series and low-ion metal transitions in quasar absorption line spectroscopy studies indicate that cold ($\sim$10$^{4}$ K) gas exists around galaxies of all types out to $\sim$300 kpc \citep{Chen:2010, Prochaska:2011, Tumlinson:2013, Keeney:2018}. Cold circumgalactic medium (CGM) gas makes up a substantial fraction (30$-$50\%) of galactic baryons \citep{Werk:2014} and may accrete onto galaxies to fuel star formation. It is then particularly intriguing that the CGM of some quenched galaxies are observed to have massive reservoirs of cold gas \citep{Thom:2012, Berg:2019}. How these galaxies remain quenched despite these large cold gas reservoirs is uncertain. It is possible that they are about to resume star formation, that this gas is incapable of cooling to the right temperatures in the interstellar medium (ISM), or perhaps that there is some mechanism preventing this gas from accreting onto its host galaxy. Furthermore, careful ionization modeling based on the COS-Halos observations suggests that the densities of cold CGM gas are at least an order-of-magnitude too low for the cold gas to be in thermal pressure equilibrium with the theoretically predicted pressure of hot CGM gas \citep{Werk:2014}. Although the density of the hot CGM gas phase is itself poorly constrained, these observations are consistent with the presence of non-thermal pressure support in the CGM. Understanding the origin and physical properties of cold CGM gas is integral to a self-consistent theory of galaxy evolution. 

There are two families of explanations for the origins of cold CGM gas: the gas formed elsewhere and was transported to the CGM, or the gas was formed in situ \citep{Hafen:2019}.  Both of these mechanisms play a role to different degrees in different galaxies. Traditionally, transporting cold gas from the ISM to the CGM has been challenging, as the survival times of cold gas embedded in a hot medium are short \citep{McKee:1975, Zhang:2017}. However, there is a wealth of multiwavelength observations supporting the existence of multiphase outflows \citep{Heckman:2000, Chen:2010, Ashley:2020, Burchett:2020, Fleutsch:2020} and theoretical works demonstrating prolonged survival and growth of cold gas in a hot wind \citep{McCourt:2015, Thompson:2016, Wiener:2017, Schneider:2018, Gronke:2018, Fielding:2020, Li:2020}.  It remains unclear if the amount of cold gas entrained in outflows is enough to explain the observed abundance of cold CGM gas. 

Alternatively, cold gas can form directly out of a hot ($\sim 10^6$ K) medium through thermal instability \citep{Field:1965}. When the cooling time of gravitationally stratified hot gas in thermal hydrostatic equilibrium is $\lesssim 10$ times the gravitational free-fall time, small isobaric perturbations can drive runaway cooling, resulting in cold gas condensation and precipitation \citep{McCourt:2012, Sharma:2012a, Voit:2015}. Although thermal instability in gravitationally stratified gas was originally invoked to explain observations of galaxy clusters \citep{Gaspari:2012, Sharma:2012b}, thermal instability is believed to also play an integral role in producing cold CGM gas and regulating the baryon cycle \citep{Fielding:2017, Voit:2017, Voit:2019a, Esmerian:2020}.  Both cosmological and idealized simulations have explored the impact of a broad range of processes on thermal instability including thermal conduction \citep{Sharma:2010, Wagh:2014}, magnetic fields \citep{Ji:2018}, turbulence \citep{Voit:2018}, the shape of the gravitational potential \citep{Meece:2015, Choudhury:2016}, and perturbation amplitude \citep{Pizzolate:2005, Singh:2015, Choudhury:2019}. While the impact of cosmic rays has largely been ignored in simulations of thermal instability, \citet{Sharma:2010} used two-dimensional simulations to show that the presence of adiabatic cosmic rays decreases the density of cold gas that forms through thermal instability.

Cosmic rays are likely an important source of energy in the CGM. In the Milky Way ISM, cosmic ray energy is roughly in equipartition with thermal and magnetic energies \citep{Ginzburg:1985, Boulares:1990}, and many recent simulations have shown that cosmic rays launch far-reaching galactic outflows \citep{Uhlig:2012, Booth:2013, Girichidis:2016, Simpson:2016, Ruszkowski:2017, Bustard:2020, Jana:2020}, altering the phase structure of the CGM \citep{Salem:2016, Butsky:2018, Buck:2020, Ji:2020}. Galaxy-scale simulations that include cosmic rays find that cosmic ray pressure may even be the dominant source of pressure in the CGM \citep{Girichidis:2018, Ji:2020}. 

There are several ways in which cosmic rays could alter the thermal instability in CGM gas. Non-thermal cosmic ray pressure supports cold gas, enabling it to cool isochorically \citep{Sharma:2010, Kempski:2020a} and may explain the inferred low densities of cold CGM gas.  Cosmic ray pressure also counteracts gravity and may prevent cold halo gas from accreting onto the galaxy \citep{Hopkins:2020:cr_galaxy}. In some transport approximations, streaming cosmic rays transfer energy to the thermal gas and provide a source of heating. In some regimes, this additional heating term can balance radiative cooling and has been shown to prevent gas from overcooling in simulations of galaxy clusters \citep{Guo:2008, Sharma:2010, Jacob:2017a, Jacob:2017b}. Linear thermal stability analysis predicts that CGM gas ($10^5 \K < T < 10^7 \K$) is likely thermally unstable in the presence of streaming cosmic rays, but the predicted thermal instability growth rate depends on the invoked cosmic ray transport mechanism \citep{Kempski:2020a}. However, linear analysis breaks down once thermal instability saturates, and there is a strong need for simulations to further understand the cold gas properties and subsequent evolution.  

In this work, we run the first three-dimensional simulations of thermal instability in a gravitationally stratified medium that include cosmic ray pressure with diffusion and streaming.  We focus our analysis on the properties of the cold gas that forms through thermal instability and its implications for observations of cold CGM gas.
The paper is organized as follows. In Section \ref{sec:methods}, we describe the initial conditions, model assumptions, and the scope of the parameter study. In Section \ref{sec:physical_expectations}, we provide physical expectations of the impact of cosmic ray pressure, transport, and heating on thermal instability. In Section \ref{sec:results}, we present the results of our simulations with a focus on the gas density, cold mass fraction, and cold mass flux. In Section \ref{sec:discussion}, we discuss the implications of these results on cold cloud sizes, galaxy accretion rates, and the CGM pressure problem. We summarize our findings in Section \ref{sec:conclusions}. In the Appendix, we demonstrate the impact of resolution (Appendix \ref{sec:appendix_resolution}), cold gas temperature (Appendix \ref{sec:appendix_Tmin}), and halo profile (Appendix \ref{sec:appendix_profile}).

\section{Methods}\label{sec:methods}
We perform our simulations with the astrophysical simulation code, {\sc enzo} \citep{Bryan:2014, Enzo:2019}, using the magnetohydrodynamics (MHD) local Lax-Friedrichs \citep[LLF;][]{Kurganov:2000} Riemann solver. This MHD solver uses the piecewise linear reconstruction method \citep[PLM;][]{VanLeer:1977} and performs divergence cleaning of the magnetic field as described in \citet{Dedner:2002} and tested in \citet{Wang:2008}. We perform all simulations on a 3-dimensional grid with uniform resolution.  

In addition to the standard Euler equations, which define the conservation of mass, momentum, and energy in hydrodynamics simulations, our simulations also conserve the magnetic flux and cosmic ray energy. The following equations describe the evolution of the gas, magnetic field, and cosmic ray fluids.

\begin{equation}\label{eqn:mass}
\frac{\partial \rho}{\partial t} + \nabla \cdot (\rho {\bf v}) = 0
\end{equation}

\begin{equation}\label{eqn:momentum}
\frac{\partial(\rho {\bf v})}{\partial t} + \nabla 
	\cdot \bigg(\rho{\bf vv}^{\rm T} -\frac{\bf B \cdot\nabla B}{4\pi}\bigg) + \nabla P_{\rm tot} = -\rho {\bf g}
\end{equation}

\begin{equation}\label{eqn:induction}
\frac{\partial \bf B}{\partial t} + \nabla\cdot({\bf Bv^{\mathrm{T}}
- vB^{\mathrm{T}}}) = \bf 0
\end{equation}

\begin{equation}\label{eqn:energy}
\frac{\partial \varepsilon_{\rm g}}{\partial t} + \nabla\cdot ({\bf v} \varepsilon_{\rm g})
= - P_{\rm g}\nabla\cdot{\bf v} - \mathcal{L} + \mathcal{H} + {{\mathcal{H}}}_\mathrm{c}
\end{equation}

\begin{equation}\label{eqn:crenergy}
\frac{\partial \varepsilon_{\rm c}}{\partial t} + \nabla\cdot {\bf F}_\mathrm{c}
=  - P_{\rm c}\nabla\cdot{\bf v}  - {{\mathcal{H}}}_\mathrm{c}
\end{equation}

\begin{equation}\label{eqn:crtransport}
{\bf F}_\mathrm{c} = \underbracket{{\bf v\varepsilon_{\mathrm{c}}}}_{\rm advection} + \underbracket{{\bf v}_\mathrm{s}(\varepsilon_{\mathrm{c}}+P_{\mathrm{c}})}_{\rm streaming}-\underbracket{\kappa_{\rm c}{\bf \hat{b}}({\bf \hat{b}\cdot\nabla}
\varepsilon_{\mathrm{c}})}_{\rm diffusion}
\end{equation}

In the equations above, $\rho$ is the gas density, ${\bf v}$ is the gas velocity vector, ${\bf g}$ is the gravitational acceleration, and $t$ is the time variable. ${\bf B}$ is the magnetic field vector and ${\bf \hat{b}}$ is the magnetic field direction, ${\bf \hat{b}} = {\bf B} / |{\bf B}|$. $\mathcal{L}$ and $\mathcal{H}$ are the gas cooling and heating terms defined in section \ref{sec:cool_heat}. $\varepsilon_{\rm g}$ and $\varepsilon_{\rm c}$ are the gas and cosmic ray energy densities (energy per volume). $P_{\rm g} = (\gamma - 1) \varepsilon_{\rm g},\, P_B = B^2/8\pi,\, P_{\rm c} = (\gamma_{\rm c}- 1)\varepsilon_{\rm c}$ are the gas, magnetic, and cosmic ray pressures. Together, they comprise the total pressure, $P_{\rm tot} = P_{\rm g} + P_B + P_{\rm c}$. The adiabatic indices are $\gamma = 5/3$, and $\gamma_{\rm c}= 4/3$ respectively. 

${\bf F}_{\rm c}$ describes the cosmic ray flux, which encompasses advection, streaming, and diffusion. In the streaming approximation, cosmic rays move along magnetic field lines at the streaming velocity,
\begin{equation}\label{eqn:stream}
\mathrm{\bf v}_s = -sgn({\bf \hat{b}}\cdot \nabla\varepsilon_{\rm c}){\bf v_A},
\end{equation}
and heat the gas at a rate proportional to the \alfven velocity, 
\begin{equation}
{{\mathcal{H}}}_{\rm c}= |{\bf v_A} \cdot \nabla P_{\rm c}|.
\end{equation}
In the equations above, $sgn$ returns the sign of the enclosed expression, and the \alfven velocity is defined as ${\bf v_A} = {\bf B}/\sqrt{4\pi\rho}$. We note that the streaming term is always positive, so that energy is only ever transferred \textit{from} the cosmic rays to heat the gas. When streaming is turned off, ${{\mathcal{H}}}_{\rm c}= 0$. In the diffusion approximation, we assume a constant cosmic ray diffusion coefficient, $\kappa_{\rm c}$. When modeling cosmic ray transport, we either invoke diffusion \textit{or} streaming. Cosmic ray advection is always turned on. For an in-depth description of the implementation and tests of the anisotropic cosmic ray physics in {\sc enzo}, see \citet{Butsky:2018}. 

\subsection{Initial Conditions}
The initial conditions model the behavior of a column of gas extending off the disk of a galaxy into the CGM. The physical domain is comprised of a gravitationally stratified medium in hydrostatic equilibrium, similar to the simulations described in \citet{McCourt:2012} and \citet{Ji:2018}. The gravitational acceleration is given by the following expression:
\begin{equation}
{\bf g} = g_0 \frac{z/a}{[1 + (z/a)^2]^{1/2}} \hat{z}.
\end{equation}
Here, $z$ is the vertical distance from the midplane, $g_0$ is a constant acceleration factor, and $a$ is the gravitational smoothing length scale. This definition ensures that the gravitational acceleration goes smoothly to zero at the midplane but is nearly constant for $|z| > a$. The corresponding gas free-fall time from a position, $z$, above the disk midplane is:
\begin{equation}
    t_{\rm ff} = \sqrt{\frac{2z}{g_0}}.
\end{equation}

Using the criterion for hydrostatic equilibrium, $dP_{tot}/dz = - \rho(z){\bf g}(z)$, we derive the total pressure profile, $P_{\rm tot}(z)$, from the gravitational acceleration profile. We initialize magnetic and cosmic ray pressures to be constant fractions of the gas pressure throughout, $\beta = P_{\rm g}/P_B$ and $\eta = P_{\rm c}/P_{\rm g}$. Therefore, the the total pressure can be expressed as a multiple of the thermal pressure: $P_{\rm tot} = (1 + \beta^{-1} + \eta)P_{\rm g}$. 

Given the derived vertical pressure profile, we can choose a variety of gas density and temperature profiles ($\rho(z), T(z)$) -- so long as they obey the ideal gas law: $P = n k_B T$. 
We consider two such halo profiles: isothermal (constant temperature)  and  ``iso-cooling'' (constant cooling time). 

In the isothermal halo, the density and temperature profiles are described by
\begin{equation}\label{eqn:isothermal}
    \rho(z) = \rho_0 \mathrm{exp}\bigg[-\frac{a}{H}\big(1 + \beta^{-1} + \eta\big)\bigg(\big[1 + (z/a)^2\big]^{1/2}\bigg)\bigg]
\end{equation}
\begin{equation}
    T(z) = T_0.
\end{equation}
We define the scale height, $H = g_0 / c_s^2$, and choose a gravitational smoothing length scale, $a = 0.1 H$. In the purely hydrostatic case ($\beta = \infty, \eta = 0$), the temperature and density profiles described above are identical to those described in \citet{McCourt:2012} and \citet{Ji:2018}. In simulations with $\beta < \infty$, the magnetic field is initialized to have a constant $\beta$ everywhere so that magnetic pressure contributes to hydrostatic equilibrium. This is different from the initial conditions described in \citet{Ji:2018}, in which the magnetic fields do not contribute to hydrostatic equilibrium and are initialized with a constant magnetic field strength throughout.\footnote{The choice of initial magnetic field configuration does not qualitatively impact the results.}

When the cosmic ray and magnetic pressures contribute to hydrostatic equilibrium, the thermal gas contributes less to balancing gravity. Therefore, in the isothermal halo profile, non-thermal pressure changes the steepness of the vertical gas density profile (Eq.~\ref{eqn:isothermal}). Since gas cooling times are proportional to the square of the density, simulations with an isothermal halo profile initialized with different amounts of non-thermal pressure will have different cooling time profiles, $t_{\rm cool}(z)$. The differences in cooling times as a function of height makes it difficult to discern the impact of non-thermal pressure support from the impact of different cooling times on the onset of thermal instability. For this reason, we consider a new hydrostatic gas profile, in which the density and temperature of the gas are calibrated such that cooling time is constant everywhere \citep[``iso-cooling'';][]{Meece:2015}. Using this new constraint, the temperature and density profiles in the ``iso-cooling'' setup are given by 

\begin{equation}\label{eqn:isocooling}
    T(z) = T_0\bigg[1 - \frac{1}{2 - \alpha}\frac{a}{H}\big(1 + \beta^{-1} + \eta\big)\bigg(\big[1 + (z/a)^2\big]^{1/2} -1\bigg)\bigg],
\end{equation}
\begin{equation}
    \rho(z) = \rho_0 \bigg(\frac{T}{T_0}\bigg)^{1-\alpha},
\end{equation}
where $\alpha$ is the power-law index of the cooling function (see Section \ref{sec:cool_heat}). 

Both the isothermal and ``iso-cooling'' profiles describe gas distributions that are homogeneous at a every height, $z$. However, thermal instability needs local inhomogeneity in order to grow. We seed the instability by introducing small isobaric perturbations with wavenumbers, $k$, such that $4 \leq (k L / 2\pi) \leq 32$, and a root-mean-square perturbation amplitude of 0.02. 

We choose the constants $g_0 = 5\times10^{-10}\ \mathrm{cm\ s^{-2}}$ (so that the gravitational free-fall time at the scale height is $7.37\times10^8$ yr), $\rho_0 = 10^{-27}\ \mathrm{g\ cm^{-3}}$, and $T_0 = 10^6 \K$. In physical units, the scale height, $H$, is 43.85 kpc. Although these constants are chosen to represent typical gas properties in the CGM of an $L^*$ galaxy, the simulations are scale-free and insensitive to the choice of physical parameter values. 

Our fiducial simulations are performed in tall, skinny boxes with resolution of 64 x 64 x 256. The vertical axis spans 4 $H$ in the ``iso-cooling'' profile and 6 $H$ in the isothermal profile. The horizontal axes span 1 $H$ in both profiles. We have confirmed that the choice of geometry does not affect the properties of the resulting thermal instability. The $\hat{x}$ and $\hat{y}$ boundaries have periodic boundary conditions, and the $\hat{z}$ boundary has ``hydrostatic'' boundary conditions. At the hydrostatic boundary, fluid values are interpolated quadratically from the horizontally-averaged values in the preceding cells, with a couple of notable exceptions: 1) the vertical velocity is set to zero and 2) the magnetic field is set to its value in the nearest vertical layer.

\subsection{Cooling and Heating}\label{sec:cool_heat}
Both cooling and heating are crucial to developing local thermal instability in a globally stable atmosphere. Without a heating mechanism, simulations suffer from a cooling catastrophe. We assume that the gas is optically thin and cools radiatively following a simple power-law function described below. 
\begin{equation}
\mathcal{L} = n^2 \Lambda(T),
\end{equation}
\begin{equation}\label{eqn:cooling}
    \Lambda(T) = \Lambda_0 T^{\alpha}\ \mathrm{erg\ cm^3\ s^{-1}},
\end{equation}
for $T_{\rm min} < T < T_{\rm max}$. For our fiducial runs, we use $\alpha = -2/3$, which is a reasonable approximation of the real cooling curve in the CGM. The truncated cooling term, $T_{\rm min}$, regulates the minimum temperature in our simulation which is $\sim T_0/20$. We use a tanh function to let the cooling rate go smoothly to zero near $T = T_{\rm min}$ and $T = T_{\rm max}$. Truncating the cooling curve at $T_{\rm min}$ allows us to  control the characteristic temperature of the cold gas. Since the size of cold gas clouds is proportional to the cooling time, the choice of $T_{\rm min}$ also prevents the simulation from prohibitively short time steps or severely under-resolved cold gas clouds. The truncation at $T_{\rm max}$ does not impact the simulation, as the cooling times in the hot phase are much longer than the duration of the simulation.  The constant $\Lambda_0$ is a free parameter that is used to change the ratio of the cooling time to free-fall time in the simulations. The cooling time of gas is given by:
\begin{equation}
    t_{\rm cool} = \frac{k_B T}{(\gamma - 1) n \Lambda(T)},
\end{equation}
where $k_B$ is the Boltzmann constant.

Observations of high column densities imply that the CGM is likely long-lived and globally stable \citep{Tumlinson:2011, Stocke:2013, Werk:2014}. We model this global stability by balancing the total cooling with a mass-weighted heating in each vertical layer, so that the heating in a given cell is
\begin{equation}
    \mathcal{H}_i = \rho_i \frac{\langle\mathcal{L}\rangle}{\langle\rho\rangle},
\end{equation}
where the angled brackets represent the volume average of the enclosed quantity in a vertical layer. Since {\sc enzo} explicitly tracks the specific energy ($e = \varepsilon_{\rm g} / \rho$), we modify the above equations so that the cooling and heating in each cell becomes
\begin{equation}
    \Delta e_{i, \rm cool} =\Delta t \frac{\mathcal{L}_i}{\rho_i}
\end{equation}
\begin{equation}
\Delta e_{i, \rm heat} = \Delta t 
\frac{\langle \mathcal{L}_i \rangle }{ \langle \rho_i \rangle }.
\end{equation}

In the equations above, $\mathcal{H}_i$ and $\mathcal{L}_{i}$ are the volumetric heating and cooling in a given cell, $i$, and $\Delta t$ is the current simulation time step. Ultimately we model global thermal equilibrium by ensuring that the total cooling in each vertical layer is exactly balanced by the total heating. The fiducial simulations with cosmic ray streaming have an additional perturbative heating term, ${{\mathcal{H}}}_{\rm c}$, that is not accounted for in $\mathcal{H}$ above. 

We turn off all cooling and heating within 0.1$H$ of the midplane to prevent unphysical runaway cooling. Similarly, we turn off all cooling and heating within 0.05$H$ of the vertical domain boundaries to prevent any unphysical precipitation seeded by the boundary conditions.  

\subsection{Parameter Survey Description}\label{sec:parameter}
We systematically vary the parameters described below.
\begin{itemize}[leftmargin=*]
    \item[] {\bf Cooling time}: We keep the gravitational free-fall time fixed and vary $\Lambda_0$ to achieve the following ratios of cooling time to free-fall time (measured at $z = H$): $\tcff \in [0.1, 0.3, 1, 3, 10]$. All simulations are evolved for $t = 10 t_{\rm cool}$. 
    
    \item[] {\bf Magnetic field strength:} For every cooling time above, we run simulations with $\beta \in [\infty, 100, 10, 3]$. Our fiducial runs have $\beta = 100.$
    \item[] {\bf Cosmic ray pressure:} For every combination of initial cooling time and magnetic field strength, we vary the initial ratio of cosmic ray pressure to gas pressure, $\eta = P_{\rm c} / P_{\rm g} \in [0, 0.01, 0.1, 1, 10]$. Additionally, we run a small subset of the parameter space with an initial $P_{\rm c}/P_{\rm g} = 3.0$. 
    \item[] {\bf Cosmic ray transport}: 
    
    {\it Advection:} The simulations with cosmic ray pressure described above all include cosmic ray advection. Simulations with additional cosmic ray transport are run for $\tcff = 0.1 - 3$. 
        
    {\it Diffusion:} For every simulation with $\beta = 100$, we run simulations with three different diffusion coefficients, $\kappa_{\rm c}\in [7.9\times10^{28}, 2.6\times10^{29}, 7.9\times10^{29}]$ cm$^2$ s$^{-1}$, corresponding to diffusion timescale ratios of $t_{\rm diff}/t_{\rm ff} \in [10, 3, 1]$. We define the diffusion timescale as $t_{\rm diff} = H^2 / \kappa_{\rm c}$. In the diffusion transport prescription, the transport velocity only depends on the direction of the magnetic field and not its strength. Therefore, our choice of $\beta = 100$ allows us to better isolate the effects of cosmic ray transport from those of magnetic fields. Simulations with cosmic ray diffusion have no cosmic ray streaming and no cosmic ray heating.
    
    {\it Streaming:} For every initial cosmic ray pressure, we turn on cosmic ray streaming in runs with $\beta \in [100, 10, 3]$. Since the cosmic ray streaming velocity depends on the magnetic field strength (Eq.~\ref{eqn:stream}), this effectively models cosmic ray streaming at three different transport rates, $t_{\rm stream} / t_{\rm ff} \in [6, 1.8, 1]$, where $t_{\rm stream} = H / {\bf v_A}$. 
    
    The fiducial simulations with cosmic ray streaming have an additional heating term, ${{\mathcal{H}}}_{\rm c}$, that is not present in cosmic ray diffusion. To separate out the impact of cosmic ray heating, we run each streaming simulation with and without the additional heating term. Runs with streaming but without cosmic ray heating are marked with the label ``no heating'' for clarity. Simulations with cosmic ray streaming have no cosmic ray diffusion ($\kappa_{\rm c}= 0$).
    
    \item[] {\bf Miscellaneous:} The parameters listed above summarize the fiducial simulation suite. However there are a few additional parameters that we vary for a subset of the above simulations. Most of the simulations described below are discussed in Section \ref{sec:dens_contrast} and in the Appendix. 
    
    {\it Resolution:} The fiducial resolution of our simulations is $64\times64\times256$ cells across a simulation domain of $1\times1\times4$H. We also run a subset of the above simulations with half and double the resolution elements.
    
    {\it Halo profile:} Our fiducial simulations have an ``iso-cooling'' profile. We also run a subset of simulations with an isothermal gas profile. 
    
    {\it Cold gas temperature:} The fiducial simulation suite has a $T_{\rm min} = 5\times10^4 \K$. We also run a handful of simulations with $T_{\rm min} = 1\times10^4 \K$.  
    
\end{itemize}

\section{Physical Expectations}\label{sec:physical_expectations}
\subsection{Classical Thermal Instability}
Classical thermal instability describes how the interplay of local cooling and heating in a globally stable gas leads to condensation and the formation of a two-phase medium. Consider an initially uniform patch of hot gas that undergoes a small density perturbation. Since heating rates are proportional to density while cooling rates are proportional to density squared, the perturbed overdense gas will cool faster than it can be heated. As the gas cools and condenses, its cooling rate continues to increase, until the cold gas reaches a new equilibrium temperature. The initially uniform single-phase medium is now transformed into a two-phase medium  with cold gas cloudlets in pressure equilibrium with the volume-filling hot gas.

The growth rate of thermally unstable gas can be derived analytically, measured by the density fluctuation:
\begin{equation}\label{eqn:dens_fluc}
    \frac{\delta \rho}{\rho} = \frac{\rho - \langle{\rho}\rangle}{\langle\rho\rangle}.
\end{equation}
By perturbing Eqs.~\ref{eqn:mass} - \ref{eqn:energy}, we can determine the dispersion relation and solve for the characteristic growth timescale of the thermal instability under different conditions. For gas with a simple power-law cooling function and mass-weighted heating, the exponential growth rate of the thermal instability is characterized by $t_{\rm TI}$ \citep{McCourt:2012, Sharma:2012a}:
\begin{equation}\label{eqn:ti}
    t_{\rm TI} = \frac{\gamma t_{\rm cool}}{(1 - d {\rm ln} \Lambda / d {\rm ln} T)}.
\end{equation}
In the expression above, $d {\rm ln} \Lambda / d {\rm ln} T = \alpha$ is the logarithmic slope of the cooling function.

In the presence of gravity, buoyancy forces (characterized by the free-fall time) become important. Assuming gas is initially in hydrostatic equilibrium, an overdense parcel of gas will begin to sink in the direction of the gravitational acceleration. As the cold gas sinks, its surroundings become progressively more dense, and the cold gas experiences compressive heating. If the cooling time is sufficiently fast, the overdense gas will form into cold cloudlets (``condensation'') and fall down the potential well towards the galaxy (``precipitation''). Alternatively, if the cooling time is slow the overdense parcel of gas will be compressively heated faster than it can cool, preventing condensation. 

The balance between radiative cooling and compressive heating is often expressed as the ratio of the gas cooling time and free-fall time, $t_{\rm cool} / t_{\rm ff}$. If $\tcff$ is less than 1, cooling is considered efficient and cold gas is expected to form. If $\tcff$ is greater than one but is less than $\sim 10-20$, cold gas may still form with the help of magnetic fields or large density perturbations \citep{Meece:2015, Voit:2017, Ji:2018, Prasad:2018, Choudhury:2019}. The median ratio of $\tcff$ in the CGM is expected to be between 5 and 20 \citep{Esmerian:2020}.

Magnetic fields enhance thermal instability in gas with $\tcff$ $\geq 1$ in two ways. Magnetic tension can directly counteract compressive heating by supporting overdense gas against gravity so that it cools in-situ. Magnetic tension can also indirectly counteract compressive heating by confining pressurized hot gas \citep{Ji:2018}. 

\subsection{Thermal Instability with Cosmic Rays}
The behavior of cosmic rays in galaxy-scale simulations is well-modeled as that of a relativistic fluid. This cosmic ray fluid interacts with the gas by providing non-thermal pressure support ($P_{\rm c}$ in Eqs.~\ref{eqn:momentum} and \ref{eqn:energy}), and, in some cases, by heating the thermal gas (${{\mathcal{H}}}_{\rm c}$ in Eqs.~\ref{eqn:energy} and \ref{eqn:crenergy}). The cosmic ray fluid moves both with the gas (advection) and relative to the gas, along magnetic field lines. The latter motion is typically approximated by either diffusion or streaming.

The impact of cosmic rays on thermal instability depends on how cosmic ray pressure scales with gas density, $P_{\rm c} \propto \rho^{\gamma_{\rm c, eff}}$. In the limit of inefficient cosmic ray transport (i.e., advection only), cosmic rays are fully coupled with the gas, and  $\gamma_{\rm c, eff} = \gamma_{\rm c} = 4/3$. Cosmic ray transport redistributes cosmic ray pressure from high density to low density regions, effectively lowering $\gamma_{\rm c, eff}$. In the limit of very efficient cosmic ray transport, cosmic ray pressure decouples from the gas and $\gamma_{\rm c, eff} \rightarrow 0$. The exact value of $\gamma_{\rm c, eff}$ depends on the balance of the efficiency of cosmic ray transport, the ratio of cosmic ray pressure to gas pressure, the gas cooling time, and the \alfven velocity.

In the discussion below, we provide a brief overview of the relevant cosmic ray physics and provide some intuition for how different aspects of the cosmic ray fluid can affect thermal instability. We first focus on the impact of cosmic ray pressure in the limit of no cosmic ray transport and make predictions for the gas density contrast and cold cloud sizes. We then discuss the expected qualitative impact of cosmic ray transport. 

\subsubsection{Thermal Instability with Cosmic Ray Pressure}
Cosmic ray pressure can convert thermal instability from an isobaric to an isochoric process. Unlike regular gas, cosmic rays do not lose energy through radiative cooling. In a purely thermal gas, a cooling cloud condenses to balance its change in temperature in order to remain in pressure equilibrium with the hot medium. In the presence of cosmic rays, as a cooling cloud contracts, its non-thermal cosmic ray pressure grows. The added pressure support lets gas cool without contracting as much, since the cooling cloud needs to be in \textit{total} pressure equilibrium, $P_{\rm tot} = P_{\rm g} + P_{\rm B} + P_{\rm c}$. With sufficiently high cosmic ray pressures, gas cools at constant density \citep[isochorically;][]{Sharma:2010, Kempski:2020a}. 

This non-thermal pressure support increases the size of cold gas clouds, $\ell_{\rm cloudlet}$, and lowers the density contrast, $\delta$, between the cold and hot gas phases,
\begin{equation}
\delta = \frac{\langle\rho_{\rm cold}\rangle}{\langle \rho_{\rm hot}\rangle}.
\end{equation}
We introduce a simple model for how these quantities depend on cosmic ray pressure below. 

Assuming that the total pressure stays constant, $(P_{\rm g} + P_{\rm B} + P_{\rm c})_{\rm cold} = (P_{\rm g} + P_{\rm B} + P_{\rm c})_{\rm hot}$, and that the magnetic and cosmic ray pressures scale with gas density, we expect the density contrast to follow 
\begin{equation}
\begin{split}
    P_{\rm B, hot} = \beta_{\rm cold}^{-1}P_{\rm g, cold} \left(\frac{\rho_{\rm hot}}{\rho_{\rm cold}}\right)^{\gamma_{\rm B}},\\
    P_{\rm c, hot} = \eta_{\rm cold}P_{\rm g, cold} \left(\frac{\rho_{\rm hot}}{\rho_{\rm cold}}\right)^{\gamma_{\rm c, eff}}, 
    \end{split}
\end{equation}
we expect the density contrast to follow:
\begin{equation}\label{eqn:dens_contrast}
\left(1 + \beta_{\rm cold}^{-1} + \eta_{\rm cold}\right) = \frac{\Theta}{\delta} + \frac{\beta_{\rm cold}^{-1}}{\delta^{\gamma_{\rm B}}} + \frac{\eta_{\rm  cold}}{\delta^{\gamma_{\rm c, eff}}},
\end{equation}
where $\eta_{\rm cold}, \beta_{\rm cold}$ are evaluated in the cold gas, and $\Theta = T_{\rm hot}/T_{\rm cold}$ is the temperature contrast between the hot and cold gas phases. The temperature contrast is set by the details of the atomic physics that determines the cooling curve and is independent of magnetic fields and cosmic rays. The constant $\gamma_{\rm B} = 4/3$ describes how magnetic pressure scales with gas density, $P_{\rm B} \propto \rho^{\gamma_{\rm B}}$, in the flux-freezing limit of ideal MHD. This is similar to the advection-only limit of cosmic ray transport. We explore the application of the predicted density contrast in Section \ref{sec:dens_contrast} and in Appendix \ref{sec:appendix_Tmin}. 

The characteristic scale of cold gas clouds is predicted to be set by the minimum of the product of the gas sound speed and cooling time\footnote{The minimum value of $c_{\rm s}t_{\rm cool}$ is expected to happen around $T \approx 10^{4.2}$K for a variety of gas pressures \citep{Liang:2020}.}, $\ell_{\rm cloudlet} \sim {\rm min}(c_{\rm s}t_{\rm cool})$ \citep{McCourt:2018}. Both the effective sound speed and the gas cooling time may be altered by non-thermal pressure support.

In the presence of magnetic fields and cosmic rays, the maximum wave speed is given by, $c_{\rm max}^2 = c_{\rm s}^2 + {\rm v}_{\rm A}^2 + c_{\rm s, c}^2$, where $c_{\rm s} = \sqrt{(\gamma P_{\rm g}/\rho)}$ is the thermal gas sound speed, $\rm v_A = \sqrt{2 P_{\rm B} / \rho}$ is the \alfven velocity, and $c_{\rm s, c} = \sqrt{\gamma_{\rm c, eff}P_{\rm c} / \rho}$ is the cosmic ray sound speed.  We can rewrite the maximum wave speed as a function of the ratios of magnetic and cosmic ray pressures to the gas pressure, $\beta$ and $\eta$:
\begin{equation}\label{eqn:ceff}
    c_{\rm max} = c_{\rm s}\bigg(1 + \frac{2}{\gamma}\beta^{-1} + \frac{\gamma_{\rm c, eff}}{\gamma}\eta\bigg)^{1/2}. 
\end{equation}

Assuming that gas cooling follows a power law (Eq.~\ref{eqn:cooling}), the gas cooling time scales as:
\begin{equation}\label{eqn:tcool_scale}
    t_{\rm cool} = \frac{\frac{3}{2} \rho k_B T}{\big(\frac{\rho^2}{\mu m_{\rm p}}\big) \Lambda_0 T^{\alpha}}  \propto \rho^{-1} T^{1-\alpha}, 
\end{equation}
where $\mu$ is the mean molecular weight and $m_{\rm p}$ is the proton mass. Since the temperature of cold gas is set by the cooling curve and isn't affected by non-thermal pressure, we expect non-thermal pressure to only alter the density in the cooling time scaling relation. Combining the expected expression for the cold cloud size with Eqs.~\ref{eqn:ceff} and \ref{eqn:tcool_scale}, we predict the non-thermal pressure-supported cold gas cloud size, $\ell_{\rm cloudlet}^*$, to scale as:
\begin{equation}\label{eqn:rcloud}
   \frac{\ell_{\rm cloudlet}^*}{\ell_{\rm cloudlet}} = \bigg(\frac{\rho_{\rm cold}}{\rho_{\rm cold}^*}\bigg)\bigg(1 + \frac{2}{\gamma}\beta_{\rm cold}^{-1} + \frac{\gamma_{\rm c, eff}}{\gamma}\eta_{\rm cold}\bigg)^{1/2}, 
\end{equation}
where $\rho_{\rm cold}^*$ is the density of non-thermal pressure-supported cold gas. In the limit of high cosmic ray pressure (Eq. \ref{eqn:dens_contrast}), the cold gas density is the same as the hot gas density, $\rho_{\rm cold}^* = \rho_{\rm hot}$, so the ratio $\rho_{\rm cold} / \rho_{\rm cold}^* \approx \Theta$ can be expressed in terms of the temperature contrast between cold and hot phases. For a cosmic ray pressure-dominated medium with inefficient transport ($\gamma_{\rm c, eff}\eta_{\rm cold} \gg 1$), we expect:
\begin{equation}\label{eqn:rcloud_cr}
       \frac{\ell_{\rm cloudlet, \eta \gg 1}^*}{\ell_{\rm cloudlet}} \approx \Theta \left(\frac{\gamma_{\rm c, eff}}{\gamma}\eta_{\rm cold}\right)^{1/2}.
\end{equation}
Following Eq.~\ref{eqn:rcloud_cr}, we expect simulations with $\Theta = 20$, $P_{g, 0} = 490\,{\rm K\, cm}^{-3}$, $\eta_{\rm cold} = 100$ to have cold cloud sizes, $\ell_{\rm cloudlet}^* \approx 200\, \ell_{\rm cloudlet} \approx 40\, {\rm kpc}$, which is in good agreement with the results discussed below (see Section~\ref{sec:results_crpressure}). In the real CGM, the temperature contrast between hot and cold phases is closer to $\Theta = 100$ and the range of gas pressures $P_{\rm g}/k_B \sim 1 - 10^3\, {\rm K \, cm}^{-3}$ gives predicted cold cloudlet sizes of $\ell_{\rm cloudlet} \sim 1-1000$ pc. Therefore, in the limit of no cosmic ray transport in a cosmic ray-pressure dominated halo, we expect cold cloud sizes to be to $\sim 1000$ times larger than predicted for cold gas in thermal pressure equilibrium. This prediction is an upper limit for cold cloud sizes in a cosmic ray pressure-dominated halo as realistic cosmic ray transport will reduce $\gamma_{\rm c, eff}$, thereby reducing the predicted cold cloud increase.

In addition to altering cold gas density and cloud size, cosmic ray pressure also contributes to hydrostatic equilibrium and changes the effective entropy profiles of the gas. This could prevent cold gas clumps from precipitating after they condense out of the hot background medium. Additionally, if the cosmic ray scale height is sufficiently large relative to the gas scale height, $H_{\rm c}/H_{\rm g} \ge 5/2$, gas becomes convectively unstable \citep{Kempski:2020a}. Our simulations are initialized with constant $\eta$, so $H_{\rm c}/H_{\rm g}$ is always close to 1.

\subsubsection{Thermal Instability with Cosmic Ray Transport}
In addition to advection, the cosmic ray fluid moves relative to the gas via some transport mechanism. Although there is no consensus as to which transport mechanism produces the most realistic results in galaxy simulations \citep{Butsky:2018, Farber:2018, Buck:2020, Hopkins:2020:cr_transport2}, cosmic ray transport has been historically modeled as either streaming or diffusion. Physically, both streaming and diffusion model how perturbations in the magnetic field lines scatter the pitch-angle of the cosmic ray orbit. The main difference lies in the assumption about what is driving the perturbations in the magnetic field, either the streaming instability (streaming) or extrinsic turbulence (diffusion).

In the fluid approximation, both cosmic ray streaming and cosmic ray diffusion model how cosmic rays move along magnetic field lines, down the cosmic ray pressure gradient. In the diffusion approximation, the transport rate is set by the constant diffusion coefficient, $\kappa_{\rm c}$, and in the streaming approximation, the transport rate is set by the local \alfven velocity. Since cosmic ray transport reduces the non-thermal pressure support in cold gas, we expect simulations with cosmic ray transport to have smaller cold cloud sizes, higher gas densities, and higher cold mass flux rates than simulations with only cosmic ray advection. However, the quantitative impact on thermal instability will be sensitive to the details of the cosmic ray transport model parameters. 

In the streaming approximation of cosmic ray transport, cosmic rays heat the gas at a rate proportional to the \alfven velocity and cosmic ray pressure gradient (${{\mathcal{H}}}_{\rm c}$ in Eqs.~\ref{eqn:energy} and \ref{eqn:crenergy}). We expect cosmic ray heating to be significant in gas with low $\beta$ and high $\eta$ \citep{Kempski:2020a}. This heating process is also more efficient in simulations with larger $\tcff$, which are evolved long enough for cosmic ray transport processes to become important.

Cosmic ray transport also alters the growth rate of the thermal instability. In the limit of no cosmic ray pressure, the growth rate is that predicted by classical thermal instability. In the limit of high cosmic ray pressure, the growth rate depends on the invoked cosmic ray transport mechanisms (see \citet{Kempski:2020a} for detailed derivations).

While the discussion above provides some intuition of the isolated impacts of cosmic ray pressure, transport, and heating, simulations are necessary to study the interplay of these effects in the non-linear regime. 

\begin{figure*}
\includegraphics[width=\textwidth]{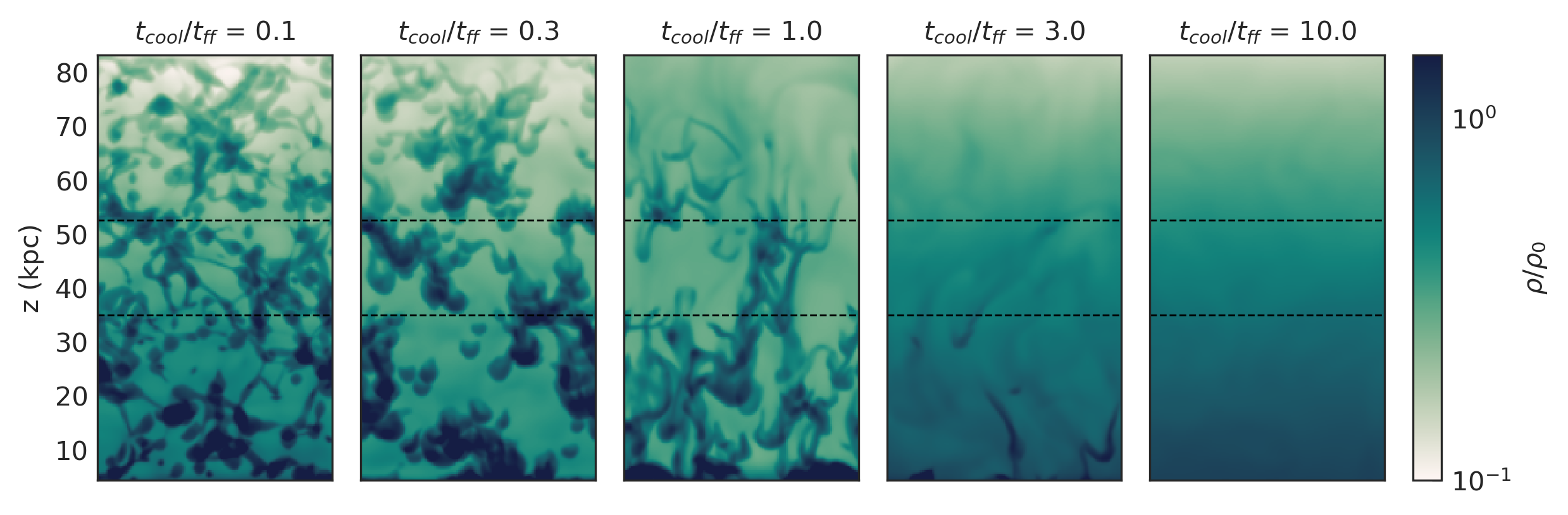}
\caption{ \footnotesize Projections of the gas density for different initial values of $\tcff$. The dimensions of each slice are 1H x 2H in the $\hat{x}$ and $\hat{z}$ directions respectively and the projection depth in $\hat{y}$ is 1$H$ (43.85 kpc). The simulations showcased above were initialized with $\beta = 100$ and no cosmic rays. Since the simulations are symmetric about the midplane, we only show the top half of the domain. The snapshots are taken at $t = 4 t_{\rm cool}$, which corresponds to longer physical times for simulations with higher $t_{\rm cool} / t_{\rm ff}$.  As expected, there is a direct relationship between the value of $\tcff$ and the condensation of dense, cold gas. Simulations evolved for longer physical times show signs of cold gas precipitating towards the midplane. However, in simulations with $\tcff \geq 3$, insufficient condensation occurs to form appreciable amounts of cold gas. }
\label{fig:dens_slice} 
\end{figure*}
\section{Results}\label{sec:results}
In the following sections, we quantify the evolution of thermal instability by measuring three quantities: the density fluctuation, the cold mass fraction, and the cold mass flux. We focus our analysis around the scale height, $0.8 H \leq |z| \leq 1.2 H$. This region is both sufficiently thick to capture the relevant effects and sufficiently removed from the simulation center and boundaries (see Figure~\ref{fig:dens_slice}). We first show the growth and saturation of thermal instability without cosmic rays for a wide range of initial $\tcff$\  values. We then show the impact of cosmic ray pressure in advection-only simulations, followed by the impact of various cosmic ray transport models.

The density fluctuation, $\delta \rho / \rho$ (Eq.~\ref{eqn:dens_fluc}), measures the variance in density values between the cold and hot phases. In the early stages of thermal instability, the density fluctuation grows following a power law that depends on the slope of the cooling function (Eq.~\ref{eqn:ti}). At late times, the density fluctuation saturates. At very late times, the measured density fluctuation may decrease as cold gas clumps precipitate from the scale height down to the simulation midplane. 

The cold mass fraction measures the total mass of cold gas as a fraction of the total gas mass ($M_{\rm cold} / M_{\rm total}$). The average temperature of cold gas in the simulation is set by $T_{\mathrm{min}}$, which is $5\times 10^4 \K$ in our fiducial simulations. However, since the temperature distribution is highly bimodal (see Section \ref{sec:discussion}), we define cold gas to be $T \leq 3\times 10^5 \K$ in our analysis. Like the density fluctuation, the cold mass fraction near the scale height can decrease at late times as cold gas clumps precipitate and fall towards the midplane.

The cold mass flux, 
\begin{equation}
    \frac{\dot{M}_{\rm cold}}{\dot{M}_{\rm ff}} = \rho_{\rm cold}v_{\rm in} \frac{t_{\rm ff}}{\rho_0 H},
\end{equation} 
measures the rate at which cold gas precipitates towards the midplane, normalized by the free-fall mass flux.\footnote{Our initial conditions satisfy hydrostatic equilibrium and there is no initial mass flux.} The quantity $v_{\rm in} = -v_z \cdot \hat{z}$ measures the infalling velocity of cold gas clouds. Since the simulations are run for a constant number of cooling times, simulations with larger $\tcff$\ have time to accelerate to larger $v_{\rm in}$ velocities. Larger inflowing velocities result in an increase in the cold mass flux (assuming the simulation formed cold gas). Later, we will demonstrate how cosmic ray pressure can suppress cold mass flux by limiting $v_{\rm in}$.

\subsection{Thermal Instability Without Cosmic Rays}
\begin{figure*}
\includegraphics[width=\textwidth]{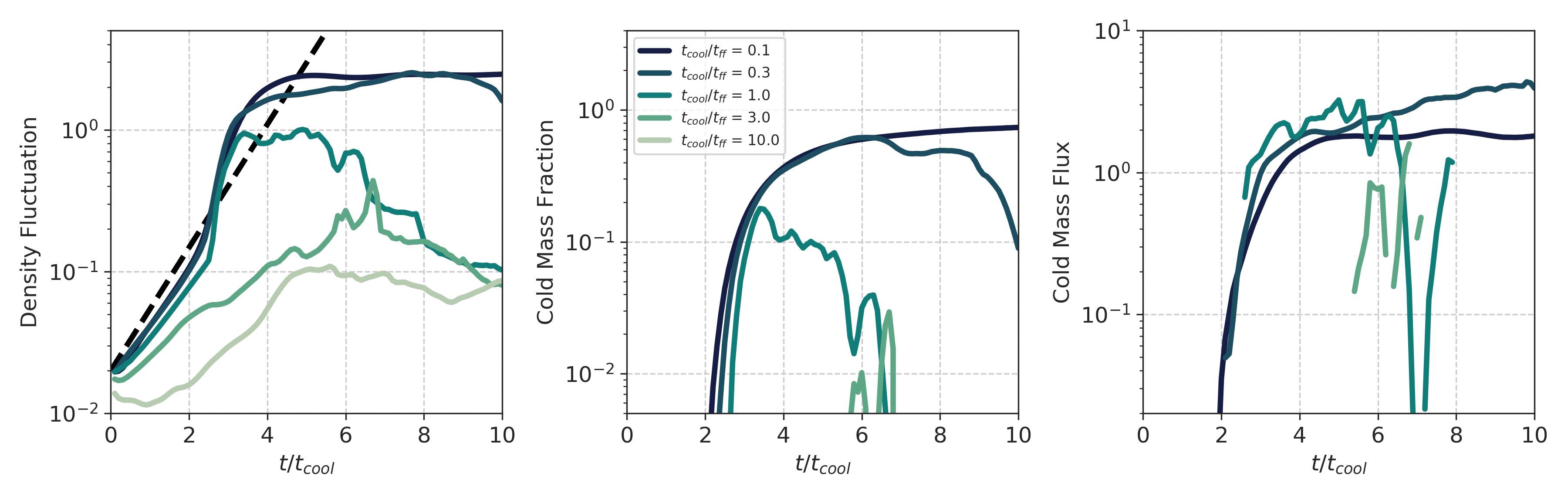}
\caption{ \footnotesize The time evolution of the density fluctuation ($\langle\delta\rho / \rho\rangle $; \textit{left}), cold mass fraction ($M_{\rm cold} / M_{\rm total}$; \textit{middle}), and infalling cold mass flux ($\dot{M}_{\rm cold} / \dot{M}_{\rm ff}$; \textit{right}).  The different lines show different initial values of the ratio of cooling time to free-fall time at the scale height. The black dashed line in the left panel shows the linear theory prediction for the density fluctuation growth with mass-weighted heating and a logarithmic cooling slope, $\alpha = -2/3$ (Eq.~\ref{eqn:cooling}). All values are measured between 0.8 and 1.2 times the scale height. The fiducial simulations presented are initialized with $\beta = 100$. The x-axis values are scaled by the cooling time of the simulation, so that simulations with higher values of $\tcff$ are evolved over longer physical times. This figure demonstrates thermal instability without cosmic rays: simulations with lower $\tcff$ show higher density fluctuations and cold mass fractions. Although the cold mass fraction is significantly higher in simulations with $\tcff = 0.1-0.3$ than in simulations with $\tcff = 1$, their cold mass flux rates are comparable shortly after the saturation of the thermal instability. This is because simulations with longer $\tcff$ have been evolved for longer physical time, and have developed a higher inflow velocity, $v_{\rm in} = -{\bf v}\cdot \hat{z}$.}
\label{fig:dens_fluc_growth} 
\end{figure*}

We first demonstrate the onset of thermal instability without cosmic ray physics. Figure~\ref{fig:dens_slice} shows density projections of simulations with $\beta = 100$, and varying initial values of $\tcff$. These snapshots are pictured after the simulations have evolved for 4 cooling times,\footnote{ Since the projections were generated after the same number of cooling cycles, this means that simulations with longer cooling times were run for longer physical times.} which is roughly when the density fluctuation saturates (see Figure~\ref{fig:dens_fluc_growth}). In simulations with short cooling times, cold gas condenses out of the background medium much faster than the gravitational free fall time. Therefore, after several cooling cycles, there is still plenty of cold gas high in the ``atmosphere''. For larger values of $\tcff$, the gravitational acceleration becomes more important, and condensed gas clumps start precipitating down towards the midplane. The size of the cold condensed gas clumps is also visibly larger in simulations with larger $\tcff$, because cloud sizes are proportional to the cooling time, and because cold gas cloudlets have had more time to coagulate \citep{Gronke:2020}. There is no sign of thermal instability in the simulation with $\tcff = 10$. 

The left panel of Figure~\ref{fig:dens_fluc_growth} quantifies the time evolution of the average density fluctuation, $\langle \delta\rho / \rho\rangle$, measured at z = 0.8-1.2H, for the five simulations pictured in Figure~\ref{fig:dens_slice}. At early times, the density fluctuation for simulations with $\tcff \leq 1$ grows at the rate predicted by linear theory (see Eq.~\ref{eqn:ti}).
The thermal instability saturates after about roughly 4 cooling times. The saturated density fluctuation value is higher for simulations with lower $\tcff$ and are generally consistent with those presented in \citet{McCourt:2012} and \citet{Ji:2018}. For $\tcff \geq 0.3$, the density fluctuation decreases at late times as cold gas precipitates from the scale height towards the midplane. 

The middle panel of Figure~\ref{fig:dens_fluc_growth} shows the time evolution of the mass fraction, $M_{\rm cold} / M_{\rm total}$, of cold gas. The cold mass fraction is nearly identical for runs with $\tcff = 0.1$ and $\tcff = 0.3$ for the majority of the simulation. At late times ($t \gtrsim 2\, \tff$), the cold mass fraction decreases as cold gas clouds begin to precipitate. There is no sign of precipitation in the simulation with $\tcff = 0.1$, since it was only evolved for $t = 10\, \tcool = 1\, \tff$. The simulation with $\tcff = 1$ only reaches a cold gas mass fraction of about 0.1 before the cold gas clumps precipitate out of the halo. Unsurprisingly, simulations with $\tcff \geq 3$ do not form substantial amounts of cold gas. 

The right panel of Figure~\ref{fig:dens_fluc_growth} shows the time evolution of the cold mass flux towards the midplane, $\dot{M}_{\rm cold} / \dot{M}_{\rm ff}$. Although the simulation with $\tcff = 1$ has a lower cold mass fraction near the scale height than simulations with $\tcff = 0.1, 0.3$, it has comparable cold mass flux values due to larger values of the inflow velocity, $v_{\rm in}$. Since simulations are evolved for a fixed number of cooling cycles, simulations with larger $\tcff$ are evolved for longer physical time and develop larger infall velocities. Without thermal instability, there is no cold mass flux. 

\begin{figure*}
\includegraphics[width=\textwidth]{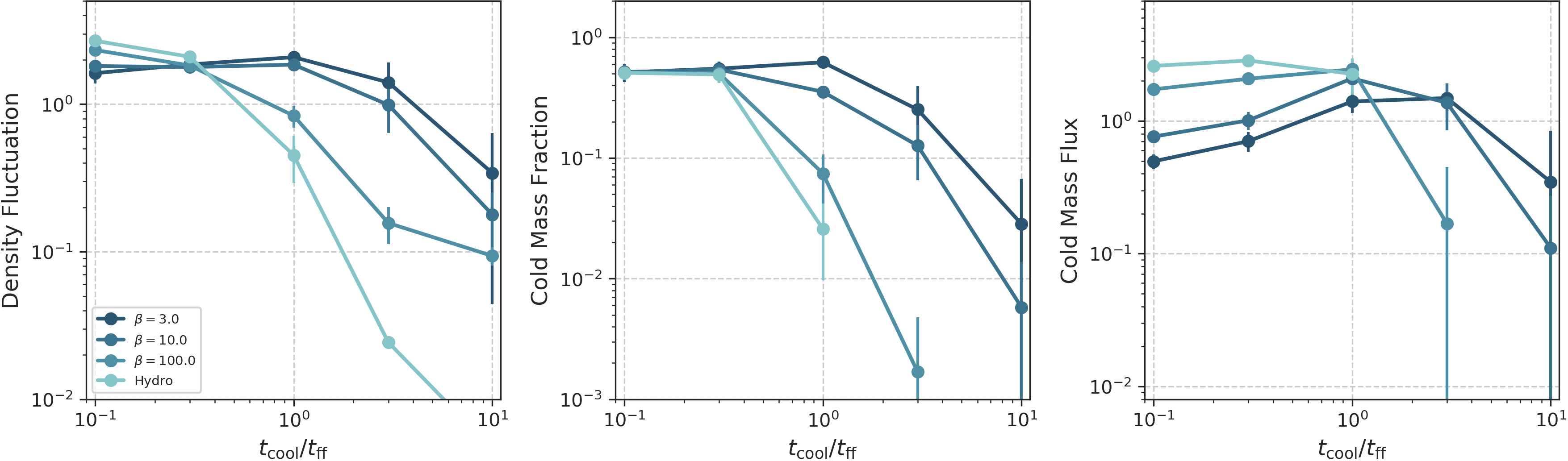}
\caption{ \footnotesize  The average density fluctuation ($\langle\delta\rho / \rho\rangle $; \textit{left}), cold mass fraction ($M_{\rm cold} / M_{\rm total}$; \textit{middle}), and cold mass flux ($\dot{M}_{\rm cold} / \dot{M}_{\rm ff}$; \textit{right}) as a function of the initial $\tcff$ for simulations with varying initial magnetic field strengths, in the absence of cosmic rays.  The measurements are taken between 0.8 and 1.2$H$ and averaged over all outputs between 4 and 6 $\tcool$, which corresponds to the saturated phase of the thermal instability (see Figure~\ref{fig:dens_fluc_growth}).  Magnetic fields suppress buoyancy oscillations, which simultaneously enhances the formation of cold gas through thermal instability for simulations with higher $\tcff$ and decreases the inflowing cold mass flux in simulations with lower $\tcff$.}
\label{fig:dens_fluc_tctf} 
\end{figure*}

Next, we demonstrate the effects of magnetic fields on thermal instability in the absence of cosmic rays. Figure~\ref{fig:dens_fluc_tctf} shows the time-averaged density fluctuation, cold mass fraction, and cold mass flux as a function of the initial $\tcff$. Each point represents the mean measurement for all simulation outputs between 4 and 6 $\tcool$, and the error bars show one standard deviation. The colors show different initial magnetic field strengths ranging from no magnetic fields ($\beta = \infty$), to strong magnetic fields ($\beta = 3$). These initial magnetic field strengths span the range of magnetic fields considered in our cosmic ray simulations and serve as a useful reference to disentangle the effects of cosmic rays from the effects of magnetic fields. 

When cooling times are short ($\tcff \le 1$), thermal instability is efficient in all simulations, independent of magnetic field strength. The density fluctuation is lower in runs with strong magnetic fields and $\tcff = 0.1$ due to the additional non-thermal pressure support lowering cold gas densities.  Although the cold mass fraction is constant between all runs with short cooling times, the cold mass flux decreases with increasing magnetic field strength due to increased non-thermal pressure support. 

When cooling times are long ($\tcff \geq 1$), magnetic fields enhance thermal instability by suppressing buoyancy oscillations \citep{Ji:2018}. The density fluctuation, cold mass fraction, and cold mass flux increase with decreased $\beta$. Our density fluctuation values are consistent with those reported in \citet{Ji:2018}. Small differences are likely due to measurements being taken at different simulation times. 

These MHD-only simulations highlight the previously demonstrated dependence of thermal instability on the gas cooling time and magnetic field strength. Simulations with short $\tcff$ produce higher density fluctuations and cold mass fractions in less physical time. Magnetic fields enhance thermal instability when cooling times are long.
\begin{figure*}
\includegraphics[width=\textwidth]{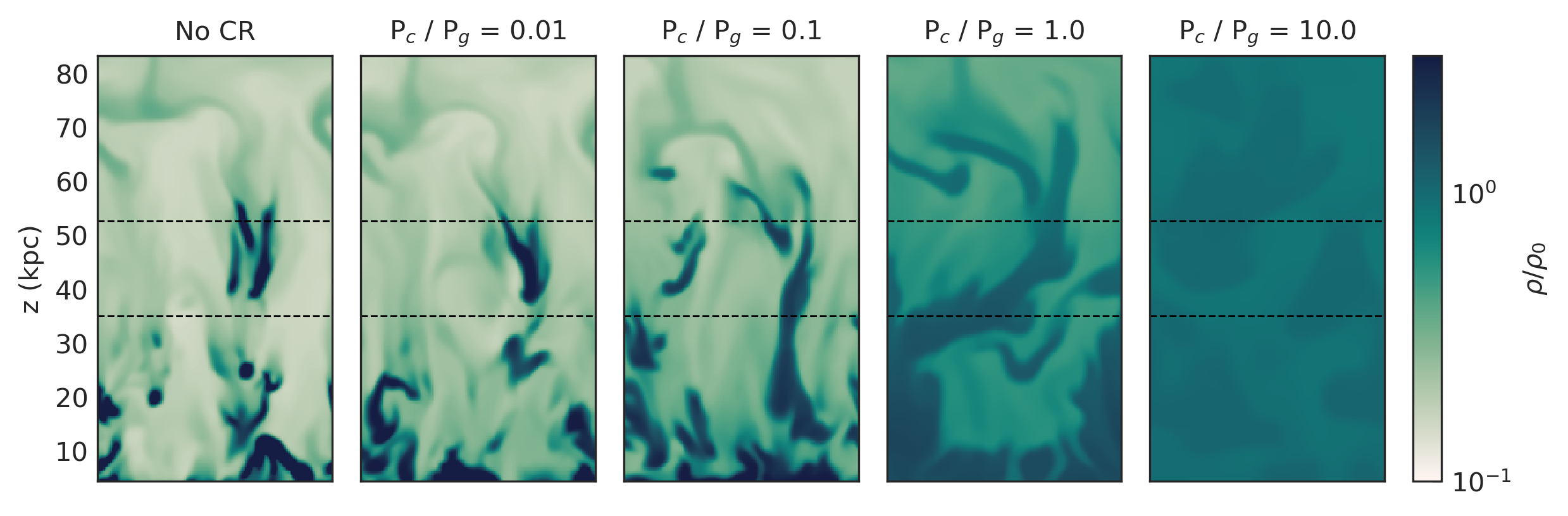}
\includegraphics[width=\textwidth]{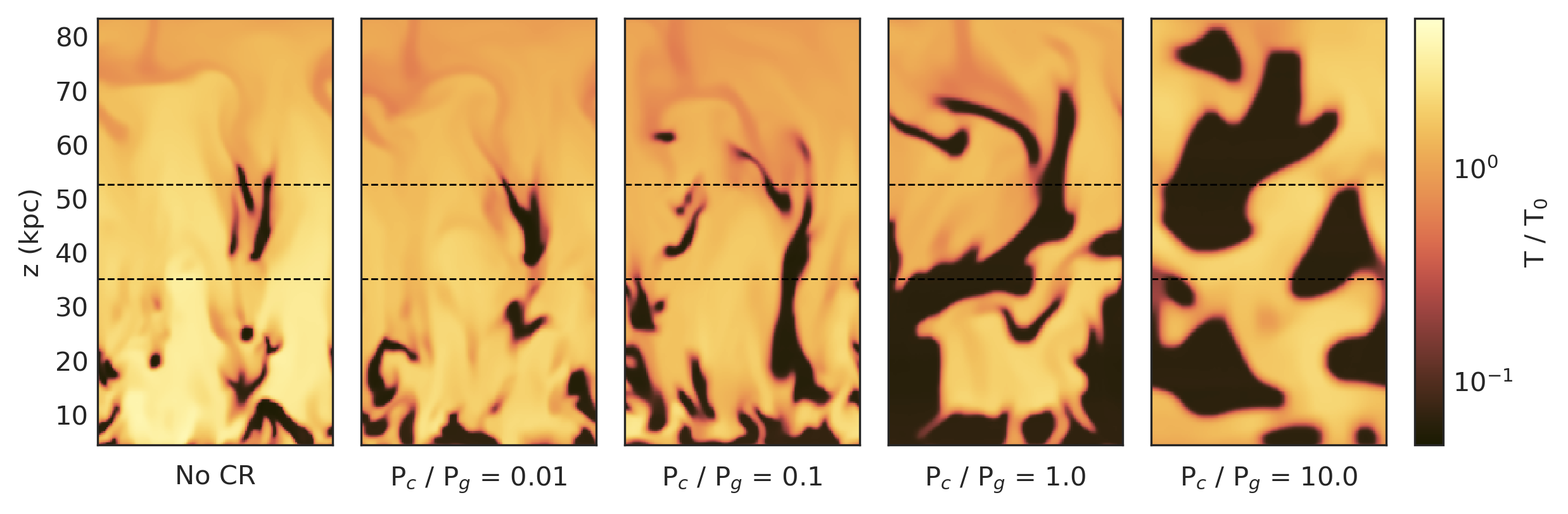}
\caption{ \footnotesize 2D slices of the 3D density (top) and temperature (bottom) for simulations with varying initial ratios of cosmic ray pressure to gas pressure, $P_{\rm c} / P_{\rm g}$. Similarly to Figure~\ref{fig:dens_slice}, the dimensions of each panel are $1H \times 1.8H$. All simulations are initialized with $\tcff = 1.0$, and the slices are taken at $t = 6\,\tcool$. Simulations with cosmic rays were run with advection as the only form of cosmic ray transport. Increased cosmic ray pressure support creates larger cold gas clouds with lower densities. The temperature of the cold gas does not change with increased cosmic ray pressure since the temperature is set by the cooling function and our choice of $T_{\rm min}$. With sufficiently high cosmic ray pressure, cold gas remains in the atmosphere and does not precipitate.}
\label{fig:dens_slice_cr} 
\end{figure*}

\vspace{1cm}
\subsection{Thermal Instability with Cosmic Ray Pressure}\label{sec:results_crpressure}
\begin{figure*}
\includegraphics[width=\textwidth]{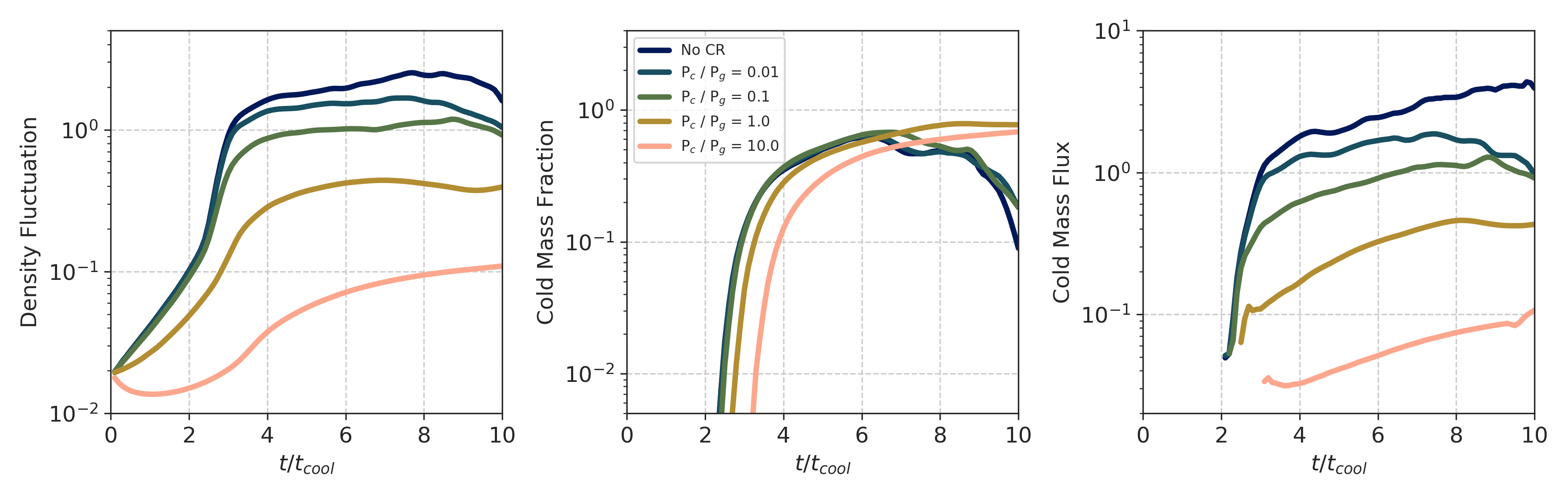}
\caption{ \footnotesize The time evolution of the density fluctuation ($\langle\delta\rho / \rho\rangle $; \textit{left}), cold mass fraction ($M_{\rm cold} / M_{\rm total}$; \textit{middle}), and cold mass flux ($\dot{M}_{\rm cold} / \dot{M}_{\rm ff}$; \textit{right}) near the scale height ($0.8 H < z < 1.2 H$). All of the depicted simulations have $\tcff = 0.3$ and $\beta = 100$ but have different initial values of $P_{\rm c} / P_{\rm g}$. Increased cosmic ray pressure decreases density fluctuation but has a modest effect on the cold mass fraction. Non-thermal cosmic ray pressure support counteracts gravity and lowers the cold mass flux towards the midplane. Simulations with $P_{\rm c} \geq P_{\rm g}$ have more cold gas near the scale height at late times. }
\label{fig:dens_fluc_growth_cr} 
\end{figure*}
\begin{figure*}
\includegraphics[width=\textwidth]{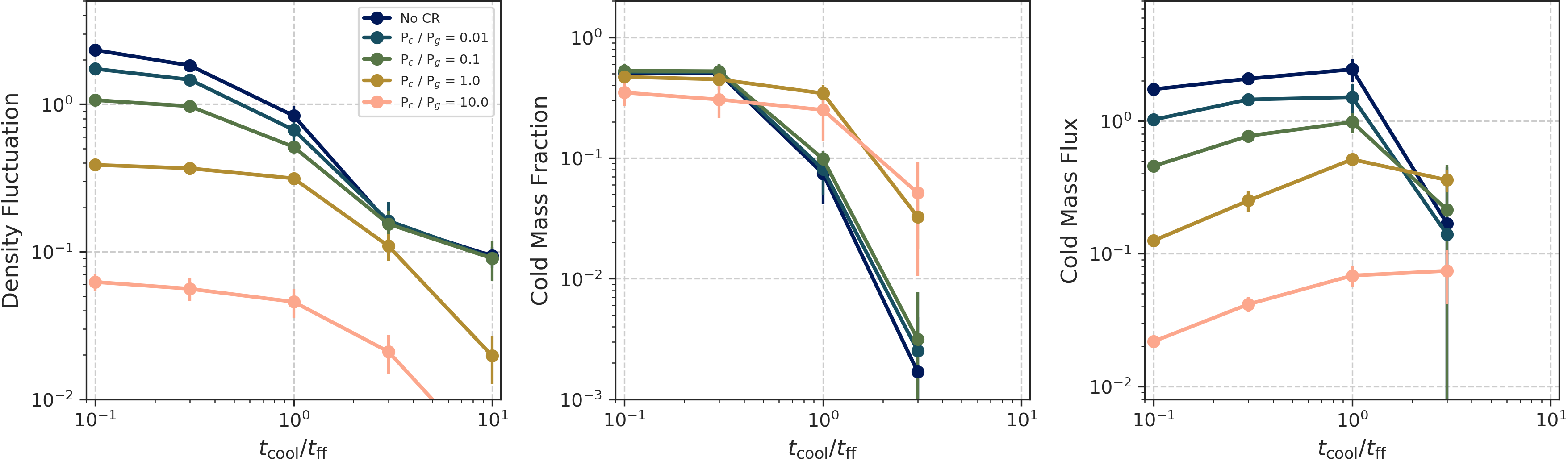}
\caption{ \footnotesize  The average density fluctuation ($\langle\delta\rho / \rho\rangle $; \textit{left}), cold mass fraction ($M_{\rm cold} / M_{\rm total}$; \textit{middle}), and cold mass flux ($\dot{M}_{\rm cold} / \dot{M}_{\rm ff}$; \textit{right}) as a function of the initial $\tcff$ for simulations with $\beta = 100$ and varying initial ratios of $P_{\rm c}/P_{\rm g}$. The measurements are taken between 0.8 and 1.2 $H$ and averaged over all outputs between 4 and 6 $t_{\rm cool}$, which corresponds to the saturated phase of the thermal instability (see Figure~\ref{fig:dens_fluc_growth_cr}). Increased cosmic ray pressure decreases the gas density fluctuation. High cosmic ray pressures increase the cold fraction in simulations with $\tcff \geq 1$ by preventing cold gas from precipitating towards the midplane. Cold gas does not form in simulations with $\tcff = 10$.}
\label{fig:dens_fluc_tctf_cr} 
\end{figure*}

In the limit that cosmic ray transport is very slow relative to the time scale of thermal instability, cosmic ray advection with the thermal gas becomes important. We isolate this phenomenon with a suite of simulations, varying the initial cosmic ray pressure fraction, $P_{\rm c} / P_{\rm g}$, for which both cosmic ray diffusion and cosmic ray streaming are turned off.

Figure~\ref{fig:dens_slice_cr} shows the impact of cosmic ray pressure on the density and temperature slices of thermally unstable gas. All simulations are initialized with $\tcff = 1.0, \beta = 100$ and have been run for 4 cooling times. The columns differ from each other by their initial value of $P_{\rm c} / P_{\rm g}$. 

Increased cosmic ray pressure alters the morphology of cold gas. Radiative gas cooling does not remove cosmic ray energy. In the limit where the cosmic ray pressure is high and dominates the total pressure, the loss of thermal pressure from radiative cooling leads to negligible compression. As a result the gas cools isochorically. This behavior is in contrast to traditional thermal instability, where as the gas cools, it loses thermal pressure, and collapses isobarically, with large density fluctuations (for more detail, see Section~\ref{sec:gas_phase}). Therefore, increased cosmic ray pressure results in larger cold gas clouds that have a lower density contrast with the background medium. When cosmic ray pressure dominates (right panel), cold cloud structures span tens of kiloparsecs, which is broadly consistent with our predictions using Eq.~\ref{eqn:rcloud_cr} in Section~\ref{sec:physical_expectations}. Additionally, cosmic ray pressure contributes to hydrostatic equilibrium and keeps cold gas at high altitudes for longer.

The temperature of the cold gas phase does not change with increased cosmic ray pressure. This is because gas temperature is set by the cooling function, which is an approximation to atomic physics and is insensitive to cosmic ray physics. 

Figure~\ref{fig:dens_fluc_growth_cr} shows the time evolution of the density fluctuation, cold mass fraction, and cold mass flux for the simulations pictured in Figure~\ref{fig:dens_slice_cr}. The different colored lines represent simulations with different initial values of $P_{\rm c} / P_{\rm g}$. All simulations have an initial $\tcff = 0.3$ and $\beta = 100$. 

The density fluctuation decreases monotonically with increased cosmic ray pressure. Remarkably, the cold mass fraction remains relatively unchanged. Even in the extreme case of $P_{\rm c}/P_{\rm g} = 10$, the cold mass fraction  only varies from the control run by a factor of 2-3, whereas the density fluctuation measurement varies by a factor of 50. At late times, runs with significant cosmic ray pressure support have more cold gas mass near the scale height since the cosmic ray pressure prevents it from precipitating towards the midplane. Although the cold mass fraction is relatively unchanged, the cold mass flux decreases substantially with increased cosmic ray pressure. Even a modest initial value of $P_{\rm c}/P_{\rm g} = 0.01$ is enough to decrease the cold mass flux by a factor of $\sim 2$. Since cosmic rays advect with the gas, cold gas clumps end up having a larger ratio of $P_{\rm c}/P_{\rm g}$. This added non-thermal pressure supports the cold gas against gravity. 

Figure~\ref{fig:dens_fluc_tctf_cr} compiles the average values of the density fluctuation, cold mass fraction, and cold mass flux measured between $t = 4 t_{\rm cool}$ and $t = 6 t_{\rm cool}$ as a function of the simulation's initial $\tcff$. Increasing the cosmic ray pressure monotonically decreases the measured density fluctuation for all initial values of $\tcff$ (left panel). As non-thermal cosmic ray pressure increases, the gas is better able to cool isochorically, decreasing the density contrast between the cold and hot phases. With increased cosmic ray pressure, the average density fluctuation also becomes less sensitive to the initial $\tcff$. 

When cooling times are short, ($\tcff \le 1$), the average cold mass fraction (middle panel) remains relatively unchanged with increased cosmic ray pressure. However, when cooling times are long ($\tcff \geq 1$), the average cold mass fraction is higher for simulations with $P_{\rm c}/P_{\rm g} \geq 1$.  \emph{The presence of cosmic rays can, therefore, significantly increase the amount of cold gas in the CGM, particularly in the outer halo where cooling times would otherwise be too long relative to free fall times}. In part, this is due to cosmic ray pressure counteracting compressive heating, similar to the role of strong magnetic fields in Figure~\ref{fig:dens_fluc_tctf} \citep[see also][]{Ji:2018}. Additionally, once the cold gas is formed, cosmic ray pressure support prevents it from precipitating by supporting the gas against gravity. This effect is prevalent when cosmic ray pressure is at least as strong as the gas pressure and when the cooling times are long, such that gas inflow velocities become important. This is likely a very relevant regime for low redshift galaxies.

\begin{figure*}
\includegraphics[width=\textwidth]{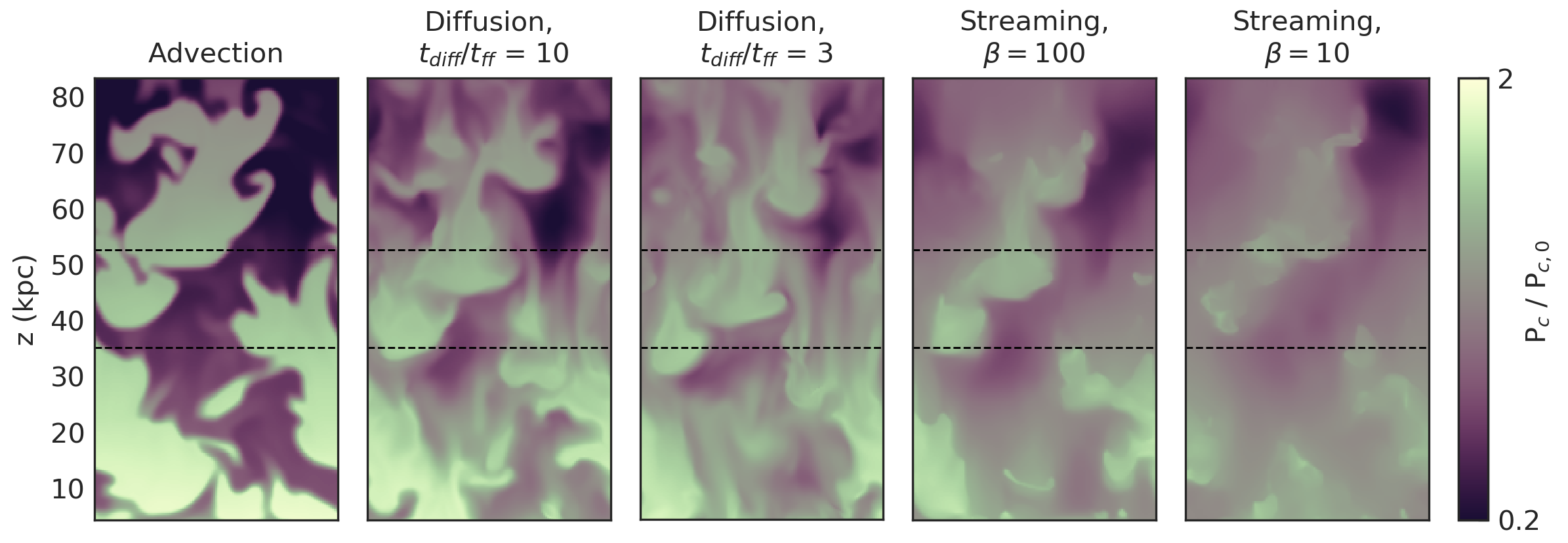}
\caption{ \footnotesize  2D slices of the 3D cosmic ray pressure for simulations with different implementations of cosmic ray transport: advection, diffusion, and streaming. Similarly to Figures~\ref{fig:dens_slice} and \ref{fig:dens_slice_cr}, the dimensions of each panel are $1H \times 1.8H$. All simulations are initialized with $\tcff = 0.3$ and $P_{\rm c}/P_{\rm g} = 1$. Except for the right-most panel, all simulations have $\beta = 100$. The slices are taken at $t = 6\,\tcool$. Cosmic ray transport redistributes cosmic ray pressure from regions of high density to low density. When cosmic ray transport is more efficient (diffusion with $t_{\rm diff} / t_{\rm ff} = 3$ or streaming with $\beta = 10$), the cosmic ray pressure approaches a more homogeneous distribution. }
\label{fig:slice_cr_transport} 
\end{figure*}

\begin{figure*}
\centering
\includegraphics[width=0.7\textwidth]{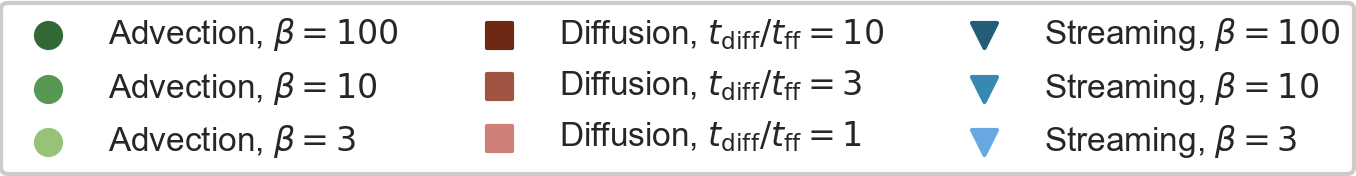}
\vspace*{3mm}
\includegraphics[width=.9\textwidth]{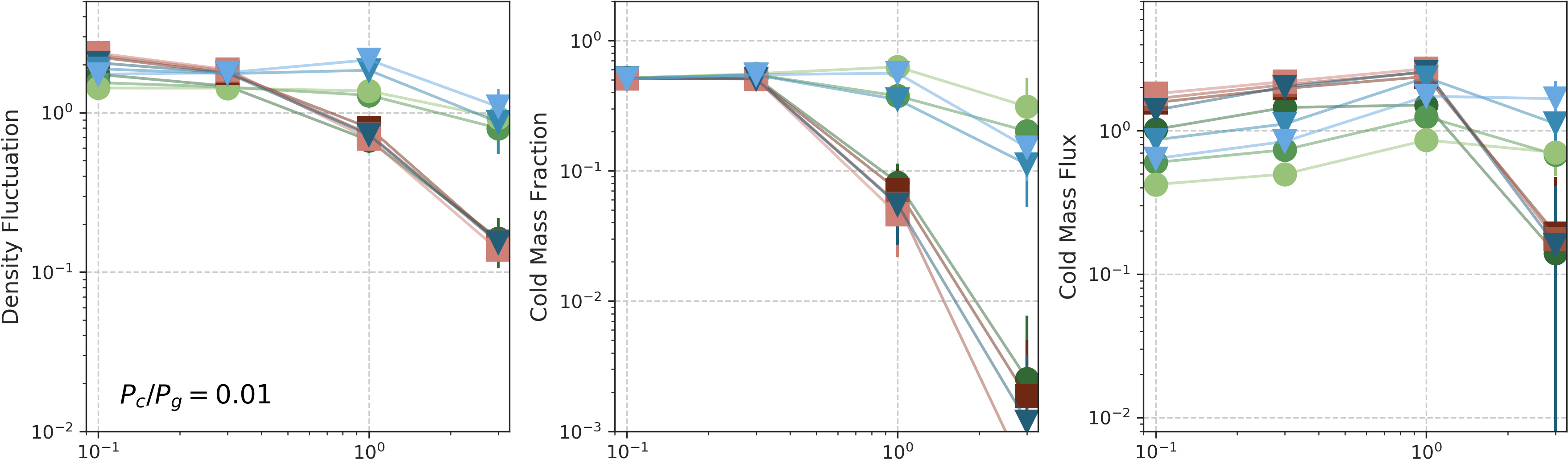}
\includegraphics[width=.9\textwidth]{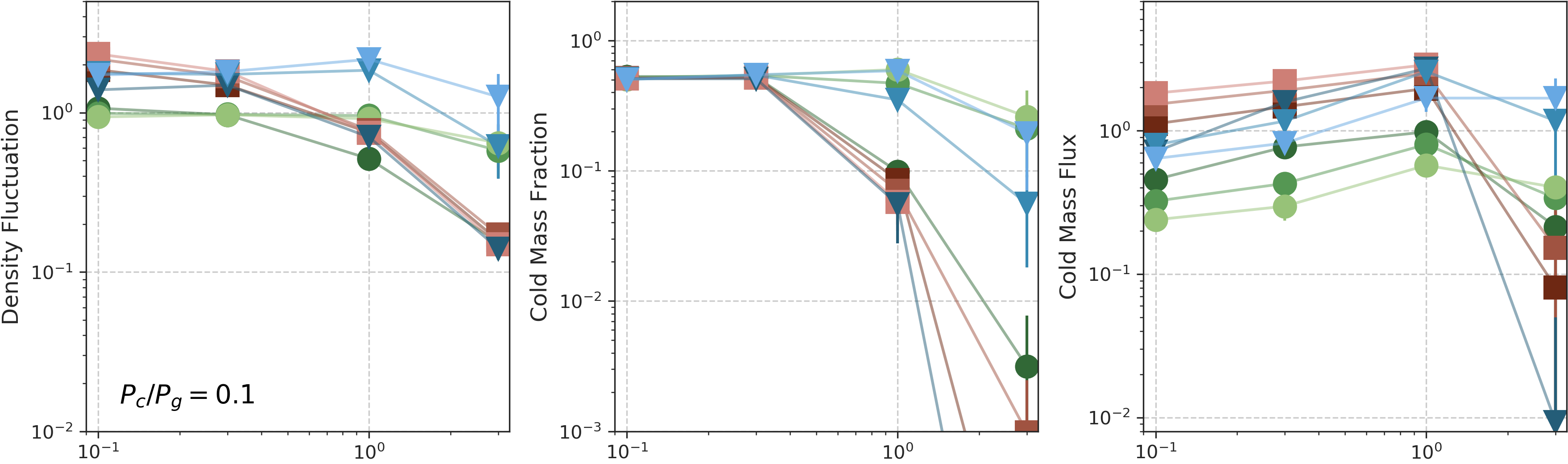}
\includegraphics[width=.9\textwidth]{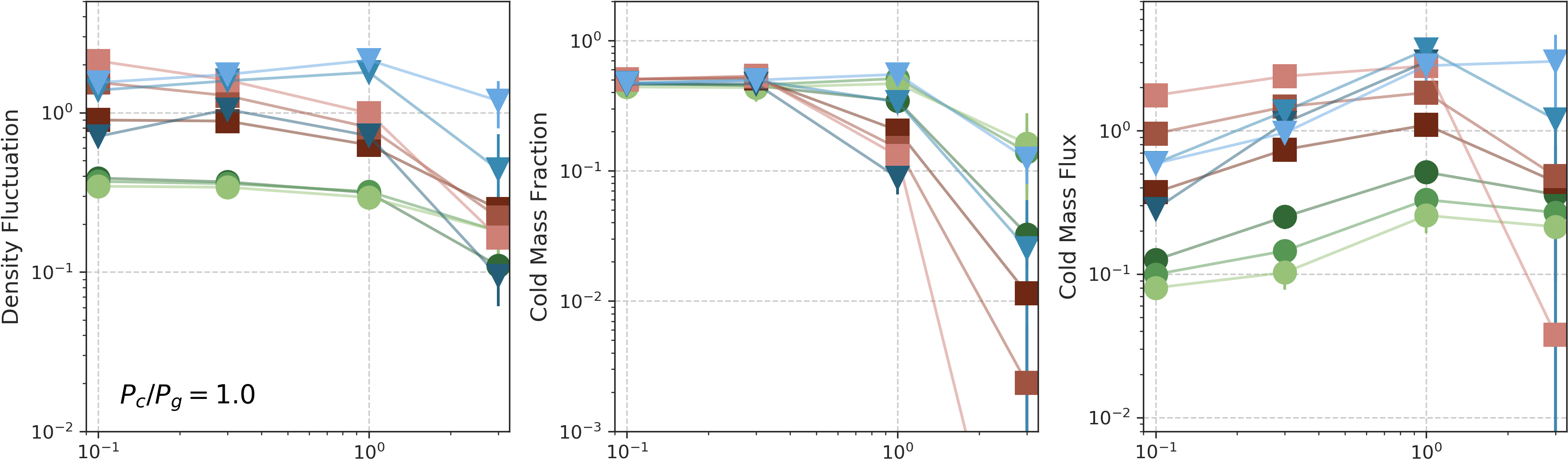}
\includegraphics[width=.9\textwidth]{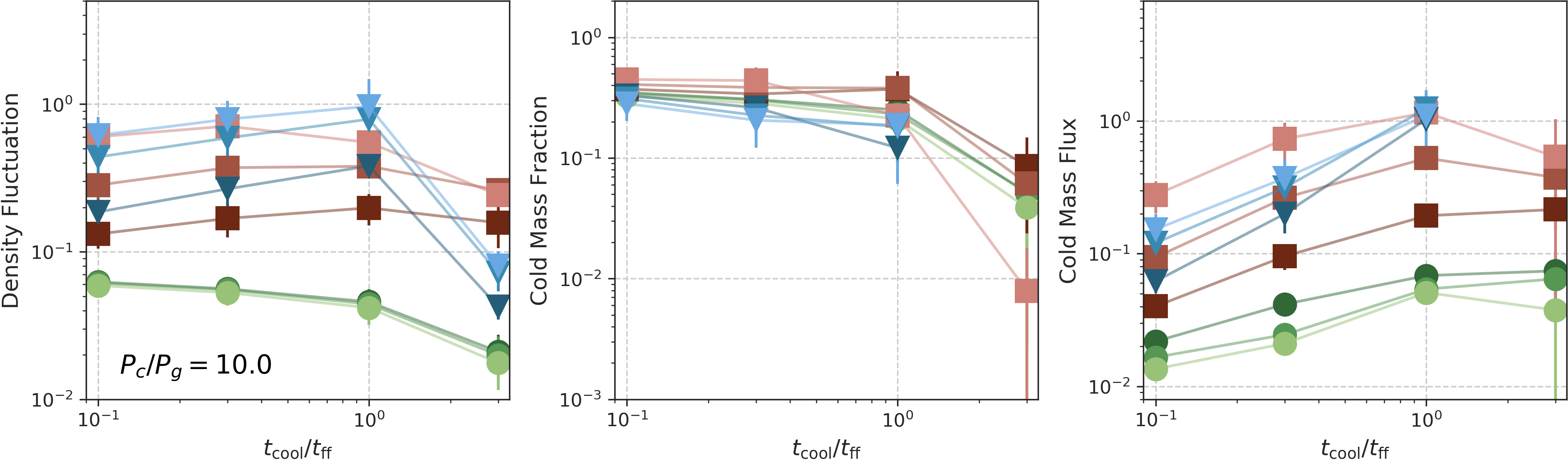}
\caption{ \footnotesize  Similar to Figures~\ref{fig:dens_fluc_tctf} and \ref{fig:dens_fluc_tctf_cr}, the panels above measure the mean density fluctuation ($\langle\delta\rho / \rho\rangle $; \textit{left}), cold mass fraction ($M_{\rm cold} / M_{\rm total}$; \textit{middle}), and cold mass flux ($\dot{M}_{\rm cold} / \dot{M}_{\rm ff}$; \textit{right}) as a function of their initial $\tcff$. The data points are measured between 0.8 and 1.2 $H$ and averaged over all simulation outputs between $4t_{\mathrm{cool}} \leq t \leq 6t_{\mathrm{cool}}$. The error bars show 1 standard deviation. Starting from the top, the rows vary the initial cosmic ray pressure ratio, $P_{\rm c} / P_{\rm g} = 0.01, 0.1, 1, 10$. The green lines show simulations with cosmic ray advection for different initial values of $\beta$, the red lines show simulations with cosmic ray diffusion for three different diffusion timescales, and the blue lines show simulations with cosmic ray streaming for three different initial values of $\beta$.  Cosmic ray transport redistributes cosmic ray pressure so that the resulting density fluctuation, cold mass fraction, and cold mass flux tend to lie between those of MHD-only simulations and simulations with only cosmic ray advection. }
\label{fig:dens_fluc_tctf_transport} 
\end{figure*}
\begin{figure*}[t]
\includegraphics[width=0.5\textwidth]{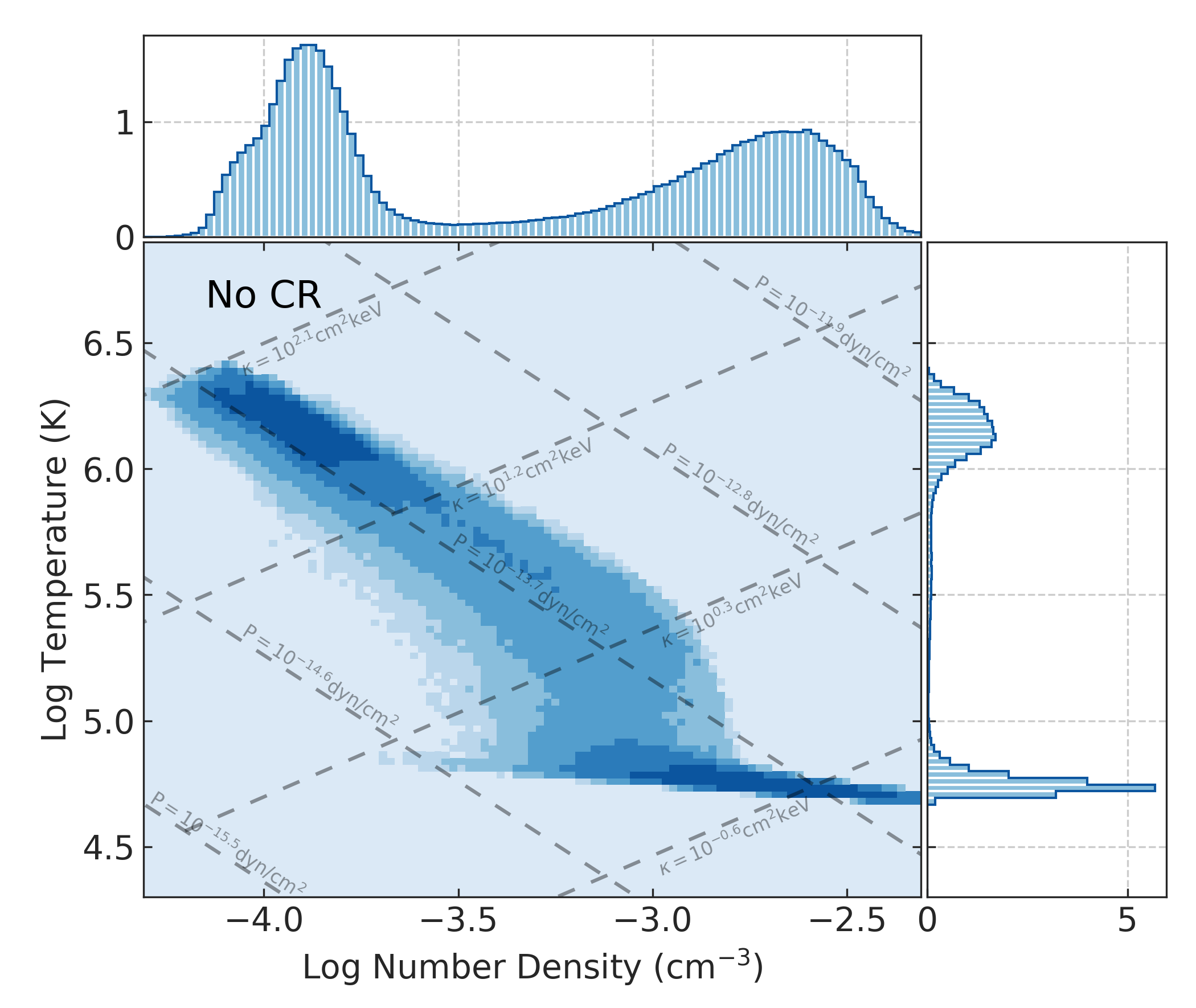}
\includegraphics[width=0.5\textwidth]{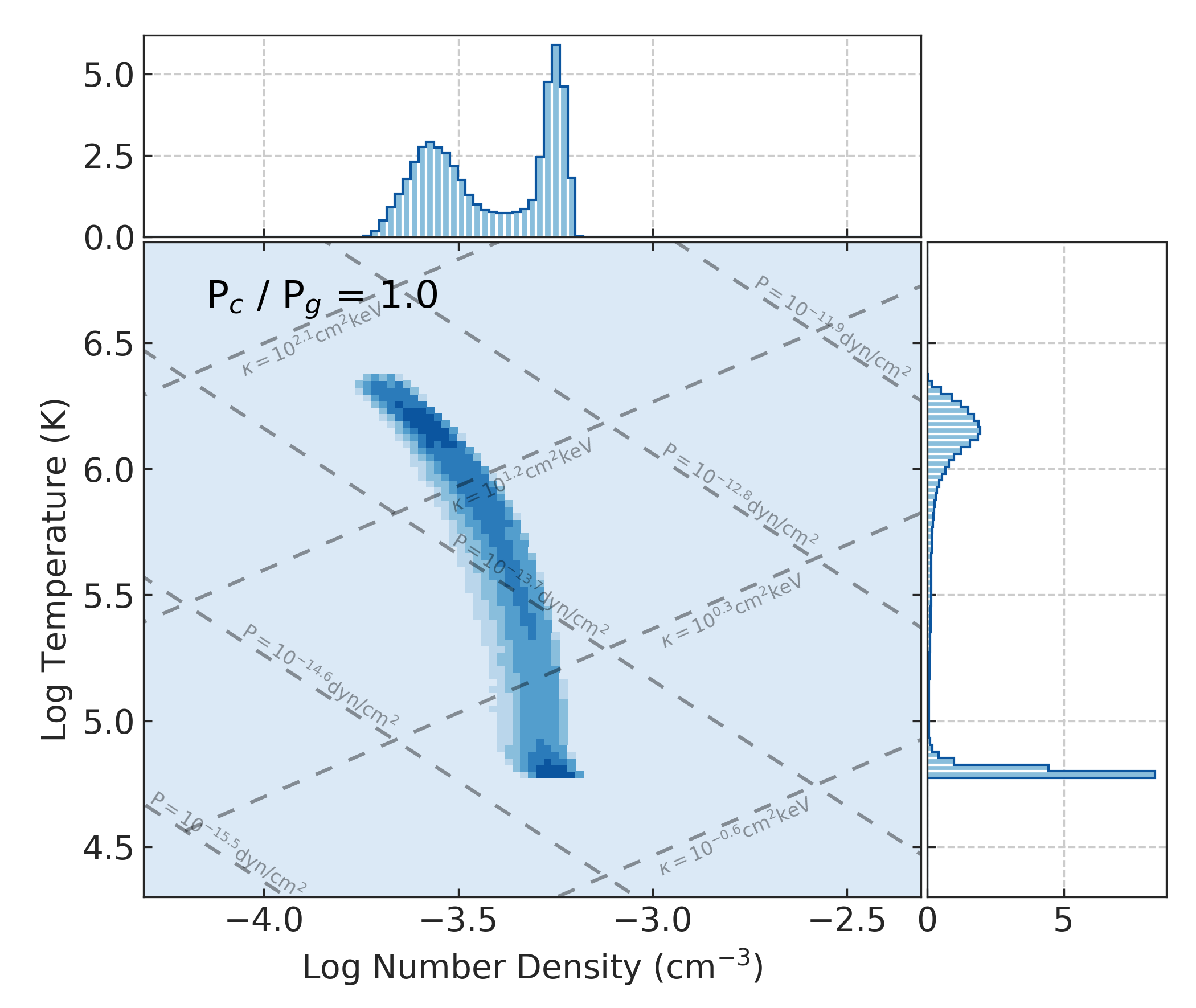}
\includegraphics[width=0.5\textwidth]{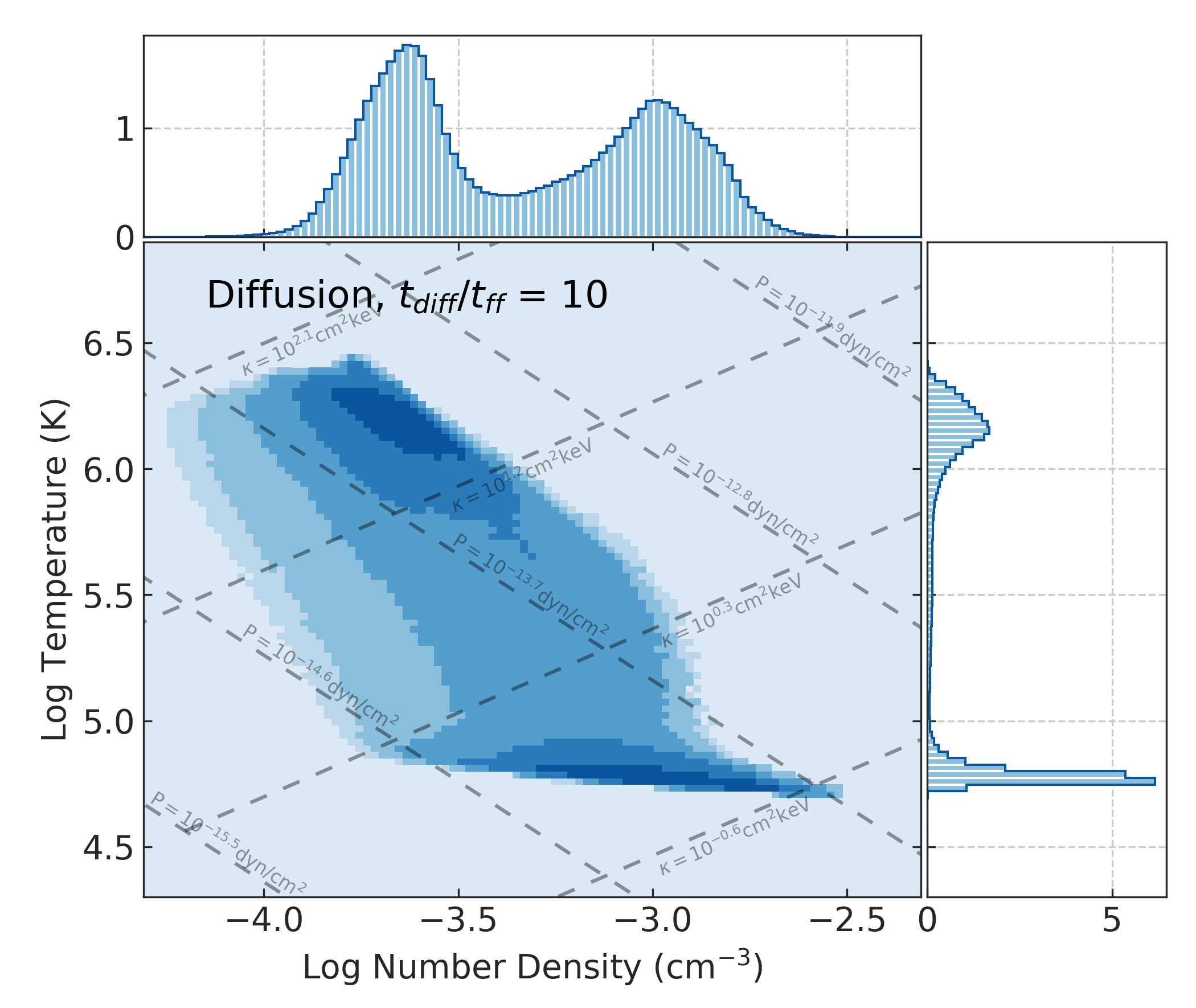}
\includegraphics[width=0.5\textwidth]{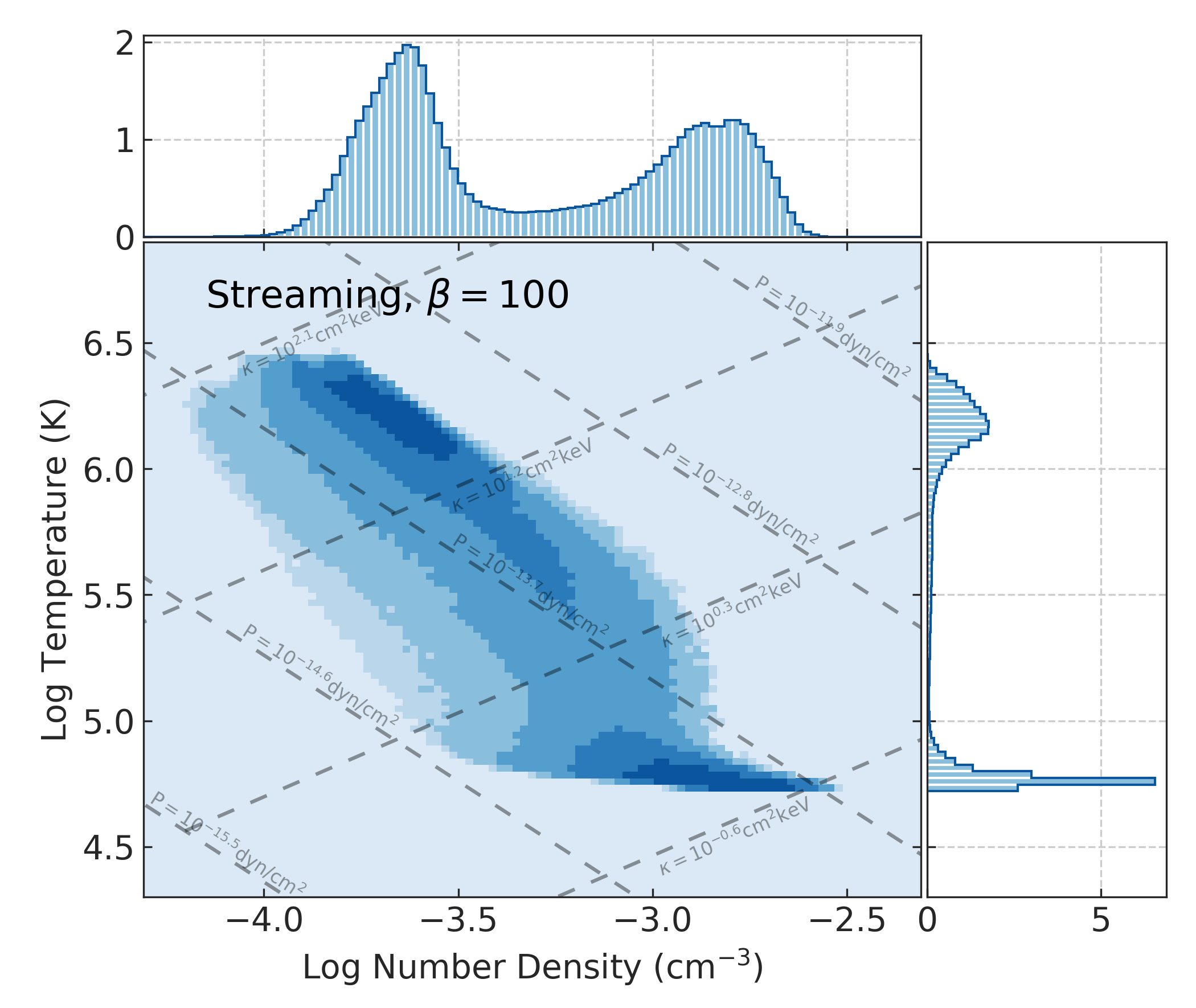}
\caption{ \footnotesize  The mass-weighted 2D histograms of density and temperature for simulations with $\tcff = 0.3$ with various cosmic ray transport prescriptions. The data presented here are an accumulation from all outputs between $t = 4-6 t_{\rm cool}$, measured between 0.8 and 1.2$H$ to minimize the influence of stochastic processes. The top-left panel shows the phase histogram of a simulation without cosmic rays and $\beta = 100$. The remaining panels all have an initial $P_{\rm c}/P_t = 1.0$ and $\beta = 100$, but differ in their models of cosmic ray transport: advection (top-right), diffusion (bottom-left), and streaming (bottom-right).  The dashed lines show contours of constant thermal pressure and entropy. Without cosmic rays, gas cools at constant thermal pressure. With sufficient cosmic ray pressure support, gas cools isochorically. Cosmic ray transport widens the gas temperature-density phase space into low-density, low-temperature regimes that are not present in either simulations without cosmic rays or simulations without efficient cosmic ray transport. }
\label{fig:phase} 
\end{figure*}

Finally,  the third panel of Figure~\ref{fig:dens_fluc_tctf_cr}, shows that the cold mass flux decreases with increased cosmic ray pressure in nearly all cases. Since cosmic ray pressure does not decrease the cold mass fraction at the midplane (see the middle panel), the decrease in cold mass flux is a result of decreased cold gas inflow velocity due to increased cosmic ray pressure support. The increased cold mass flux at $\tcff = 3$, $P_{\rm c}/P_{\rm g} = 1$ is an exception to this trend and is due to the relative increase in the average cold mass fraction. 

Increased cosmic ray pressure does not inhibit the formation of cold gas through thermal instability. However, cosmic ray pressure changes the saturation of the instability from forming small, dense cloudlets that precipitate readily to large, diffuse clouds that remain static. 

\subsection{Thermal Instability and Cosmic Ray Transport}
In the previous section, we built intuition for the impact of cosmic ray pressure on thermal instability in the limit of no cosmic ray transport. Next, we will focus on the impact of cosmic ray transport on the formation of cold gas and its properties. 

As predicted by \citep{Kempski:2020a}, we find that the growth rate of density fluctuations depends on the cosmic ray transport model, so that models with cosmic ray transport have growth rates in-between those without cosmic rays and without cosmic ray transport. In the limit of efficient cosmic ray transport or low cosmic ray pressures, the growth rates approach that of the traditional thermal instability (Eq. \ref{eqn:ti}). However, in this work, we focus on the nonlinear evolution of thermal instability at late times.

As there is no consensus on the dominant cosmic ray transport mechanisms, we investigate both cosmic ray diffusion and cosmic ray streaming at three different transport rates.
In the case of cosmic ray diffusion, we adopt diffusion coefficients, $\kappa_{\rm c}$, such that the diffusion time, $t_{\rm diff} = \kappa_{\rm c} / H^2$, is a constant fraction of the free-fall time: $t_{\rm diff} / t_{\rm ff} = [10, 3, 1]$. Given the free-fall time in our simulations is $t_{\rm ff} = 7.4\times10^8\, \mathrm{yr}$ at the scale height, this corresponds to diffusion coefficients of $\kappa_{\rm c}= [7.9\times10^{28}, 2.6\times10^{29}, 7.9\times10^{29}]\, \mathrm{cm}^2 \mathrm{s}^{-1}$.

In the case of cosmic ray streaming, we define the streaming time, $t_{\rm stream} = H / {\rm v}_{\rm A}$. Since cosmic ray streaming is proportional to the strength of the magnetic field, we simulate cosmic ray streaming for three different initial magnetic field strengths ($\beta = 100, 10, 3$) as a proxy for varying the cosmic ray transport rate. This roughly corresponds to the following ratios of streaming time to free-fall time: $t_{\rm stream} / t_{\rm ff} = [6, 1.8, 1]$. Simulations with cosmic ray streaming also have a perturbative heating term, ${{\mathcal{H}}}_{\rm c}$. Although this heating term is implemented in \textit{addition} to the global heating model, $\mathcal{H}$, heating due to cosmic ray streaming is expected to be much smaller than the global heating model.

Figure~\ref{fig:slice_cr_transport} shows the 2D slices of the cosmic ray pressure in simulations with different models of cosmic ray transport. All simulations have $\tcff = 0.3$,  $P_{\rm c}/P_{\rm g} = 1$, and are pictured after $t = 6 t_{\rm cool}$. Cosmic ray transport redistributes cosmic ray pressure from regions of high cosmic ray pressure to regions of low cosmic ray pressure. The more more efficient the cosmic ray transport (diffusion with $t_{\rm diff} / t_{\rm ff} = 3$ or streaming with $\beta = 10$), the more uniform the distribution of cosmic ray pressure. Consequently simulations with more efficient cosmic ray transport provide less non-thermal pressure support to cold gas, and form smaller cold gas structures. Simulations with cosmic ray diffusion produce relatively large cold gas clouds with long connecting filaments. Simulations with cosmic ray streaming have smaller, more compact cold gas clouds. 

Figure~\ref{fig:dens_fluc_tctf_transport} shows the average density fluctuation, cold mass fraction, and inflowing cold mass flux as a function of the initial $\tcff$. Consistent with previous figures, each point represents the average quantity and its standard deviation measured between 0.8 and 1.2 $H$, from all outputs between $t = 4-6 t_{\mathrm{cool}}$. The rows are organized by the initial cosmic ray pressure ratio, ranging from $P_{\rm c} / P_{\rm g} = 0.01$ (top) to $P_{\rm c} / P_{\rm g} = 10.0$ (bottom). The different lines are colored by the cosmic ray transport model: advection (green), diffusion (red), and streaming (blue).  We remind the reader that for the majority of the lines, the corresponding MHD-only run has $\beta = 100$ (see Figure~\ref{fig:dens_fluc_tctf}). However, two runs with cosmic ray advection and streaming should be compared against runs with $\beta = 10$ and $\beta = 3$. 

For the most part, simulations with cosmic ray transport lie between simulations without cosmic rays and simulations with only cosmic ray advection. Runs with relatively slow cosmic ray streaming and diffusion exhibit many of the same characteristics as runs with only cosmic ray advection: cold gas has lower densities and less precipitation than predicted by purely thermal pressure support. The more efficient the cosmic ray transport, the more closely it resembles the MHD-only thermal instability. This is because cosmic ray transport redistributes cosmic ray pressure from regions of high cosmic ray density to regions of low cosmic ray density. When cosmic ray transport is very efficient relative to the cooling timescale, the cosmic ray pressure profile becomes uniform (see the discussion in Section \ref{sec:gas_phase}). As cosmic ray pressure decreases in the cold gas clouds, the cold gas density and cold mass flux increase.

We note that when cooling times are long relative to the free-fall time, the temporal variation of a single model can be larger than the variance between models. This makes it difficult to draw detailed comparisons between models with different cosmic ray physics. Although the amount of cold gas formed is comparable, the biggest difference is that with sufficiently strong non-thermal pressure support, any cold gas that forms at the scale height remains at the scale height. For runs with cosmic ray streaming, the perturbative heating term results in a decrease in the cold mass fraction and an increase in its flux towards the midplane. This perturbative heating destroys cold gas in simulations with high cosmic ray pressures ($P_{\rm c} / P_{\rm g} \geq 1$) that are evolved for longer physical times ($\tcff \ge 1$).

\begin{figure*}
\includegraphics[width=0.95\textwidth]{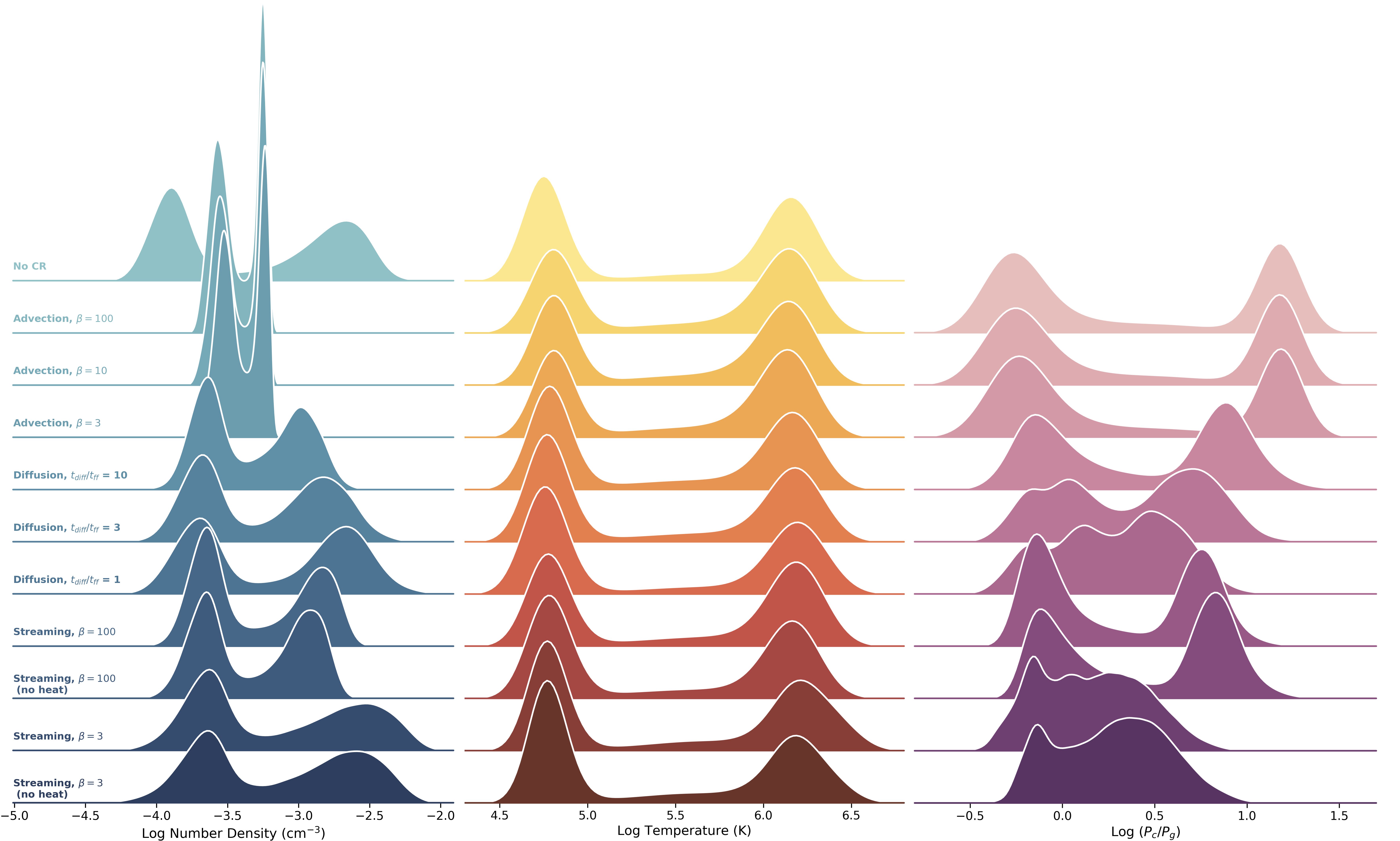}
\includegraphics[width=0.95\textwidth]{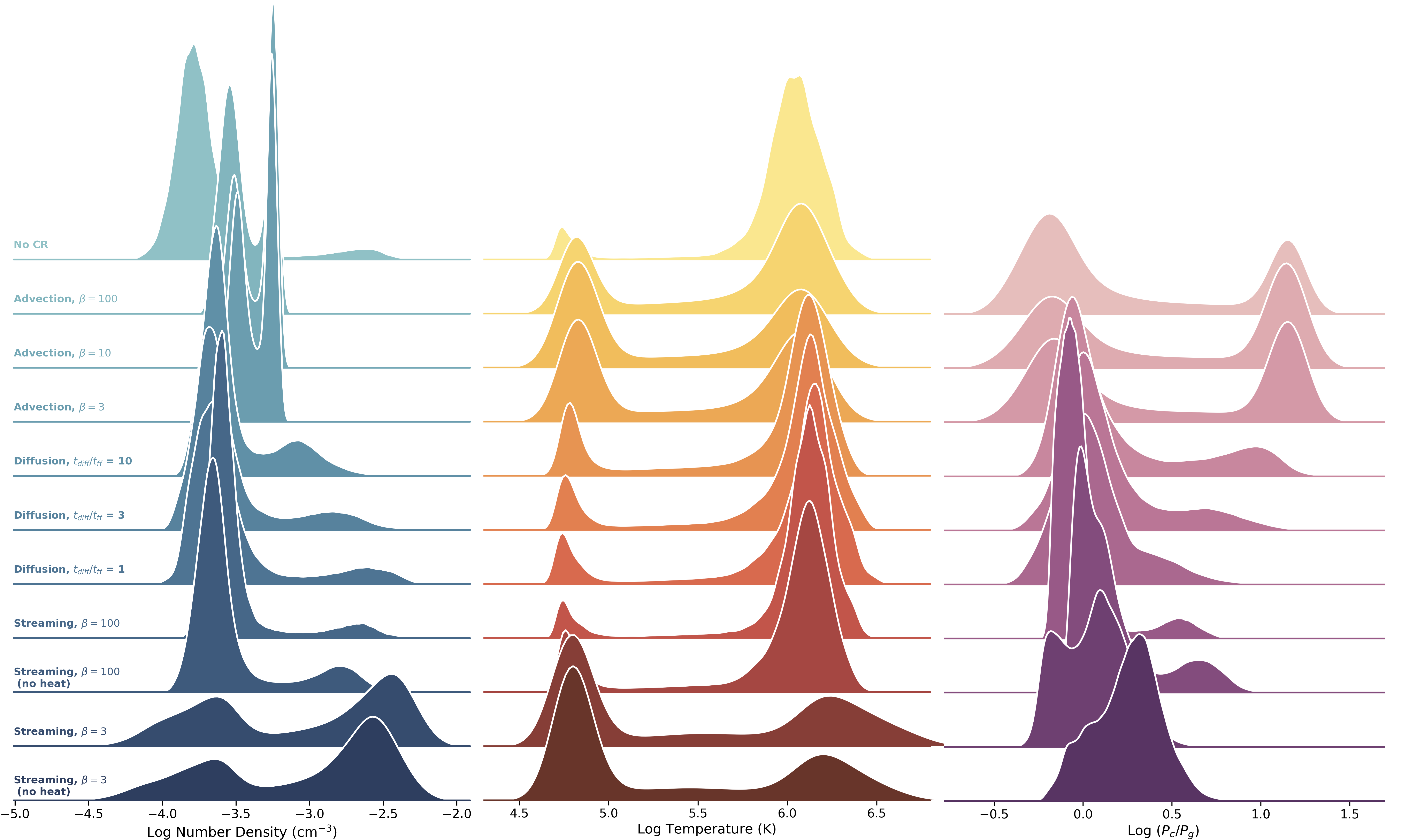}
\caption{ \footnotesize The mass-weighted probability density functions (PDFs) of the gas density, temperature, and cosmic ray pressure ratio for simulations with $P_{\rm c} / P_{\rm g} = 1$ and $\tcff = 0.3$ (top) and $\tcff = 1.0$ (bottom). The data in the PDFs is an aggregate from all simulation outputs between $t = 4-6 t_{\rm cool}$, measured between 0.8 and 1.2 scale heights. The PDFs are normalized so that the area under each distribution integrates to 1. Different implementations of cosmic ray transport change the distribution of the ratio of cosmic ray pressure to gas pressure, which in turn sets the distribution of gas densities. When cold gas has high cosmic ray pressures, the density contrast between the cold and hot gas phases decreases. When cooling times are short ($\tcff \le 1$), cosmic ray physics does not have a significant impact on the distribution of gas temperatures. However, when cooling times are long enough for the free-fall time to be important ($\tcff \gtrsim 1$), cosmic ray pressure prevents precipitation and keeps cold gas near the scale height. This effect is strongest for simulations with cosmic ray advection and modest in simulations with efficient cosmic ray transport. }
\label{fig:joy_division} 
\end{figure*}

\subsection{Impact of Cosmic Rays on Gas Phase}\label{sec:gas_phase}
Figure~\ref{fig:phase} shows 2D histograms of gas density and temperature, comparing simulations with $\tcff = 0.3$ and different cosmic ray transport. The colors show probability density, which is quantified on the top and right of each panel. The dashed lines show contours of constant thermal pressure and entropy. 

Without cosmic rays (top left), the gas first cools at constant thermal pressure before dipping to lower pressures once it reaches the most rapid cooling regime, around $T\sim10^5\K$. This pressure decrement is likely the result of insufficient spatial resolution to capture the cooling length scale $c_{\rm s}t_{\rm cool}$ \citep[e.g.,][]{Fielding:2020, Fielding:2020b}. Although the cooling curve cuts off at $T = 5\times10^4\K$ in these simulations, we expect the general shape of the phase diagram to be the same if a lower cutoff temperature were adopted. The loss of thermal energy due to cooling leads to compression, which maintains pressure equilibrium (when the cooling length is resolved). 

With added cosmic ray pressure (top right), the loss of thermal energy does not lead to a significant \emph{total} pressure gradient, so there is negligible compression and cooling proceeds at nearly constant density. In the simulation highlighted here, cosmic ray pressure is initialized to be equal to the gas pressure, which is sufficiently strong to inhibit compression and cause the gas to cool isochorically. 

Cosmic ray transport (bottom row) redistributes cosmic ray pressure so that the resulting gas temperatures and densities bridge the gap between simulations without cosmic rays and simulations without cosmic ray transport. However, cosmic ray transport also significantly broadens the temperature-density phase distribution of gas. Specifically, simulations with cosmic ray transport have more intermediate-temperature gas at low densities, which is more likely to hold photoionized (rather than collisionally ionized) gas. 

In Figure~\ref{fig:joy_division}, we compare the probability distribution functions (PDFs) of the gas density, temperature, and the ratio of $P_{\rm c} / P_{\rm g}$ as a function of the initial cosmic ray pressure ratio and cosmic ray transport mechanism. The top panels show simulations with an initial $\tcff = 0.3$ and the bottom panels show simulations with an initial $\tcff = 1.0$. Both sets of simulations have an initial $P_{\rm c}/P_{\rm g} = 1$. Each row shows the properties of simulations with different cosmic ray transport models. The data in the PDFs represents gas near the scale height ($0.8 H \leq z \leq 1.2 H$) and is accumulated from all outputs between 4 and 6 $\tcool$. 

For simulations with $\tcff = 0.3$, the distribution of gas temperatures is similar for a variety of cosmic ray transport prescriptions. However, the density profiles vary substantially. Without cosmic rays, the gas forms a two-phase medium where the density profiles complement the temperature profiles such that gas is always close to thermal pressure equilibrium. Cosmic ray pressure support enables the cold and hot gas to have a lower density contrast. This effect is most pronounced in simulations with advection as the only form of cosmic ray transport, which have distinctly bimodal distributions of cosmic ray pressure between the hot and cold gas phases. Cosmic ray diffusion and streaming move cosmic ray pressure out of the regions where it is concentrated, thus removing the impact of non-thermal pressure support and thereby allowing the cold clouds to compress. 

For simulations with high $\tcff = 1$, the temperature distributions are affected by the choice of cosmic ray physics. There are several factors at play. High cosmic ray pressures not only alter the gas density, but they can also keep low-entropy gas from precipitating. This effect becomes relevant if the cooling time is equal to or greater than the dynamical time (see Figures~\ref{fig:dens_fluc_tctf_cr} and \ref{fig:dens_fluc_tctf_transport}). In this case, the MHD-only run has less cold gas at the scale height because that cold gas has precipitated out. Simulations with fast streaming also have strong magnetic fields which alter the amount of cold gas formed. For example, although both simulations with fast cosmic ray diffusion ($t_{\rm diff}/t_{\rm ff} = 1$) and fast cosmic ray streaming ($\beta = 3, t_{\rm stream} / t_{\rm ff} = 1$) have the same cosmic ray transport time scale, simulations with stronger magnetic fields have a significantly larger fraction of cold gas.

The distribution of the cosmic ray pressure ratio, $P_{\rm c} / P_{\rm g}$ is a function of the transport mechanism. Simulations with only cosmic ray advection have a distinctly bimodal distribution with the cold gas receiving substantially more pressure support than the hot gas. In the presence of both streaming and diffusion, the bimodality in cosmic ray pressure ratios shrinks. The more efficient the cosmic ray transport, the more the cosmic ray pressure ratio distribution converges on the initial value. In simulations evolved for a longer physical time ($\tcff \geq 1$), the convergence on a single value is more pronounced since the cosmic rays have had more time to propagate. Most notably, the bimodality distribution of cosmic ray pressure ratios is inversely proportional to the bimodality of gas densities. Simulations in which cold gas builds up higher cosmic ray pressure than hot gas can cool isochorically. Conversely, simulations where the cosmic ray pressure ratio is roughly constant have a larger density contrast between the cold and hot phases.

\subsection{Impact of Cosmic Ray Heating}
The transport model for cosmic ray streaming traditionally includes a heating term through which cosmic rays transfer energy to the thermal gas. The simulations labeled ``streaming'' described so far all have this perturbative heating source. However, since our simulation set-up and model for gas cooling and heating is very idealized, we also include simulations with cosmic ray streaming \textit{without} the additional cosmic-ray heating term, ${{\mathcal{H}}}_{\rm c}$. This models the scenario in which cosmic ray heating is just one of the many potential sources that contributes to the overall heating that ultimately balances cooling in the simulations. 

We include simulations with cosmic ray streaming but no perturbative heating in Figure~\ref{fig:joy_division}, labeled with the appended ``no heat''. Overall, we find that the inclusion of the cosmic ray heating term has a modest impact on the gas phase. Since cosmic ray heating is a function of both the cosmic ray pressure gradient and the \alfven velocity, we expect it to be important when the magnetic field is strong and the cosmic ray pressure gradient is steep. That said, even the relatively weak magnetic fields in our fiducial simulations ($\beta = 100$) are enough to have an observable effect due to cosmic ray heating. 

Without the added cosmic ray heating term, simulations with cosmic ray streaming have less of a density contrast between hot and cold gas phases and a narrower distribution in the hot gas phase than simulations that do include cosmic ray heating. Furthermore, the distribution of the cosmic ray pressure ratio, $P_{\rm c} / P_{\rm g}$ is shifted towards higher values when the cosmic ray heating term is omitted. This is expected behavior as the cosmic ray heating term is expected to remove cosmic ray energy and use it to heat the gas. 

Although cosmic ray heating has modest effects on the global temperature and density profiles it is likely important for setting the local cold gas properties. The effect of cosmic ray heating is strongest at the cold cloud edge, where the cosmic ray pressure gradient is large. In this case, cosmic ray heating can broaden the boundary between the cold and hot gas phases, with interesting implications for the observed ion abundances and kinematics \citep{Wiener:2017}. Like \citep{Heintz:2020}, we find that the effects of cosmic ray heating are spatially offset from regions with the shortest cooling times: cosmic ray heating predominantly affects diffuse cool gas, by shifting it to higher temperatures. This effect is present in the lack of low temperature, low density gas in the bottom right panel of Figure \ref{fig:phase}.  While the effects of cosmic ray heating are modest in Figure \ref{fig:joy_division}, they primarily affect the temperature of the warm gas (by directly heating it) and the density of the cold gas (indirectly, by removing the available cosmic pressure that would have supported the cold gas). Changing the cloud boundary may also alter the mass and momentum transfer between the cold and hot gas phases \citep{Fielding:2020}.

\begin{figure*}
\centering
\includegraphics[width=0.7\textwidth]{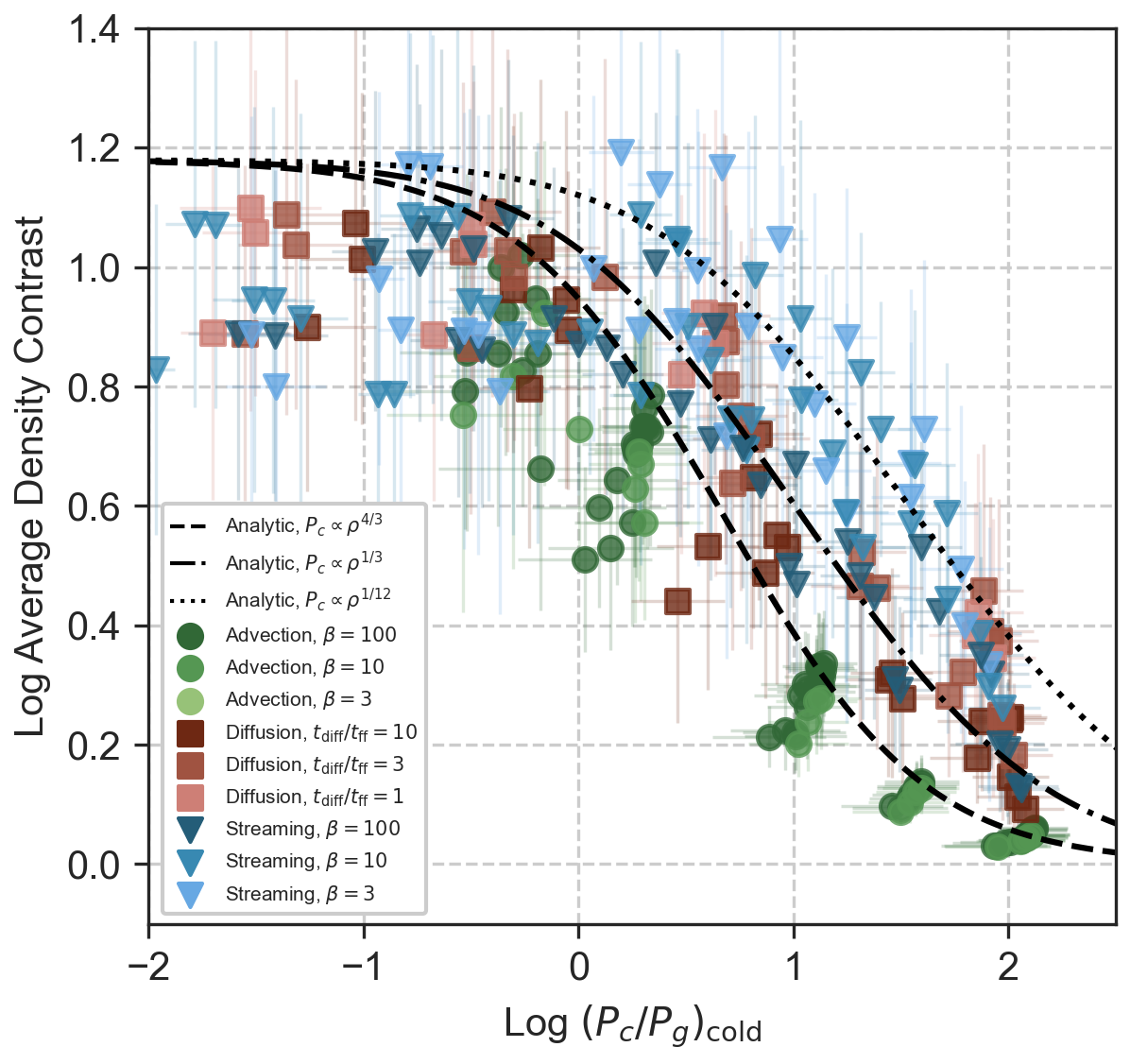}
\caption{ \footnotesize The average density contrast between the cold and hot gas phases ($\langle\rho_{\rm cold}\rangle / \langle \rho_{\rm hot}\rangle$) as a function of the average ratio of $P_{\rm c} / P_{\rm g}$ in cold gas clumps. The points show data from all simulations described in the fiducial parameter suite (Section \ref{sec:parameter}) that form cold gas, both at fiducial and high resolution. The data were measured between 0.8 and 1.2 scale heights and averaged over 20 outputs between $t = 4-6 t_{\rm cool}$. The different colors indicate simulations with different prescriptions of cosmic ray transport: advection (green), diffusion (red), and streaming (blue). The black lines show predictions for the density contrast for various possible scalings of cosmic ray pressure with gas density, assuming $\Theta = 20$ and $\beta_{\rm cold} = 3$ (Eq.~\ref{eqn:dens_contrast}). In all cases, the density contrast between the cold and hot gas phases decreases with increased cosmic ray pressure. However, the rate of that decrease depends on the transport model, which effectively changes the degree to which cosmic ray pressure scales with gas density.}
\label{fig:family_portrait} 
\end{figure*}

\subsection{Cold Gas Density Contrast} \label{sec:dens_contrast}
Ultimately, the impact of the wide range of cosmic ray transport model parameters can be summarized by the distribution of cosmic ray pressure between the cold and hot phases and its impact on the gas phase. 

In Figure~\ref{fig:family_portrait}, we show the gas density contrast, $\langle\rho_{\rm cold}\rangle / \langle\rho_{\rm hot}\rangle$, as function of the cosmic ray pressure ratio ($P_{\rm c} / P_{\rm g}$) in the cold gas. This figure includes all simulations with cosmic ray physics that form cold gas, at both fiducial and high-resolution. Both the average density contrast and cosmic ray pressure are measured in a region near the midplane, $0.8H \leq |z| \leq 1.2H$. In order to mitigate stochastic effects between outputs, the points show data averaged over 20 outputs between $t = 4t_{\rm cool}$ and $t = 6t_{\rm cool}$. The green circles show simulations with cosmic ray advection for $\beta = 100$ (light green), $\beta = 10$ (medium green), and $\beta = 3$ (light green). The red squares show simulations with cosmic ray diffusion with $t_{\rm diff}/t_{\rm ff} = 10$ (dark red), $t_{\rm diff}/t_{\rm ff} = 3$ (medium red), and $t_{\rm diff}/t_{\rm ff} = 1$ (light red). The blue triangles show simulations with cosmic ray streaming with $\beta = 100$ (dark blue), $\beta = 10$ (medium blue), and $\beta = 3$ (light blue). The black lines show predictions for density contrast profiles (Eq.~\ref{eqn:dens_contrast}) as a function of how cosmic ray pressure scales with gas density, $P_{\rm c} \propto \rho^{\gamma_{\rm c, eff}}$. In these predicted profiles, we use a temperature contrast of $\Theta = 20$, corresponding to $T_{\rm min} = 5\times10^4$ K. 

We assume a fixed $\beta_{\rm cold} = 3$, which is roughly the average $\beta$ in cold gas for simulations with an initial $\beta = 100$. Although the magnetic field is initialized with $\beta = 100$ everywhere in most simulations, the magnetic field strength is amplified in dense gas due to the flux-freezing assumption in ideal MHD. Furthermore, the magnetic field becomes tangled on large scales, which effectively slows cosmic ray transport. Simulations with higher initial magnetic field strengths have more magnetic pressure support in cold clouds than accounted for in the analytic lines. 

Overall, the average density contrast between the cold and hot gas phases decreases with increasing cosmic ray pressure (Section \ref{sec:results_crpressure}). However, the slope of that decrease depends on the cosmic ray transport prescription. In the pure advection case, we expect $\gamma_{\rm c, eff} = \gamma_{\rm c}= 4/3$, which corresponds to the dashed black line. The simulated data follow the expected density contrast relation. The scatter in the advection simulations at low cosmic ray pressures is likely due to a combination of resolution effects, different initial values of $\tcff$, and the relative importance of magnetic fields (which are not accounted for in the plotted analytic model). Further investigations of the effects of resolution and the choice of density contrast, are discussed in Appendices \ref{sec:appendix_resolution} and \ref{sec:appendix_Tmin}. 

Since cosmic ray transport redistributes cosmic ray pressure from overdense regions, we expect simulations with cosmic ray transport to have a weaker scaling of cosmic ray pressure and gas density than in simulations with only cosmic ray advection ($\gamma_{\rm c, \rm eff} < 4/3$). However the exact value of $\gamma_{\rm c, \rm eff}$ depends on the details of cosmic ray transport and assumptions about the impact of cosmic ray heating. For simulations with cosmic ray transport (diffusion with $t_{\rm diff}/t_{\rm ff} = 10$ and streaming with $\beta = 100$), we empirically find that $\gamma_{\rm c, \rm eff} \simeq 1/3$, which is modeled by the dash-dotted line in Figure~\ref{fig:family_portrait}. For context, \citet{Wiener:2017} predict $\gamma_{\rm c, \rm eff} = 2/3$ in the case of cosmic ray streaming where cosmic ray heating balances gas cooling. Simulations with relatively faster cosmic ray transport have much weaker scalings of cosmic ray pressure with gas density. In the extreme case that cosmic ray transport is so efficient that $\gamma_{\rm c, \rm eff} \rightarrow 0$, we expect the predicted density contrast to be unchanged as a function of $P_{\rm c}/P_{\rm g}$.

\section{Discussion}\label{sec:discussion}

\subsection{The CGM Pressure Problem}
Based upon the original work of \cite{Mo:1996}, numerous two-phase models of the CGM envision cold clouds condensing out of a hot medium as a result of hydrostatic instabilities. Such models have successfully reproduced many of the observed properties of high-velocity clouds in the Milky Way halo \citep[e.g][]{Maller:2004, Putman:2012}. 

Photoionization modeling of observed low-ionization-state metal-ion column densities derives cold CGM gas volume densities that are at least an order of magnitude lower than predicted by traditional two-phase models \citep[e.g.][]{Werk:2014, Stern:2016}. There are two possible explanations for such low densities of cold CGM gas: 1) the cold gas is in thermal equilibrium with the hot CGM gas, which also has a much lower density than predicted by traditional two-phase models, or 2) the cold gas is out of thermal pressure equilibrium with the hot gas phase. 

If cold gas is out of thermal pressure equilibrium, the lack of pressure support would imply very short lifetimes of cold clouds, in stark contrast to the $\sim$100 Myr lifetimes required by simulations of the CGM that predict a cold gas origin in recycled ISM material \citep[in which halo gas is largely in hydrostatic equilibrium;][]{Ford:2014, Angles:2017, Tumlinson:2017}. In order to bring observations in agreement with simulations, there either needs to be some physical process capable of continually generating the observed abundance of cold gas or some additional source of non-thermal pressure.

Simulations show that the average magnetic field in the CGM can be as high as $\sim 0.1 - 0.5 \mu G$ \citep{Pakmor:2020, Nelson:2020, vandeVoort:2020}. These magnetic fields both enhance thermal instability and provide pressure support to the cold gas that forms. Due to flux freezing, the cold gas clumps are dominated by magnetic pressure ($\beta \ll 1$) even though the background halo has relatively low magnetic pressure ($\beta \geq 100$) \citep{Nelson:2020}. Since there are few observational constraints of magnetic fields in galactic halos, it is unclear if such high magnetic field values are typical. However, these simulations fall well within the upper limit of magnetic field strength (2$\mu G$) determined by \citep{Lan:2020}, and provides another mechanism through which non-thermal pressure support can alleviate the pressure problem.

Galaxy simulations that include cosmic ray physics also find that cosmic ray pressure supports cold CGM gas at low densities \citep{Salem:2016, Butsky:2018, Buck:2020}. In a cosmic ray pressure-dominated halo, cool, warm, and hot gas phases can exist at the same gas density \citep{Ji:2020}. Although there are few observational constraints of cosmic ray pressures in the CGM, simulations predict that a cosmic ray pressure-dominated halo could be consistent with existing $\gamma$-ray observations \citep{Chan:2019, Ji:2020}. 

In our simulations, cosmic ray pressure (and in some cases, magnetic pressure) supplements thermal pressure to support low-density cold gas. In simulations with low initial cosmic ray pressures, cold cloudlets are supported by cosmic ray pressure while the background medium is dominated by thermal pressure. Simulations with high initial values of cosmic ray pressure allow gas to cool isochorically and form large, diffuse cold gas clouds. Cosmic ray transport complicates this behavior by redistributing cosmic ray pressure from the cold gas to the surrounding medium. \emph{Overall, our findings highlight that even modest cosmic ray pressures can resolve the perceived pressure problem with two-phase halo models by providing a viable physical mechanism to maintain the cold gas while out of thermal pressure equilibrium.} 

\subsection{Cold Cloud Sizes}
The sizes of cold gas clouds remains an unresolved issue within the CGM community; CGM cloud sizes (given as path lengths) derived from absorption-line spectroscopy at low and high redshift range from 1 pc $-$ 1 Mpc \citep[e.g.][]{Rauch:2001, Prochaska:2009, Stocke:2013, Chen:2014, Werk:2014, Crighton:2015, Stern:2016, Zahedy:2019}. The challenge of this measurement is essentially two-fold: absorption-line probes of the CGM rely heavily on model assumptions for a length-scale estimate (e.g. strength and nature of ionization source) and most observations are limited to a single sightline per galaxy. 

Nonetheless, recent studies have used additional, novel techniques to place constraints on the physical scales of structures in the CGM.  \citet{Rubin:2018} analyzed the CGM of 27 star-forming galaxies along spatially-resolved background-galaxy lines of sight to determine that the coherence scale of Mg II absorbers is $> 1.9$ kpc. Specifically, cold gas clouds are either  $> 2$ kpc in radius themselves or represent collections of much smaller clouds with similar properties. \citet{Rudie:2019} used observations of a galaxy halo probed by a lensed background quasi-stellar object (QSO) to show that while warm (O VI bearing) gas has structures that are $> 400$ pc,  lower-ionization cold gas exhibits significant variations on the same scales, implying that cold gas clouds have sizes $<$ 400 pc. \citet{Werk:2019} studied multiphase absorption at the Milky Way disk-halo interface, and showed that warm gas clouds were larger than 1 kpc, whereas cold gas was clumpy on scales down to 10 pc, possibly implying cold clouds are smaller than 10 pc \citep[see also:][]{Bish:2019}. \citet{Zahedy:2019} find that the cold clump thickness in luminous red galaxies (LRGs) is between 10 pc and 1 kpc, with a median of 120 pc.  

Theoretically, cold gas is predicted to have a characteristic cloudlet size $\ell_{\rm cloudlet} \sim \mathrm{min}(c_s t_{\mathrm{cool}})$ \citep{McCourt:2018}. Using this model, a cold cloud in typical CGM conditions  can have a characteristic size somewhere between $\ell_{\rm cloudlet} \sim 1 {\rm pc} - 1 \rm {kpc}$, depending on the gas pressure, metallicity, and UV background radiation. Additionally, depending on the conditions of the surrounding medium, the true size of cold gas clouds may be significantly larger, as clouds can coagulate to form larger structures \citep{Gronke:2020}. 

Resolving sub-parsec scales is currently impossible in galaxy-scale simulations and prohibitively expensive in even idealized meso-scale simulations. Recently, a number of groups have demonstrated that enhancing the spatial resolution in and beyond the galactic halo results in more cold gas due to increased accuracy in the formation and retention of cool gas structures \citep{vandeVoort:2019, Hummels:2019, Mandelker:2019, Peeples:2019, Suresh:2019}. However, the size of the best-resolved CGM cells in these simulations is still hundreds of times larger than the expected cold gas cloud size. Indeed the properties of cold CGM gas in galaxy simulations have not converged with resolution, implying that current state-of-the art simulations are not resolving the true cold gas physics. Likewise, \citet{Nelson:2020} find cold cloud sizes of 0.5 - 1 kpc in TNG50, but also warn that the cloud sizes do not converge, even at their highest resolution.

In our simulations, cosmic ray pressure increases cloud sizes. The cloud size (and density) is a function of the degree of cosmic ray pressure support in the cold cloud. By altering gas densities, cosmic ray pressure alters the predicted cloud-size ansatz, $\ell_{\rm cloudlet} \sim {\rm min}(c_s t_{\mathrm{cool}})$. If cosmic ray pressure support is modest, the resulting gas will still likely be a mist comprised of numerous tiny cloudlets, albeit with larger cloudlet sizes. However, if cosmic ray pressure dominates over thermal pressure in halo gas, then we can expect that cold gas has similar gas densities to the hot gas phase.  In the extreme case, if the cold gas has the same density as the hot gas, we can expect that cold cloud sizes would be roughly $\sim 900$ times larger than those predicted assuming purely thermal pressure balance (assuming a temperature contrast of 100 and $\eta = 100$ in Eq.~\ref{eqn:rcloud_cr}).

With this generous boost in cloud size, we would expect the maximum cold cloud size in a cosmic ray pressure-dominated CGM, with inefficient cosmic ray transport to be $\ell_{\rm cloudlet} \sim 1 - 1000$ kpc. If the characteristic cold cloud size is $\gtrsim 10$ kpc, then existing high-resolution galaxy simulations are capable of fully resolving cold CGM gas. However, even if low-redshift galaxies have a cosmic ray pressure-dominated CGM, this regime is an unlikely description of the high-redshift universe and simulations would still need infeasibly high resolution to model cold gas evolution at early times.  

\subsection{Cold Mass Accretion Rates}
The galactic fountain, through which gas is expelled from and accreted onto the galactic disk, is an integral component of galaxy evolution. The mass accretion of cold gas in the Milky Way is measured to be somewhere between 0.0002 - 0.006 \flux \citep{Fox:2019, Werk:2019}.

Throughout this work, we have measured the mass flux (at the scale height) in the right panel of Figures~\ref{fig:dens_fluc_growth}, \ref{fig:dens_fluc_tctf}, \ref{fig:dens_fluc_growth_cr}, \ref{fig:dens_fluc_tctf_cr}, and \ref{fig:dens_fluc_tctf_transport}. The values are normalized by the free-fall cold mass flux, $\rho_0 H/t_{\rm ff} = 0.00088$ \flux . Fiducial MHD-only simulations with $\beta = 100$ therefore predict mass  flux rates of 0.0018 - 0.0026 \flux after $t = 4 t_{\mathrm{cool}}$. Cosmic ray pressure counteracts gravity and reduces the mass influx rate, so that for runs with $P_{\rm c}/P_{\rm g} = 1$, the mass accretion rate is roughly an order of magnitude lower. This is consistent with results from \citet{Su:2020}, that show that cosmic rays can suppress cooling flows even with modest cosmic ray pressures. 

However, our simulations show that the predicted mass accretion rate is sensitive to both the cosmic ray pressure and the invoked cosmic ray transport. Differences in initial cosmic ray pressures and cosmic ray transport models can vary the mass accretion rate by an order of magnitude. Therefore, thermal instability with cosmic rays can simultaneously explain both high mass accretion rates and the inferred low mass accretion rates of quenched galaxies with large reservoirs of cold CGM gas.

\subsection{Limitations of Idealized Setup}
The highly idealized nature of our simulations is helpful for isolating the impact of cosmic ray pressure and transport on thermal instability. However, the simulations are missing a physically realistic context that may impact the formation and survival of cold gas. 

The cooling and heating models in our simulations crudely approximate an isolated, globally stable, long-lived CGM by explicitly balancing the total cooling and total heating at each vertical layer. A more realistic simulation would explicitly model the local processes (e.g. gas accretion, stellar formation and feedback, mergers) that contribute to this global equilibrium.  These processes would be particularly important in simulations with long cooling times which are evolved for at least 500 million years. On those timescales, the assumption of an isolated CGM breaks down, as the CGM is likely to be perturbed by either feedback from its host galaxy or accretion from the surrounding medium.

Additionally, the isolated nature of the simulations means that once cold gas has precipitated out of the hot  gas,  the  cooling time of the remaining gas is too long to create any more cold gas. In a realistic galaxy halo, we expect outflows or inflows to replenish thermally unstable gas. These processes interfere with the prediction for the long term state of cold gas, especially for simulations with high $\tcff$ in the ``isolated'' thermal instability studied here. 

The cosmic ray transport models considered in this work are simplified approximations of two different regimes: pure streaming or pure diffusion. Realistically, both streaming and diffusion should happen, to different degrees, simultaneously and there are recent algorithms that can handle the two self-consistently \citep{Jiang:2018, Chan:2019, Thomas:2019}. Furthermore, unresolved microphysics alter the local cosmic ray diffusion coefficient \citep{Farber:2018} and the dominant scattering mechanisms that determine cosmic ray streaming parameters (see \citealt{Hopkins:2020:cr_transport1} and references therein for a detailed comparison). Additionally, we do not explicitly include cosmic ray ``cooling'' due to hadronic and Coulomb losses. While this process is important for alleviating cosmic ray pressure in the interstellar medium (ISM), the typical densities in the CGM (even within the condensed cold clouds) are too low for this process to be significant. Despite the various caveats to our model described herein, the \textit{quantitative} details of our results may fluctuate to some degree, but we expect the \textit{qualitative} results to be robust \citep{Hopkins:2020:cr_transport2}.

\section{Conclusions}\label{sec:conclusions}
Thermal instability is an important mechanism through which cold gas forms in the CGM. In this work, we used idealized simulations of thermal instability in a gravitationally stratified medium to study the formation and evolution of cold ($T \simeq 10^4 \K$) gas in the presence of cosmic ray pressure and transport. We systematically varied several key parameters, including the ratio of the cooling time to the free-fall time, initial magnetic field strength, initial cosmic ray pressure, and cosmic ray transport parameters (assuming either cosmic ray diffusion or streaming). Our results are summarized below. 
\begin{enumerate}[leftmargin = 0.5cm]

\item Cosmic rays change the morphology of cold gas that forms through thermal instability by providing non-thermal pressure support. Increased cosmic ray pressure prevents cold gas from compressing, allowing it to cool while maintaining lower densities (Figures~\ref{fig:dens_fluc_growth_cr} and \ref{fig:dens_fluc_tctf_cr}). When cosmic ray pressure dominates, thermally unstable gas cools at constant density (isochorically). Our results demonstrate how the inclusion of cosmic rays as a non-thermal pressure source can explain the apparent lack of thermal pressure equilibrium between observed cold and hot gas phases in the CGM. We make predictions for the density contrast between cold and hot gas as a function of magnetic and cosmic ray pressures in the cold gas  (Eq.~\ref{eqn:dens_contrast} and Figure~\ref{fig:family_portrait}).

\item Simulations that include cosmic rays form larger cold gas clouds. If cosmic ray pressure is relatively low, cold gas can still form a ``mist'', with cosmic ray pressure-supported cloudlets that are larger than the predicted sub-parsec cloudlets in the absence of cosmic rays. However, in a cosmic ray pressure-dominated halo in the limit of inefficient cosmic ray transport, the characteristic size of cold gas cloudlets may be up to $\sim 1000$ times larger than those expected in a purely thermal medium. We make predictions for how the characteristic size of cold gas scales with magnetic and cosmic ray pressures (Eqs.~\ref{eqn:rcloud} and \ref{eqn:rcloud_cr}). 
\item  In some cases, cosmic ray pressure can increases the cold mass fraction in the halo by preventing the formed cold gas from precipitating towards the galaxy (Figures~\ref{fig:dens_fluc_growth_cr} and \ref{fig:dens_fluc_tctf_cr}). If cosmic ray pressure contributes to maintaining hydrostatic equilibrium, it enables gas to have a shallower entropy profile as a function of height above the galactic midplane. Consequently, halos with substantial cosmic ray pressure support ($P_{\rm c}/P_{\rm g} \gtrsim 1$) have significantly reduced cold mass flux. This effect may explain observations of quenched galaxies with an abundance of cold gas in their CGM. Cosmic ray pressure does not change the average temperature of the cold gas phase (Figure~\ref{fig:joy_division}). 
\item Simulations that include realistic transport mechanisms for cosmic rays (e.g., diffusion or streaming) demonstrate behavior that bridges the gap between simulations lacking cosmic rays and simulations that only include cosmic rays as an adiabatic non-thermal pressure term (e.g., advection only). Cosmic ray transport redistributes cosmic ray pressure from regions of high cosmic ray pressure in cold clouds to areas of low cosmic ray pressure in the hot background medium (Figure~\ref{fig:joy_division}). This pressure redistribution decreases cold cloud sizes and increases the cold gas density and cold mass flux compared to simulations with cosmic ray advection but without cosmic ray transport (Figure~\ref{fig:dens_fluc_tctf_transport}). Simulations with cosmic ray transport span a larger temperature-density phase space than either simulations without cosmic rays or simulations with only cosmic ray advection (Figure~\ref{fig:phase}). Cosmic ray transport is most effective when the cosmic ray transport is short relative to gas cooling times. 
\end{enumerate}
Cold CGM gas plays a uniquely important role in driving galaxy evolution: cold gas that accretes onto the galaxy fuels star formation, which in turn shapes the CGM through feedback. In this work, we have demonstrated that cosmic rays have the potential to dramatically alter the CGM gas morphology, gas phase, and kinematics. However, many details about cosmic ray physics remain poorly constrained, including expected cosmic ray pressures and magnetic field strengths in the CGM and robust models for cosmic ray transport in hydrodynamic simulations. More constraining models for cosmic ray physics are therefore crucial for understanding galaxy evolution in its entirety. 

\acknowledgements
The authors thank Greg Bryan, Yan-Fei Jiang, Philip Hopkins, Philipp Kempski, Suoqing Ji, Adam Jermyn, Peng Oh, and the anonymous referee for stimulating discussions. IB was supported by the Simons Foundation through the Flatiron Institute's Pre-Doctoral Research Fellowship and by her tenure as a Blue Waters Graduate Fellow. The Blue Waters sustained-petascale computing project is supported by the National Science Foundation (Grants No. OCI-0725070 and No. ACI-1238993) and the State of Illinois. We performed the bulk of our analysis using {\sc yt} \citep{Turk:2011}.

\bibliography{main}

\begin{thebibliography}{}
\expandafter\ifx\csname natexlab\endcsname\relax\def\natexlab#1{#1}\fi
\providecommand{\url}[1]{\href{#1}{#1}}
\providecommand{\dodoi}[1]{doi:~\href{http://doi.org/#1}{\nolinkurl{#1}}}
\providecommand{\doeprint}[1]{\href{http://ascl.net/#1}{\nolinkurl{http://ascl.net/#1}}}
\providecommand{\doarXiv}[1]{\href{https://arxiv.org/abs/#1}{\nolinkurl{https://arxiv.org/abs/#1}}}

\bibitem[{{Angl{\'e}s-Alc{\'a}zar} {et~al.}(2017){Angl{\'e}s-Alc{\'a}zar},
  {Faucher-Gigu{\`e}re}, {Kere{\v{s}}}, {Hopkins}, {Quataert}, \&
  {Murray}}]{Angles:2017}
{Angl{\'e}s-Alc{\'a}zar}, D., {Faucher-Gigu{\`e}re}, C.-A., {Kere{\v{s}}}, D.,
  {et~al.} 2017, \mnras, 470, 4698, \dodoi{10.1093/mnras/stx1517}

\bibitem[{{Ashley} {et~al.}(2020){Ashley}, {Fox}, {Jenkins}, {Wakker},
  {Bordoloi}, {Lockman}, {Savage}, \& {Karim}}]{Ashley:2020}
{Ashley}, T., {Fox}, A.~J., {Jenkins}, E.~B., {et~al.} 2020, arXiv e-prints,
  arXiv:2006.13254.
\newblock \doarXiv{2006.13254}

\bibitem[{{Berg} {et~al.}(2019){Berg}, {Howk}, {Lehner}, {Wotta}, {O'Meara},
  {Bowen}, {Burchett}, {Peeples}, \& {Tejos}}]{Berg:2019}
{Berg}, M.~A., {Howk}, J.~C., {Lehner}, N., {et~al.} 2019, \apj, 883, 5,
  \dodoi{10.3847/1538-4357/ab378e}

\bibitem[{{Bish} {et~al.}(2019){Bish}, {Werk}, {Prochaska}, {Rubin}, {Zheng},
  {O'Meara}, \& {Deason}}]{Bish:2019}
{Bish}, H.~V., {Werk}, J.~K., {Prochaska}, J.~X., {et~al.} 2019, \apj, 882, 76,
  \dodoi{10.3847/1538-4357/ab3414}

\bibitem[{{Booth} {et~al.}(2013){Booth}, {Agertz}, {Kravtsov}, \&
  {Gnedin}}]{Booth:2013}
{Booth}, C.~M., {Agertz}, O., {Kravtsov}, A.~V., \& {Gnedin}, N.~Y. 2013,
  \apjl, 777, L16, \dodoi{10.1088/2041-8205/777/1/L16}

\bibitem[{{Boulares} \& {Cox}(1990)}]{Boulares:1990}
{Boulares}, A., \& {Cox}, D.~P. 1990, \apj, 365, 544, \dodoi{10.1086/169509}

\bibitem[{{Brummel-Smith} {et~al.}(2019){Brummel-Smith}, {Bryan}, {Butsky},
  {Corlies}, {Emerick}, {Forbes}, {Fujimoto}, {Goldbaum}, {Grete}, {Hummels},
  {Kim}, {Koh}, {Li}, {Li}, {Li}, {OShea}, {Peeples}, {Regan}, {Salem},
  {Schmidt}, {Simpson}, {Smith}, {Tumlinson}, {Turk}, {Wise}, {Abel},
  {Bordner}, {Cen}, {Collins}, {Crosby}, {Edelmann}, {Hahn}, {Harkness},
  {Harper-Clark}, {Kong}, {Kritsuk}, {Kuhlen}, {Larrue}, {Lee}, {Meece},
  {Norman}, {Oishi}, {Paschos}, {Peruta}, {Razoumov}, {Reynolds}, {Silvia},
  {Skillman}, {Skory}, {So}, {Tasker}, {Wagner}, {Wang}, {Xu}, \&
  {Zhao}}]{Enzo:2019}
{Brummel-Smith}, C., {Bryan}, G., {Butsky}, I., {et~al.} 2019, The Journal of
  Open Source Software, 4, 1636, \dodoi{10.21105/joss.01636}

\bibitem[{{Bryan} {et~al.}(2014){Bryan}, {Norman}, {O'Shea}, {Abel}, {Wise},
  {Turk}, {Reynolds}, {Collins}, {Wang}, {Skillman}, {Smith}, {Harkness},
  {Bordner}, {Kim}, {Kuhlen}, {Xu}, {Goldbaum}, {Hummels}, {Kritsuk}, {Tasker},
  {Skory}, {Simpson}, {Hahn}, {Oishi}, {So}, {Zhao}, {Cen}, {Li}, \& {Enzo
  Collaboration}}]{Bryan:2014}
{Bryan}, G.~L., {Norman}, M.~L., {O'Shea}, B.~W., {et~al.} 2014, \apjs, 211,
  19, \dodoi{10.1088/0067-0049/211/2/19}

\bibitem[{{Buck} {et~al.}(2020){Buck}, {Pfrommer}, {Pakmor}, {Grand}, \&
  {Springel}}]{Buck:2020}
{Buck}, T., {Pfrommer}, C., {Pakmor}, R., {Grand}, R. J.~J., \& {Springel}, V.
  2020, \mnras, \dodoi{10.1093/mnras/staa1960}

\bibitem[{{Burchett} {et~al.}(2020){Burchett}, {Rubin}, {Prochaska}, {Coil},
  {Rickards Vaught}, \& {Hennawi}}]{Burchett:2020}
{Burchett}, J.~N., {Rubin}, K. H.~R., {Prochaska}, J.~X., {et~al.} 2020, arXiv
  e-prints, arXiv:2005.03017.
\newblock \doarXiv{2005.03017}

\bibitem[{{Bustard} {et~al.}(2020){Bustard}, {Zweibel}, {D'Onghia},
  {Gallagher}, \& {Farber}}]{Bustard:2020}
{Bustard}, C., {Zweibel}, E.~G., {D'Onghia}, E., {Gallagher}, J.~S., I., \&
  {Farber}, R. 2020, \apj, 893, 29, \dodoi{10.3847/1538-4357/ab7fa3}

\bibitem[{{Butsky} \& {Quinn}(2018)}]{Butsky:2018}
{Butsky}, I.~S., \& {Quinn}, T.~R. 2018, \apj, 868, 108,
  \dodoi{10.3847/1538-4357/aaeac2}

\bibitem[{{Chan} {et~al.}(2019){Chan}, {Kere{\v{s}}}, {Hopkins}, {Quataert},
  {Su}, {Hayward}, \& {Faucher-Gigu{\`e}re}}]{Chan:2019}
{Chan}, T.~K., {Kere{\v{s}}}, D., {Hopkins}, P.~F., {et~al.} 2019, \mnras, 488,
  3716, \dodoi{10.1093/mnras/stz1895}

\bibitem[{{Chen} {et~al.}(2014){Chen}, {Gauthier}, {Sharon}, {Johnson}, {Nair},
  \& {Liang}}]{Chen:2014}
{Chen}, H.-W., {Gauthier}, J.-R., {Sharon}, K., {et~al.} 2014, \mnras, 438,
  1435, \dodoi{10.1093/mnras/stt2288}

\bibitem[{{Chen} {et~al.}(2010){Chen}, {Helsby}, {Gauthier}, {Shectman},
  {Thompson}, \& {Tinker}}]{Chen:2010}
{Chen}, H.-W., {Helsby}, J.~E., {Gauthier}, J.-R., {et~al.} 2010, \apj, 714,
  1521, \dodoi{10.1088/0004-637X/714/2/1521}

\bibitem[{{Choudhury} \& {Sharma}(2016)}]{Choudhury:2016}
{Choudhury}, P.~P., \& {Sharma}, P. 2016, \mnras, 457, 2554,
  \dodoi{10.1093/mnras/stw152}

\bibitem[{{Choudhury} {et~al.}(2019){Choudhury}, {Sharma}, \&
  {Quataert}}]{Choudhury:2019}
{Choudhury}, P.~P., {Sharma}, P., \& {Quataert}, E. 2019, \mnras, 488, 3195,
  \dodoi{10.1093/mnras/stz1857}

\bibitem[{{Crighton} {et~al.}(2015){Crighton}, {Hennawi}, {Simcoe}, {Cooksey},
  {Murphy}, {Fumagalli}, {Prochaska}, \& {Shanks}}]{Crighton:2015}
{Crighton}, N. H.~M., {Hennawi}, J.~F., {Simcoe}, R.~A., {et~al.} 2015, \mnras,
  446, 18, \dodoi{10.1093/mnras/stu2088}

\bibitem[{{Dedner} {et~al.}(2002){Dedner}, {Kemm}, {Kr{\"o}ner}, {Munz},
  {Schnitzer}, \& {Wesenberg}}]{Dedner:2002}
{Dedner}, A., {Kemm}, F., {Kr{\"o}ner}, D., {et~al.} 2002, Journal of
  Computational Physics, 175, 645, \dodoi{10.1006/jcph.2001.6961}

\bibitem[{{Esmerian} {et~al.}(2020){Esmerian}, {Kravtsov}, {Hafen},
  {Faucher-Giguere}, {Quataert}, {Stern}, {Keres}, \& {Wetzel}}]{Esmerian:2020}
{Esmerian}, C.~J., {Kravtsov}, A.~V., {Hafen}, Z., {et~al.} 2020, arXiv
  e-prints, arXiv:2006.13945.
\newblock \doarXiv{2006.13945}

\bibitem[{{Farber} {et~al.}(2018){Farber}, {Ruszkowski}, {Yang}, \&
  {Zweibel}}]{Farber:2018}
{Farber}, R., {Ruszkowski}, M., {Yang}, H. Y.~K., \& {Zweibel}, E.~G. 2018,
  \apj, 856, 112, \dodoi{10.3847/1538-4357/aab26d}

\bibitem[{{Field}(1965)}]{Field:1965}
{Field}, G.~B. 1965, \apj, 142, 531, \dodoi{10.1086/148317}

\bibitem[{{Fielding} {et~al.}(2017){Fielding}, {Quataert}, {McCourt}, \&
  {Thompson}}]{Fielding:2017}
{Fielding}, D., {Quataert}, E., {McCourt}, M., \& {Thompson}, T.~A. 2017,
  \mnras, 466, 3810, \dodoi{10.1093/mnras/stw3326}

\bibitem[{{Fielding} {et~al.}(2020{\natexlab{a}}){Fielding}, {Ostriker},
  {Bryan}, \& {Jermyn}}]{Fielding:2020}
{Fielding}, D.~B., {Ostriker}, E.~C., {Bryan}, G.~L., \& {Jermyn}, A.~S.
  2020{\natexlab{a}}, \apjl, 894, L24, \dodoi{10.3847/2041-8213/ab8d2c}

\bibitem[{{Fielding} {et~al.}(2020{\natexlab{b}}){Fielding}, {Tonnesen},
  {DeFelippis}, {Li}, {Su}, {Bryan}, {Kim}, {Forbes}, {Somerville},
  {Battaglia}, {Schneider}, {Li}, {Choi}, {Hayward}, \&
  {Hernquist}}]{Fielding:2020b}
{Fielding}, D.~B., {Tonnesen}, S., {DeFelippis}, D., {et~al.}
  2020{\natexlab{b}}, arXiv e-prints, arXiv:2006.16316.
\newblock \doarXiv{2006.16316}

\bibitem[{{Fluetsch} {et~al.}(2020){Fluetsch}, {Maiolino}, {Carniani},
  {Arribas}, {Belfiore}, {Bellocchi}, {Cazzoli}, {Cicone}, {Cresci}, {Fabian},
  {Gallagher}, {Ishibashi}, {Mannucci}, {Marconi}, {Perna}, {Sturm}, \&
  {Venturi}}]{Fleutsch:2020}
{Fluetsch}, A., {Maiolino}, R., {Carniani}, S., {et~al.} 2020, arXiv e-prints,
  arXiv:2006.13232.
\newblock \doarXiv{2006.13232}

\bibitem[{{Ford} {et~al.}(2014){Ford}, {Dav{\'e}}, {Oppenheimer}, {Katz},
  {Kollmeier}, {Thompson}, \& {Weinberg}}]{Ford:2014}
{Ford}, A.~B., {Dav{\'e}}, R., {Oppenheimer}, B.~D., {et~al.} 2014, \mnras,
  444, 1260, \dodoi{10.1093/mnras/stu1418}

\bibitem[{{Fox} {et~al.}(2019){Fox}, {Richter}, {Ashley}, {Heckman}, {Lehner},
  {Werk}, {Bordoloi}, \& {Peeples}}]{Fox:2019}
{Fox}, A.~J., {Richter}, P., {Ashley}, T., {et~al.} 2019, \apj, 884, 53,
  \dodoi{10.3847/1538-4357/ab40ad}

\bibitem[{{Gaspari} {et~al.}(2012){Gaspari}, {Ruszkowski}, \&
  {Sharma}}]{Gaspari:2012}
{Gaspari}, M., {Ruszkowski}, M., \& {Sharma}, P. 2012, \apj, 746, 94,
  \dodoi{10.1088/0004-637X/746/1/94}

\bibitem[{{Ginzburg} \& {Ptuskin}(1985)}]{Ginzburg:1985}
{Ginzburg}, V.~L., \& {Ptuskin}, V.~S. 1985, \apspr, 4, 161

\bibitem[{{Girichidis} {et~al.}(2018){Girichidis}, {Naab}, {Hanasz}, \&
  {Walch}}]{Girichidis:2018}
{Girichidis}, P., {Naab}, T., {Hanasz}, M., \& {Walch}, S. 2018, \mnras, 479,
  3042, \dodoi{10.1093/mnras/sty1653}

\bibitem[{{Girichidis} {et~al.}(2016){Girichidis}, {Naab}, {Walch}, {Hanasz},
  {Mac Low}, {Ostriker}, {Gatto}, {Peters}, {W{\"u}nsch}, {Glover}, {Klessen},
  {Clark}, \& {Baczynski}}]{Girichidis:2016}
{Girichidis}, P., {Naab}, T., {Walch}, S., {et~al.} 2016, \apjl, 816, L19,
  \dodoi{10.3847/2041-8205/816/2/L19}

\bibitem[{{Gronke} \& {Oh}(2018)}]{Gronke:2018}
{Gronke}, M., \& {Oh}, S.~P. 2018, \mnras, 480, L111,
  \dodoi{10.1093/mnrasl/sly131}

\bibitem[{{Gronke} \& {Oh}(2020)}]{Gronke:2020}
---. 2020, \mnras, 494, L27, \dodoi{10.1093/mnrasl/slaa033}

\bibitem[{{Guo} \& {Oh}(2008)}]{Guo:2008}
{Guo}, F., \& {Oh}, S.~P. 2008, \mnras, 384, 251,
  \dodoi{10.1111/j.1365-2966.2007.12692.x}

\bibitem[{{Hafen} {et~al.}(2019){Hafen}, {Faucher-Gigu{\`e}re},
  {Angl{\'e}s-Alc{\'a}zar}, {Stern}, {Kere{\v{s}}}, {Hummels}, {Esmerian},
  {Garrison-Kimmel}, {El-Badry}, {Wetzel}, {Chan}, {Hopkins}, \&
  {Murray}}]{Hafen:2019}
{Hafen}, Z., {Faucher-Gigu{\`e}re}, C.-A., {Angl{\'e}s-Alc{\'a}zar}, D.,
  {et~al.} 2019, \mnras, 488, 1248, \dodoi{10.1093/mnras/stz1773}

\bibitem[{{Heckman} {et~al.}(2000){Heckman}, {Lehnert}, {Strickland }, \&
  {Armus}}]{Heckman:2000}
{Heckman}, T.~M., {Lehnert}, M.~D., {Strickland }, D.~K., \& {Armus}, L. 2000,
  \apjs, 129, 493, \dodoi{10.1086/313421}

\bibitem[{{Heintz} {et~al.}(2020){Heintz}, {Bustard}, \&
  {Zweibel}}]{Heintz:2020}
{Heintz}, E., {Bustard}, C., \& {Zweibel}, E.~G. 2020, \apj, 891, 157,
  \dodoi{10.3847/1538-4357/ab7453}

\bibitem[{{Hopkins} {et~al.}(2020{\natexlab{a}}){Hopkins}, {Chan}, {Squire},
  {Quataert}, {Ji}, {Keres}, \& {Faucher-Giguere}}]{Hopkins:2020:cr_transport1}
{Hopkins}, P.~F., {Chan}, T.~K., {Squire}, J., {et~al.} 2020{\natexlab{a}},
  arXiv e-prints, arXiv:2004.02897.
\newblock \doarXiv{2004.02897}

\bibitem[{{Hopkins} {et~al.}(2020{\natexlab{b}}){Hopkins}, {Squire}, {Chan},
  {Quataert}, {Ji}, {Keres}, \& {Faucher-Giguere}}]{Hopkins:2020:cr_transport2}
{Hopkins}, P.~F., {Squire}, J., {Chan}, T.~K., {et~al.} 2020{\natexlab{b}},
  arXiv e-prints, arXiv:2002.06211.
\newblock \doarXiv{2002.06211}

\bibitem[{{Hopkins} {et~al.}(2020{\natexlab{c}}){Hopkins}, {Chan},
  {Garrison-Kimmel}, {Ji}, {Su}, {Hummels}, {Kere{\v{s}}}, {Quataert}, \&
  {Faucher-Gigu{\`e}re}}]{Hopkins:2020:cr_galaxy}
{Hopkins}, P.~F., {Chan}, T.~K., {Garrison-Kimmel}, S., {et~al.}
  2020{\natexlab{c}}, \mnras, 492, 3465, \dodoi{10.1093/mnras/stz3321}

\bibitem[{{Hummels} {et~al.}(2019){Hummels}, {Smith}, {Hopkins}, {O'Shea},
  {Silvia}, {Werk}, {Lehner}, {Wise}, {Collins}, \& {Butsky}}]{Hummels:2019}
{Hummels}, C.~B., {Smith}, B.~D., {Hopkins}, P.~F., {et~al.} 2019, \apj, 882,
  156, \dodoi{10.3847/1538-4357/ab378f}

\bibitem[{{Jacob} \& {Pfrommer}(2017{\natexlab{a}})}]{Jacob:2017a}
{Jacob}, S., \& {Pfrommer}, C. 2017{\natexlab{a}}, \mnras, 467, 1449,
  \dodoi{10.1093/mnras/stx131}

\bibitem[{{Jacob} \& {Pfrommer}(2017{\natexlab{b}})}]{Jacob:2017b}
---. 2017{\natexlab{b}}, \mnras, 467, 1478, \dodoi{10.1093/mnras/stx132}

\bibitem[{{Jana} {et~al.}(2020){Jana}, {Gupta}, \& {Nath}}]{Jana:2020}
{Jana}, R., {Gupta}, S., \& {Nath}, B.~B. 2020, \mnras, 497, 2623,
  \dodoi{10.1093/mnras/staa2025}

\bibitem[{{Ji} {et~al.}(2018){Ji}, {Oh}, \& {McCourt}}]{Ji:2018}
{Ji}, S., {Oh}, S.~P., \& {McCourt}, M. 2018, \mnras, 476, 852,
  \dodoi{10.1093/mnras/sty293}

\bibitem[{{Ji} {et~al.}(2020){Ji}, {Chan}, {Hummels}, {Hopkins}, {Stern},
  {Kere{\v{s}}}, {Quataert}, {Faucher-Gigu{\`e}re}, \& {Murray}}]{Ji:2020}
{Ji}, S., {Chan}, T.~K., {Hummels}, C.~B., {et~al.} 2020, \mnras,
  \dodoi{10.1093/mnras/staa1849}

\bibitem[{{Jiang} \& {Oh}(2018)}]{Jiang:2018}
{Jiang}, Y.-F., \& {Oh}, S.~P. 2018, \apj, 854, 5,
  \dodoi{10.3847/1538-4357/aaa6ce}

\bibitem[{{Keeney} {et~al.}(2018){Keeney}, {Stocke}, {Pratt}, {Davis},
  {Syphers}, {Danforth}, {Shull}, {Froning}, {Green}, {Penton}, \&
  {Savage}}]{Keeney:2018}
{Keeney}, B.~A., {Stocke}, J.~T., {Pratt}, C.~T., {et~al.} 2018, \apjs, 237,
  11, \dodoi{10.3847/1538-4365/aac727}

\bibitem[{{Kempski} \& {Quataert}(2020)}]{Kempski:2020a}
{Kempski}, P., \& {Quataert}, E. 2020, \mnras, 493, 1801,
  \dodoi{10.1093/mnras/staa385}

\bibitem[{{Kurganov} \& {Tadmor}(2000)}]{Kurganov:2000}
{Kurganov}, A., \& {Tadmor}, E. 2000, Journal of Computational Physics, 160,
  241, \dodoi{10.1006/jcph.2000.6459}

\bibitem[{{Lan} \& {Prochaska}(2020)}]{Lan:2020}
{Lan}, T.-W., \& {Prochaska}, J.~X. 2020, \mnras, 496, 3142,
  \dodoi{10.1093/mnras/staa1750}

\bibitem[{{Li} {et~al.}(2020){Li}, {Hopkins}, {Squire}, \& {Hummels}}]{Li:2020}
{Li}, Z., {Hopkins}, P.~F., {Squire}, J., \& {Hummels}, C. 2020, \mnras, 492,
  1841, \dodoi{10.1093/mnras/stz3567}

\bibitem[{{Liang} \& {Remming}(2020)}]{Liang:2020}
{Liang}, C.~J., \& {Remming}, I. 2020, \mnras, 491, 5056,
  \dodoi{10.1093/mnras/stz3403}

\bibitem[{{Maller} \& {Bullock}(2004)}]{Maller:2004}
{Maller}, A.~H., \& {Bullock}, J.~S. 2004, \mnras, 355, 694,
  \dodoi{10.1111/j.1365-2966.2004.08349.x}

\bibitem[{{Mandelker} {et~al.}(2019){Mandelker}, {van den Bosch}, {Springel},
  \& {van de Voort}}]{Mandelker:2019}
{Mandelker}, N., {van den Bosch}, F.~C., {Springel}, V., \& {van de Voort}, F.
  2019, \apjl, 881, L20, \dodoi{10.3847/2041-8213/ab30cb}

\bibitem[{{McCourt} {et~al.}(2018){McCourt}, {Oh}, {O'Leary}, \&
  {Madigan}}]{McCourt:2018}
{McCourt}, M., {Oh}, S.~P., {O'Leary}, R., \& {Madigan}, A.-M. 2018, \mnras,
  473, 5407, \dodoi{10.1093/mnras/stx2687}

\bibitem[{{McCourt} {et~al.}(2015){McCourt}, {O'Leary}, {Madigan}, \&
  {Quataert}}]{McCourt:2015}
{McCourt}, M., {O'Leary}, R.~M., {Madigan}, A.-M., \& {Quataert}, E. 2015,
  \mnras, 449, 2, \dodoi{10.1093/mnras/stv355}

\bibitem[{{McCourt} {et~al.}(2012){McCourt}, {Sharma}, {Quataert}, \&
  {Parrish}}]{McCourt:2012}
{McCourt}, M., {Sharma}, P., {Quataert}, E., \& {Parrish}, I.~J. 2012, \mnras,
  419, 3319, \dodoi{10.1111/j.1365-2966.2011.19972.x}

\bibitem[{{McKee} \& {Cowie}(1975)}]{McKee:1975}
{McKee}, C.~F., \& {Cowie}, L.~L. 1975, \apj, 195, 715, \dodoi{10.1086/153373}

\bibitem[{{Meece} {et~al.}(2015){Meece}, {O'Shea}, \& {Voit}}]{Meece:2015}
{Meece}, G.~R., {O'Shea}, B.~W., \& {Voit}, G.~M. 2015, \apj, 808, 43,
  \dodoi{10.1088/0004-637X/808/1/43}

\bibitem[{{Mo} \& {Miralda-Escude}(1996)}]{Mo:1996}
{Mo}, H.~J., \& {Miralda-Escude}, J. 1996, \apj, 469, 589,
  \dodoi{10.1086/177808}

\bibitem[{{Nelson} {et~al.}(2020){Nelson}, {Sharma}, {Pillepich}, {Springel},
  {Pakmor}, {Weinberger}, {Vogelsberger}, {Marinacci}, \&
  {Hernquist}}]{Nelson:2020}
{Nelson}, D., {Sharma}, P., {Pillepich}, A., {et~al.} 2020, arXiv e-prints,
  arXiv:2005.09654.
\newblock \doarXiv{2005.09654}

\bibitem[{{Pakmor} {et~al.}(2019){Pakmor}, {van de Voort}, {Bieri}, {Gomez},
  {Grand}, {Guillet}, {Marinacci}, {Pfrommer}, {Simpson}, \&
  {Springel}}]{Pakmor:2020}
{Pakmor}, R., {van de Voort}, F., {Bieri}, R., {et~al.} 2019, arXiv e-prints,
  arXiv:1911.11163.
\newblock \doarXiv{1911.11163}

\bibitem[{{Peeples} {et~al.}(2019){Peeples}, {Corlies}, {Tumlinson}, {O'Shea},
  {Lehner}, {O'Meara}, {Howk}, {Earl}, {Smith}, {Wise}, \&
  {Hummels}}]{Peeples:2019}
{Peeples}, M.~S., {Corlies}, L., {Tumlinson}, J., {et~al.} 2019, \apj, 873,
  129, \dodoi{10.3847/1538-4357/ab0654}

\bibitem[{{Pizzolato} \& {Soker}(2005)}]{Pizzolate:2005}
{Pizzolato}, F., \& {Soker}, N. 2005, \apj, 632, 821, \dodoi{10.1086/444344}

\bibitem[{{Prasad} {et~al.}(2018){Prasad}, {Sharma}, \& {Babul}}]{Prasad:2018}
{Prasad}, D., {Sharma}, P., \& {Babul}, A. 2018, \apj, 863, 62,
  \dodoi{10.3847/1538-4357/aacce8}

\bibitem[{{Prochaska} \& {Hennawi}(2009)}]{Prochaska:2009}
{Prochaska}, J.~X., \& {Hennawi}, J.~F. 2009, \apj, 690, 1558,
  \dodoi{10.1088/0004-637X/690/2/1558}

\bibitem[{{Prochaska} {et~al.}(2011){Prochaska}, {Weiner}, {Chen}, {Mulchaey},
  \& {Cooksey}}]{Prochaska:2011}
{Prochaska}, J.~X., {Weiner}, B., {Chen}, H.-W., {Mulchaey}, J., \& {Cooksey},
  K. 2011, \apj, 740, 91, \dodoi{10.1088/0004-637X/740/2/91}

\bibitem[{{Putman} {et~al.}(2012){Putman}, {Peek}, \& {Joung}}]{Putman:2012}
{Putman}, M.~E., {Peek}, J.~E.~G., \& {Joung}, M.~R. 2012, \araa, 50, 491,
  \dodoi{10.1146/annurev-astro-081811-125612}

\bibitem[{{Rauch} {et~al.}(2001){Rauch}, {Sargent}, \& {Barlow}}]{Rauch:2001}
{Rauch}, M., {Sargent}, W. L.~W., \& {Barlow}, T.~A. 2001, \apj, 554, 823,
  \dodoi{10.1086/321402}

\bibitem[{{Rubin} {et~al.}(2018){Rubin}, {Diamond-Stanic}, {Coil}, {Crighton},
  \& {Stewart}}]{Rubin:2018}
{Rubin}, K. H.~R., {Diamond-Stanic}, A.~M., {Coil}, A.~L., {Crighton}, N.
  H.~M., \& {Stewart}, K.~R. 2018, \apj, 868, 142,
  \dodoi{10.3847/1538-4357/aad566}

\bibitem[{{Rudie} {et~al.}(2019){Rudie}, {Steidel}, {Pettini}, {Trainor},
  {Strom}, {Hummels}, {Reddy}, \& {Shapley}}]{Rudie:2019}
{Rudie}, G.~C., {Steidel}, C.~C., {Pettini}, M., {et~al.} 2019, \apj, 885, 61,
  \dodoi{10.3847/1538-4357/ab4255}

\bibitem[{{Ruszkowski} {et~al.}(2017){Ruszkowski}, {Yang}, \&
  {Zweibel}}]{Ruszkowski:2017}
{Ruszkowski}, M., {Yang}, H.-Y.~K., \& {Zweibel}, E. 2017, \apj, 834, 208,
  \dodoi{10.3847/1538-4357/834/2/208}

\bibitem[{{Salem} {et~al.}(2016){Salem}, {Bryan}, \& {Corlies}}]{Salem:2016}
{Salem}, M., {Bryan}, G.~L., \& {Corlies}, L. 2016, \mnras, 456, 582,
  \dodoi{10.1093/mnras/stv2641}

\bibitem[{{Schneider} {et~al.}(2018){Schneider}, {Robertson}, \&
  {Thompson}}]{Schneider:2018}
{Schneider}, E.~E., {Robertson}, B.~E., \& {Thompson}, T.~A. 2018, \apj, 862,
  56, \dodoi{10.3847/1538-4357/aacce1}

\bibitem[{{Sharma} \& {Nath}(2012)}]{Sharma:2012b}
{Sharma}, M., \& {Nath}, B.~B. 2012, \apj, 750, 55,
  \dodoi{10.1088/0004-637X/750/1/55}

\bibitem[{{Sharma} {et~al.}(2012){Sharma}, {McCourt}, {Quataert}, \&
  {Parrish}}]{Sharma:2012a}
{Sharma}, P., {McCourt}, M., {Quataert}, E., \& {Parrish}, I.~J. 2012, \mnras,
  420, 3174, \dodoi{10.1111/j.1365-2966.2011.20246.x}

\bibitem[{{Sharma} {et~al.}(2010){Sharma}, {Parrish}, \&
  {Quataert}}]{Sharma:2010}
{Sharma}, P., {Parrish}, I.~J., \& {Quataert}, E. 2010, \apj, 720, 652,
  \dodoi{10.1088/0004-637X/720/1/652}

\bibitem[{{Simpson} {et~al.}(2016){Simpson}, {Pakmor}, {Marinacci}, {Pfrommer},
  {Springel}, {Glover}, {Clark}, \& {Smith}}]{Simpson:2016}
{Simpson}, C.~M., {Pakmor}, R., {Marinacci}, F., {et~al.} 2016, \apjl, 827,
  L29, \dodoi{10.3847/2041-8205/827/2/L29}

\bibitem[{{Singh} \& {Sharma}(2015)}]{Singh:2015}
{Singh}, A., \& {Sharma}, P. 2015, \mnras, 446, 1895,
  \dodoi{10.1093/mnras/stu2264}

\bibitem[{{Stern} {et~al.}(2016){Stern}, {Hennawi}, {Prochaska}, \&
  {Werk}}]{Stern:2016}
{Stern}, J., {Hennawi}, J.~F., {Prochaska}, J.~X., \& {Werk}, J.~K. 2016, \apj,
  830, 87, \dodoi{10.3847/0004-637X/830/2/87}

\bibitem[{{Stocke} {et~al.}(2013){Stocke}, {Keeney}, {Danforth}, {Shull},
  {Froning}, {Green}, {Penton}, \& {Savage}}]{Stocke:2013}
{Stocke}, J.~T., {Keeney}, B.~A., {Danforth}, C.~W., {et~al.} 2013, \apj, 763,
  148, \dodoi{10.1088/0004-637X/763/2/148}

\bibitem[{{Su} {et~al.}(2020){Su}, {Hopkins}, {Hayward}, {Faucher-Gigu{\`e}re},
  {Kere{\v{s}}}, {Ma}, {Orr}, {Chan}, \& {Robles}}]{Su:2020}
{Su}, K.-Y., {Hopkins}, P.~F., {Hayward}, C.~C., {et~al.} 2020, \mnras, 491,
  1190, \dodoi{10.1093/mnras/stz3011}

\bibitem[{{Suresh} {et~al.}(2019){Suresh}, {Nelson}, {Genel}, {Rubin}, \&
  {Hernquist}}]{Suresh:2019}
{Suresh}, J., {Nelson}, D., {Genel}, S., {Rubin}, K. H.~R., \& {Hernquist}, L.
  2019, \mnras, 483, 4040, \dodoi{10.1093/mnras/sty3402}

\bibitem[{{Thom} {et~al.}(2012){Thom}, {Tumlinson}, {Werk}, {Prochaska},
  {Oppenheimer}, {Peeples}, {Tripp}, {Katz}, {O'Meara}, {Ford}, {Dav{\'e}},
  {Sembach}, \& {Weinberg}}]{Thom:2012}
{Thom}, C., {Tumlinson}, J., {Werk}, J.~K., {et~al.} 2012, \apjl, 758, L41,
  \dodoi{10.1088/2041-8205/758/2/L41}

\bibitem[{{Thomas} \& {Pfrommer}(2019)}]{Thomas:2019}
{Thomas}, T., \& {Pfrommer}, C. 2019, \mnras, 485, 2977,
  \dodoi{10.1093/mnras/stz263}

\bibitem[{{Thompson} {et~al.}(2016){Thompson}, {Quataert}, {Zhang}, \&
  {Weinberg}}]{Thompson:2016}
{Thompson}, T.~A., {Quataert}, E., {Zhang}, D., \& {Weinberg}, D.~H. 2016,
  \mnras, 455, 1830, \dodoi{10.1093/mnras/stv2428}

\bibitem[{{Tumlinson} {et~al.}(2017){Tumlinson}, {Peeples}, \&
  {Werk}}]{Tumlinson:2017}
{Tumlinson}, J., {Peeples}, M.~S., \& {Werk}, J.~K. 2017, \araa, 55, 389,
  \dodoi{10.1146/annurev-astro-091916-055240}

\bibitem[{{Tumlinson} {et~al.}(2011){Tumlinson}, {Werk}, {Thom}, {Meiring},
  {Prochaska}, {Tripp}, {O'Meara}, {Okrochkov}, \& {Sembach}}]{Tumlinson:2011}
{Tumlinson}, J., {Werk}, J.~K., {Thom}, C., {et~al.} 2011, \apj, 733, 111,
  \dodoi{10.1088/0004-637X/733/2/111}

\bibitem[{{Tumlinson} {et~al.}(2013){Tumlinson}, {Thom}, {Werk}, {Prochaska},
  {Tripp}, {Katz}, {Dav{\'e}}, {Oppenheimer}, {Meiring}, {Ford}, {O'Meara},
  {Peeples}, {Sembach}, \& {Weinberg}}]{Tumlinson:2013}
{Tumlinson}, J., {Thom}, C., {Werk}, J.~K., {et~al.} 2013, \apj, 777, 59,
  \dodoi{10.1088/0004-637X/777/1/59}

\bibitem[{{Turk} {et~al.}(2011){Turk}, {Smith}, {Oishi}, {Skory}, {Skillman},
  {Abel}, \& {Norman}}]{Turk:2011}
{Turk}, M.~J., {Smith}, B.~D., {Oishi}, J.~S., {et~al.} 2011, \apjs, 192, 9,
  \dodoi{10.1088/0067-0049/192/1/9}

\bibitem[{{Uhlig} {et~al.}(2012){Uhlig}, {Pfrommer}, {Sharma}, {Nath},
  {En{\ss}lin}, \& {Springel}}]{Uhlig:2012}
{Uhlig}, M., {Pfrommer}, C., {Sharma}, M., {et~al.} 2012, \mnras, 423, 2374,
  \dodoi{10.1111/j.1365-2966.2012.21045.x}

\bibitem[{{van de Voort} {et~al.}(2020){van de Voort}, {Bieri}, {Pakmor},
  {G{\'o}mez}, {Grand}, \& {Marinacci}}]{vandeVoort:2020}
{van de Voort}, F., {Bieri}, R., {Pakmor}, R., {et~al.} 2020, arXiv e-prints,
  arXiv:2008.07537.
\newblock \doarXiv{2008.07537}

\bibitem[{{van de Voort} {et~al.}(2019){van de Voort}, {Springel}, {Mandelker},
  {van den Bosch}, \& {Pakmor}}]{vandeVoort:2019}
{van de Voort}, F., {Springel}, V., {Mandelker}, N., {van den Bosch}, F.~C., \&
  {Pakmor}, R. 2019, \mnras, 482, L85, \dodoi{10.1093/mnrasl/sly190}

\bibitem[{{van Leer}(1977)}]{VanLeer:1977}
{van Leer}, B. 1977, Journal of Computational Physics, 23, 276,
  \dodoi{10.1016/0021-9991(77)90095-X}

\bibitem[{{Voit}(2018)}]{Voit:2018}
{Voit}, G.~M. 2018, \apj, 868, 102, \dodoi{10.3847/1538-4357/aae8e2}

\bibitem[{{Voit} {et~al.}(2015){Voit}, {Donahue}, {Bryan}, \&
  {McDonald}}]{Voit:2015}
{Voit}, G.~M., {Donahue}, M., {Bryan}, G.~L., \& {McDonald}, M. 2015, \nat,
  519, 203, \dodoi{10.1038/nature14167}

\bibitem[{{Voit} {et~al.}(2019){Voit}, {Donahue}, {Zahedy}, {Chen}, {Werk},
  {Bryan}, \& {O'Shea}}]{Voit:2019a}
{Voit}, G.~M., {Donahue}, M., {Zahedy}, F., {et~al.} 2019, \apjl, 879, L1,
  \dodoi{10.3847/2041-8213/ab2766}

\bibitem[{{Voit} {et~al.}(2017){Voit}, {Meece}, {Li}, {O'Shea}, {Bryan}, \&
  {Donahue}}]{Voit:2017}
{Voit}, G.~M., {Meece}, G., {Li}, Y., {et~al.} 2017, \apj, 845, 80,
  \dodoi{10.3847/1538-4357/aa7d04}

\bibitem[{{Wagh} {et~al.}(2014){Wagh}, {Sharma}, \& {McCourt}}]{Wagh:2014}
{Wagh}, B., {Sharma}, P., \& {McCourt}, M. 2014, \mnras, 439, 2822,
  \dodoi{10.1093/mnras/stu138}

\bibitem[{{Wang} {et~al.}(2008){Wang}, {Abel}, \& {Zhang}}]{Wang:2008}
{Wang}, P., {Abel}, T., \& {Zhang}, W. 2008, \apjs, 176, 467,
  \dodoi{10.1086/529434}

\bibitem[{{Werk} {et~al.}(2014){Werk}, {Prochaska}, {Tumlinson}, {Peeples},
  {Tripp}, {Fox}, {Lehner}, {Thom}, {O'Meara}, {Ford}, {Bordoloi}, {Katz},
  {Tejos}, {Oppenheimer}, {Dav{\'e}}, \& {Weinberg}}]{Werk:2014}
{Werk}, J.~K., {Prochaska}, J.~X., {Tumlinson}, J., {et~al.} 2014, \apj, 792,
  8, \dodoi{10.1088/0004-637X/792/1/8}

\bibitem[{{Werk} {et~al.}(2019){Werk}, {Rubin}, {Bish}, {Prochaska}, {Zheng},
  {O{\textquoteright}Meara}, {Lenz}, {Hummels}, \& {Deason}}]{Werk:2019}
{Werk}, J.~K., {Rubin}, K.~H.~R., {Bish}, H.~V., {et~al.} 2019, \apj, 887, 89,
  \dodoi{10.3847/1538-4357/ab54cf}

\bibitem[{{Wiener} {et~al.}(2017){Wiener}, {Pfrommer}, \& {Peng
  Oh}}]{Wiener:2017}
{Wiener}, J., {Pfrommer}, C., \& {Peng Oh}, S. 2017, \mnras, 467, 906,
  \dodoi{10.1093/mnras/stx127}

\bibitem[{{Zahedy} {et~al.}(2019){Zahedy}, {Chen}, {Johnson}, {Pierce},
  {Rauch}, {Huang}, {Weiner}, \& {Gauthier}}]{Zahedy:2019}
{Zahedy}, F.~S., {Chen}, H.-W., {Johnson}, S.~D., {et~al.} 2019, \mnras, 484,
  2257, \dodoi{10.1093/mnras/sty3482}

\bibitem[{{Zhang} {et~al.}(2017){Zhang}, {Thompson}, {Quataert}, \&
  {Murray}}]{Zhang:2017}
{Zhang}, D., {Thompson}, T.~A., {Quataert}, E., \& {Murray}, N. 2017, \mnras,
  468, 4801, \dodoi{10.1093/mnras/stx822}

\end{thebibliography}
\begin{figure*}[ht]
\centering
\includegraphics[width=\textwidth]{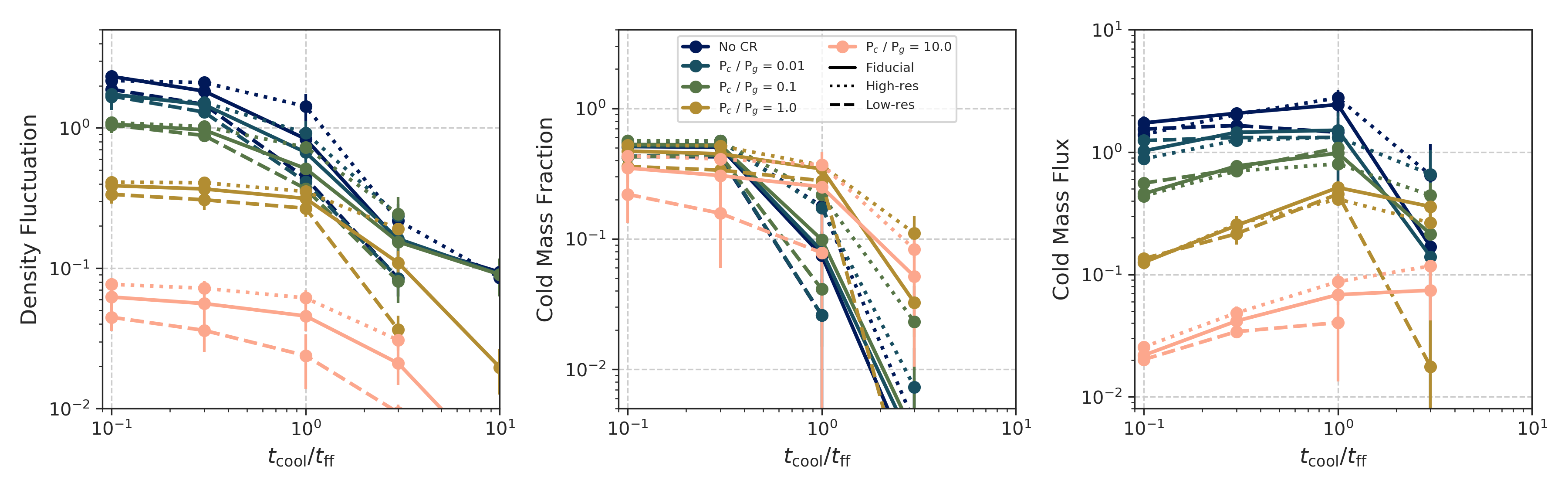}
\caption{ \footnotesize The average density fluctuation ($\langle\delta\rho / \rho\rangle $; \textit{left}), cold mass fraction ($M_{\rm cold} / M_{\rm total}$; \textit{middle}), and cold mass flux ($\dot{M}_{\rm cold} / \dot{M}_{\rm ff}$; \textit{right}) as a function of the initial $\tcff$ for simulations with varying ratios of $P_{\rm c}/P_{\rm g}$.  The solid lines, repeated from Figure~\ref{fig:dens_fluc_tctf_cr}, show simulations with the fiducial resolution ($64\times64\times256$). The dashed lines show simulations with half the fiducial resolution, and the dotted lines show simulations with double the fiducial resolution. The points show each quantity averaged between $t = 4-6t_{\rm cool}$ and measured between 0.8 and 1.2 scale heights. All simulations have an initial $\beta = 100$. The properties of thermal instability are reasonably well converged in our fiducial simulations.}
\label{fig:dens_fluc_tctf_resolution} 
\end{figure*}
\appendix
\section{Impact of Resolution}\label{sec:appendix_resolution}

\begin{figure}
\centering
\includegraphics[width=0.5\textwidth]{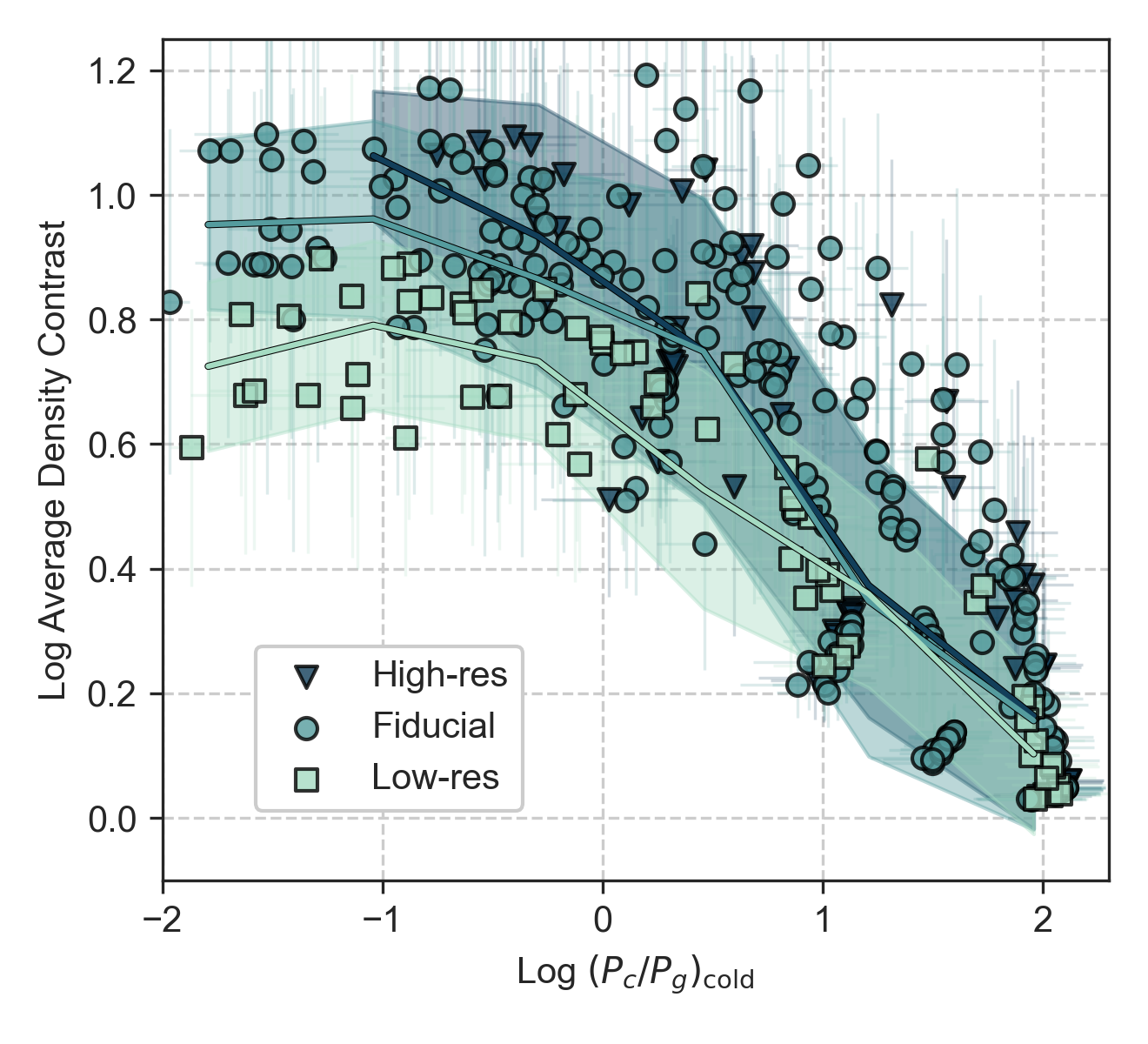}
\caption{ \footnotesize The average density contrast ($\langle\rho_{\rm cold}\rangle / \langle \rho_{\rm hot}\rangle$) as a function of the ratio of cosmic ray pressure to gas pressure in cold gas. The scattered points show all simulations with cosmic ray physics (advection, diffusion, and streaming) run with different resolutions: high resolution (triangles), fiducial resolution (circles) and low resolution (squares). The lines and shaded regions show the corresponding median and one standard deviation. When cosmic ray pressure dominates, cold cloud structure is sufficiently resolved, even with our lowest resolution simulations. }
\label{fig:resolution} 
\end{figure}

The fiducial simulation suite presented in this work resolved the simulation domain of $1 \times 1 \times 4 H$ with $64 \times 64 \times 256$ cells, which corresponds to $\Delta x = 685.2$ pc. Additionally, we have run a subset of the simulations with half and double the resolution. Figure~\ref{fig:dens_fluc_tctf_resolution} shows the density contrast, cold mass fraction, and cold mass flux as a function of the initial $\tcff$ for simulations with different initial cosmic ray pressures. The solid lines show the quantities from simulations with fiducial resolutions (as seen in Figure~\ref{fig:dens_fluc_tctf_cr}), the dashed lines show the same simulations with low resolution, and the dotted lines show simulations with higher resolution. Simulations with our fiducial resolution are reasonably well converged.

Figure~\ref{fig:resolution} shows the density contrast as a function of the cosmic ray pressure ratio $P_{\rm c}/ P_{\rm g}$ in the cold gas for three different resolutions. At low cosmic ray pressures, the density contrast does not converge with resolution. This is expected as cold gas scales ($\ell_{\rm cloudlet} \sim {\rm min}(c_s t_{\rm cool})$) are below our resolution, even with a modified minimum temperature, $T_{\rm min} = 5 \times 10^4\K$. The differences in the median profiles between the fiducial and high-resolution simulations are smaller than those between the fiducial and low-resolution simulations and become negligible when cosmic ray pressure is equal to or greater than the gas pressure. When cosmic ray pressure is ten times the gas pressure, all three resolutions converge. In this case, the low density contrasts correspond to large cloud sizes which are sufficiently resolved, even in our lowest resolution simulation, which has a cell resolution of 1.3 kpc. This result suggests that current galaxy-scale simulations may be able to resolve cold CGM gas in a cosmic ray pressure-dominated halo.

\section{Impact of $T_{\rm min}$}\label{sec:appendix_Tmin}

\begin{figure}
\centering
\includegraphics[width=0.5\textwidth]{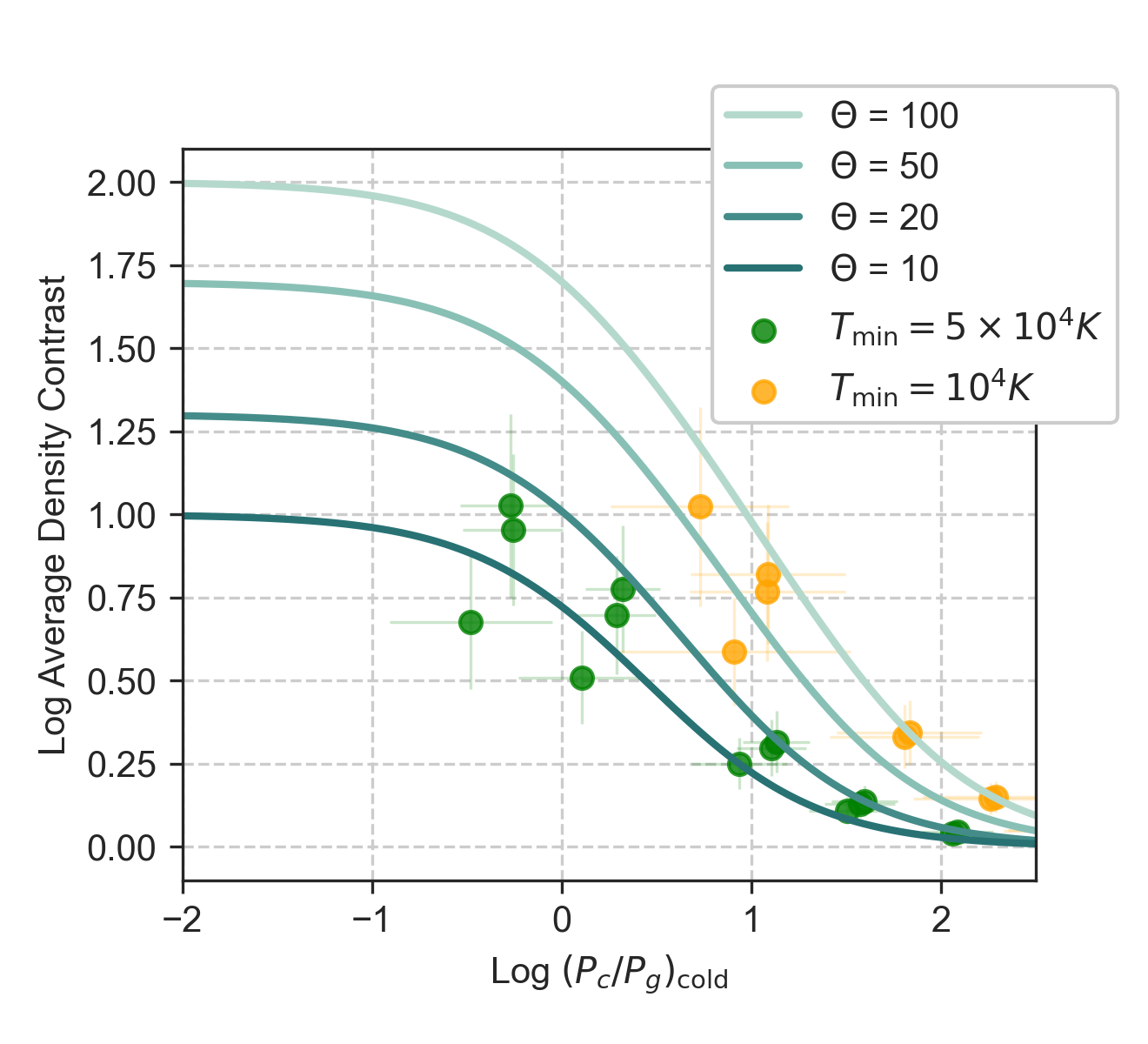}
\caption{ \footnotesize The average density contrast between the cold and hot gas phases ($\langle\rho_{\rm cold}\rangle / \langle \rho_{\rm hot}\rangle$) as a function of the average ratio of $P_{\rm c} / P_{\rm g}$ in cold gas clumps. The points show data averaged from 20 outputs between $t = 4-6 t_{\rm cool}$, measured between 0.8 and 1.2 scale heights. The different colors indicate simulations with different minimum temperatures allowed by the idealized cooling curve:  $T_{\rm min} = 5\times 10^4 \K$ (green), and $T_{\rm min} = 10^4\K$ (orange), corresponding to expected temperature contrasts of $\Theta = T_{\rm cold} / T_0$ of 20 and 50 respectively.  The different curves show predictions for the density contrast using Eq.~\ref{eqn:dens_contrast}, assuming $P_{\rm c} \propto \rho^{4/3}$, for different possible values of the temperature contrast, $\Theta$. Our simulations exactly match the predicted density contrast profiles at high cosmic ray pressures but underpredict the high density contrast needed for purely thermal pressure equilibrium.}
\label{fig:Tmin} 
\end{figure}

The density contrast between cold and hot gas phases depends on the details of the gas cooling curve. Without cosmic rays or magnetic fields, the density contrast between the cold and hot gas phases in an ideal gas will be set by the temperature contrast, $\Theta = T_{\rm hot} / T_{\rm cold}$. In our fiducial simulations, we set the minimum temperature of the gas to be $T_{\rm min} = 5\times 10^4\K$ which roughly corresponds to $\Theta = 20$. Artificially setting the minimum temperature of cold gas allowed us to keep our simulations at a modest resolution, which is necessary for a large-scale parameter study.  However, we do expect that a different choice of $T_{\rm min}$ would change predictions for the density contrast as a function of cosmic ray pressure  presented in Figure~\ref{fig:family_portrait}. We explore the impact of $T_{\rm min}$ on the predicted density contrast in this section. 

Figure~\ref{fig:Tmin} shows the density contrast as a function of the cosmic ray pressure ratio, $P_{\rm c}/P_{\rm g}$ in cold gas. The green lines show the predicted density contrast for a variety of different temperature contrasts, $\Theta$, assuming $P_{\rm c} \propto \rho^{4/3}$ (i.e., inefficient cosmic ray transport; see Eq.~\ref{eqn:dens_contrast}). The scattered points show simulation data for runs with cosmic ray advection, averaged between $4-6t_{\rm cool}$, measured between 0.8 and 1.2 scale heights. Green points show simulations with our fiducial choice of $T_{\rm min} = 5\times 10^4\K$ and orange points show simulations with $T_{\rm min} = 10^4\K$, corresponding to $\Theta \simeq 100$. These simulations were run with no magnetic fields. 

At high cosmic ray pressures, the simulation data points follow their theoretically predicted tracks well.  However, the simulation points underpredict the density contrast with decreasing cosmic ray pressure, especially the orange points which require significantly higher resolution. 

In all cases, the relevant trend is that cosmic rays only have a significant effect on the density contrast when cosmic ray pressure dominates over thermal pressure in the cold gas, $(P_{\rm c}/P_{\rm g})_{\rm cold} \gtrsim 1$. The density contrast approaches zero when $P_{\rm c}/P_{\rm g} \gtrsim 100$. In the intermediate regime, the density contrast decreases following a power law with a slope between -1/3 and -1. Higher temperature contrast between the cold and hot phases necessitates higher cosmic ray pressures to achieve the same decrease in density contrast.

\section{Impact of halo profile}\label{sec:appendix_profile}
\begin{figure*}
\centering
\includegraphics[width=\textwidth]{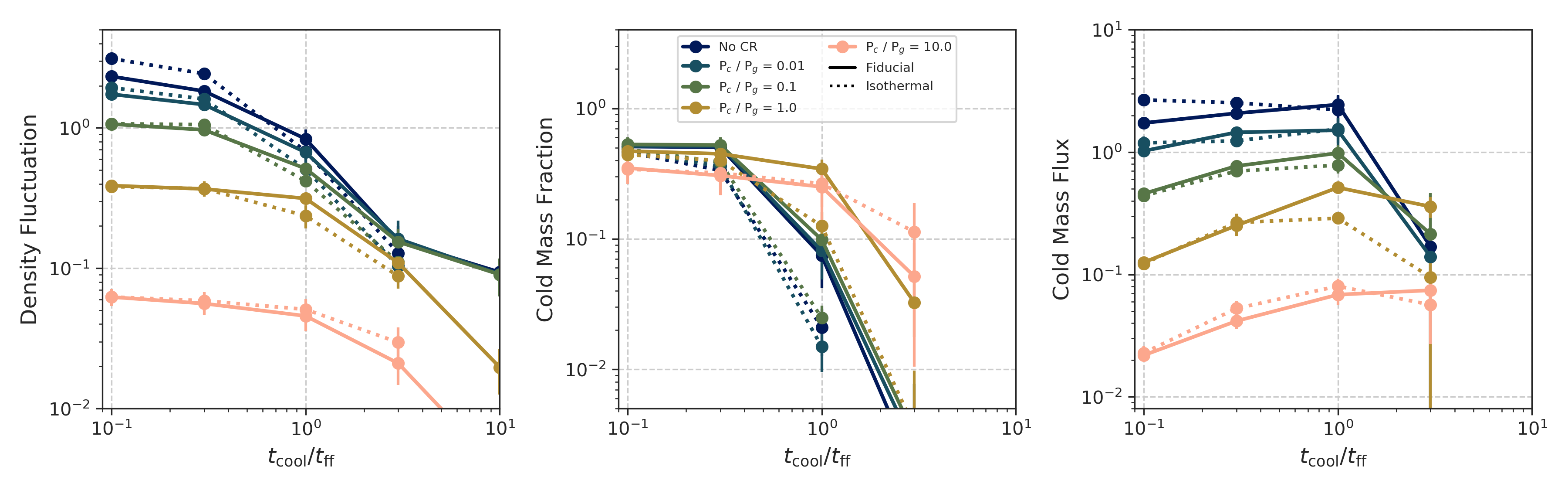}
\caption{ \footnotesize The average density fluctuation ($\langle\delta\rho / \rho\rangle $; \textit{left}), cold mass fraction ($M_{\rm cold} / M_{\rm total}$; \textit{middle}), and cold mass flux ($\dot{M}_{\rm cold} / \dot{M}_{\rm ff}$; \textit{right}) as a function of the initial $\tcff$ for simulations with varying ratios of $P_{\rm c}/P_{\rm g}$.  The solid lines are repeated from Figure~\ref{fig:dens_fluc_tctf_cr} and the dotted lines show the same simulations run with an isothermal gas density and temperature profile. The points show each quantity averaged between $t = 4-6t_{\rm cool}$ and measured between 0.8 and 1.2 scale heights. All simulations have an initial $\beta = 100$. The choice of halo profile does not qualitatively impact our results.}
\label{fig:isothermal} 
\end{figure*}

Our fiducial suite of simulations all had an ``iso-cooling'' profile: the gas density and temperature as a function of height above the midplane follow Eq.~\ref{eqn:isocooling} to ensure that gas cooling time is constant throughout. Although keeping the cooling time constant is advantageous in the context of an idealized parameter study, we by no means expect that all halos follow this narrow range of densities and temperatures.  

To test the effect of the choice of gas profile, we run a subset of the simulations with isothermal initial conditions and compare them against the fiducial simulations in Figure~\ref{fig:isothermal}. Overall, we find that the choice of profile has little effect on the qualitative density fluctuation, cold mass fraction, or cold mass flux. The largest quantitative  deviation from the fiducial simulations is in the cold mass fractions for simulations with $\tcff \geq 1$. This is likely because the gas cooling times in the isothermal profile are shortest near the midplane and cooling happens from the inside-out. Therefore, at late times, we can expect less gas near the midplane of simulations with an isothermal profile. 
\end{document}